%%%%%%%%%%%% Plain TeX %%%%%%%%%%%%%%%%%%%%%%%%%%%%%%%%%%%
\magnification=\magstep1 \overfullrule=0pt 
\advance\hoffset by -0.35truecm   %\baselineskip=18pt
%%%%%%%%%%%%%%%%%%%%%%%%%%%%%%%%%%%%%%%%%%%%%%%%%%%%%%%%%%
%%% The following FONTS are used to create ``blackboard %%
%%% symbols'' for the complex numbers etc.   %%%%%%%%%%%%% 
\font\tenmsb=msbm10       \font\sevenmsb=msbm7
\font\fivemsb=msbm5       \newfam\msbfam
\textfont\msbfam=\tenmsb  \scriptfont\msbfam=\sevenmsb
\scriptscriptfont\msbfam=\fivemsb
\def\Bbb#1{{\fam\msbfam\relax#1}}
\def\R{{\Bbb R}}\def\Z{{\Bbb Z}}
\def\C{{\Bbb C}}\def\N{{\Bbb N}}\def\T{{\Bbb T}}
%%%%%%%%%%%%%%%%%%%%%%%%%%%%%%%%%%%%%%%%%%%%%%%%%%%%%%%%%%
%%% If the fonts are not loadable, one can use the %%%%%%%
%%% following less fance alternatives              %%%%%%%  
%\def\Q{{\bf Q}}\def\R{{\bf R}}\def\Z{{\bf Z}}     %%%%%%% 
%\def\C{{\bf C}}\def\N{{\bf N}}\def\T{{\bf T}}     %%%%%%%
%%%%%%%%%%%%%%%%%%%%%%%%%%%%%%%%%%%%%%%%%%%%%%%%%%%%%%%%%%
%%%  Further fonts used %%%%%%%%%%%%%%%%%%%%%%%%%%%%%%%%%%

\font\gross=cmr10 scaled \magstep2

\font\kap=cmcsc10     
\font\tpwrt=cmtt8 scaled \magstep2
\font\foon=cmr9 scaled \magstep0
%%%%%%%%%%%%%%%%%%%%%%%%%%%%%%%%%%%%%%%%
\font\fatma=cmbxti10
%%%%%%%%%%%%%%%%%%%%%%%%%%%%%%%%%%%%%%%%%%%%%%%%%%%%%%%%%%%
\def\qed{{\vrule height4pt width4pt depth1pt}}
\def\lb{\lbrack}\def\rb{\rbrack}  \def\q#1{$\lb{\sl #1}\,\rb$}
\def\bn{\bigskip\noindent} \def\mn{\medskip\smallskip\noindent}
\def\sn{\smallskip\noindent} 
\def\hbn{\hfil\break\noindent}
\def\lra{\longrightarrow} 
\def\lla{\longleftarrow}  
\def\a{\hbox{$\cal A$}}   \def\b{\hbox{$\cal B$}}
\def\m{\hbox{$\cal M$}}   
\def\h{\hbox{$\cal H$}}   \def\e{\hbox{$\cal E$}}
   \def\S{{\cal S}}
\def\cald{{\cal D}}       \def\calbd{\overline{{\cal D}}}
\def\one{{\bf 1}}          
       
     \def\Lim{\mathop{\rm Lim}}

\def\d{\hbox{$\cal D$}}   \def\bard{\overline{\hbox{$\cal D$}}}
\def\ttd{\hbox{{\tt d}}}  \def\ttdst{\hbox{${\tt d}^*$}}
\def\td{\hbox{$\widetilde{{\tt d}}$}}
\def\tdst{\hbox{$\widetilde{\tt d}{}^*$}}
\def\dd{{\tt d}}
\def\df#1{\hbox{$\Omega^{#1}_D(\a_0)$}}
\def\uf#1{\hbox{$\Omega^{#1}(\a_0)$}}

\def\barint{{\int \kern-9pt {\rm -} \ }}
\def\Barint{{\int \kern-10pt - \ }}
\def\barintbeta{{\int_{{\scriptscriptstyle \beta}} 
     \kern-11pt {\rm -} \ }}
\def\Barintbeta{{\int_{\beta} \kern-12pt - \ }}
\def\Neqone{\hbox{{\bf{\fatma N}$\;$=$\;$1\ }}}
\def\Noneone{\hbox{{\bf{\fatma N}$\;$=$\;$(1,1)\ }}}

\def\Ntwotwo{\hbox{{\bf{\fatma N}$\;$=$\;$(2,2)\ }}}
\def\Nfourfour{\hbox{{\bf{\fatma N}$\;$=$\;$(4,4)\ }}}
\def\MM{\hbox{$M$} \kern -14.5pt \hbox{$M$}\ }
\def\CC{\hbox{${}^c$} \kern -7.5pt \hbox{${}^c$}}
\def\CCVV{\hbox{$^c(V)$} \kern -24pt \hbox{$^c(V)$}}
\def\oddots{\mathinner{\mkern1mu\raise1pt\vbox{\kern7pt\hbox{.}}
  \mkern2mu\raise4pt\hbox{.}\mkern2mu\raise7pt\hbox{.}\mkern1mu}}

\def\sqr#1#2{{\vcenter{\vbox{\hrule height.#2pt\hbox{\vrule width.#2pt 
height #1pt \kern#1pt \vrule width.#2pt} \hrule height.#2pt}}}}
\def\square{\mathchoice\sqr{5.5}{35}\sqr{5.5}{35}\sqr{2.1}3\sqr{1.5}3}
\def\llcorner{\raise1.7pt\vbox{
    \hbox{\vrule height1.9pt width.3pt $\underline{\phantom{n}}\,$}}}
\setbox1=\hbox{$\lra$}
\def\frarrow{\raise0.36pt\hbox{$\lra$}\hskip-\wd1
\raise0.18pt\hbox{$\lra$}\hskip-\wd1\hbox{$\lra$}\hskip-\wd1
\raise-0.18pt\hbox{$\lra$}\hskip-\wd1\raise-0.35pt\hbox{$\lra$}}
\def\fuarrow{\hbox{$\uparrow$}\kern-4.55pt\hbox{$\uparrow$} 
\kern-4.56pt\hbox{$\uparrow$}\kern-4.5pt\raise.17pt\hbox{$|$}}
\def\frtlarrow{\raise-2.5pt\hbox{$\frarrow$}
\hskip-\wd1{\raise2.5pt\hbox{$\lla$}}}

\def\sqr#1#2{{\vcenter{\vbox{\hrule height.#2pt
       \hbox{\vrule width.#2pt height #1pt \kern#1pt 
         \vrule width.#2pt}\hrule  height.#2pt}}}}
\def\square{\mathchoice\sqr{5.5}{35}\sqr{5.5}{35}\sqr{2.1}3\sqr{1.5}3}

\def\Omti{\hbox{$\widetilde{\Omega}^k({\cal A})$} 
  \kern -32pt\hbox{$\widetilde{\Omega}^k({\cal A})$}\ }
\def\Omtipar{\hbox{$\widetilde{\Omega}^{\ 1}_{\partial,
   \overline{\partial}}  ({\cal A})$} 
  \kern -38.7pt\hbox{$\widetilde{\Omega}^{\ 1}_{\partial,
  \overline{\partial}}({\cal A})$}\ }

\def\oddots{\mathinner{\mkern1mu\raise1pt\vbox{\kern7pt\hbox{.}}
  \mkern2mu\raise4pt\hbox{.}\mkern2mu\raise7pt\hbox{.}\mkern1mu}}
\def\ttR{\hbox{{\tpwrt R}}}\def\ttT{\hbox{{\tpwrt T}}}
\def\ttRic{\hbox{{\tpwrt Ric}}}\def\ttr{\hbox{{\tpwrt r}}}
\def\omdt{{\raise-1.65pt\hbox{$\widetilde{\phantom{\widetilde{\Omega}}
  }$}\kern - 7.4pt\hbox{$\widetilde{\Omega}$}}}
\def\nadt{{\raise-1.65pt\hbox{$\widetilde{\phantom{\widetilde{\nabla}}
 }$}\kern - 8.5pt\hbox{$\widetilde{\nabla}$}}}
\def\pabt{{\raise-1.2pt\hbox{$\widetilde{\phantom{\overline{\partial}}
 }$}\kern - 6.2pt\hbox{$\overline{\partial}$}}}
\def\scpbp{{\scriptscriptstyle \!\partial\!,\!\bar\partial}} 
\def\c{\hbox{{\tpwrt c}}}
\def\cedille#1{\setbox0=\hbox{#1}\ifdim\ht0=1ex \accent'30 #1%
 \else{\ooalign{\hidewidth\char'30\hidewidth\crcr\unbox0}}\fi}
\def\gaw{Gaw\cedille edzki}
\def\qdq{\quad\dotfill\quad}
\def\phtitle{\centerline}
\def\cinftyg{C^{\infty}(G)}
\def\Omeinsalpha{\hbox{${\Omega}^{1}_{D}({\cal A}_{\alpha})$} 
  \kern -35.65pt\hbox{${\Omega}^{1}_{D}({\cal A}_\alpha)$}\ }
\def\Omeinsnull{\hbox{${\Omega}^{1}_{D}({\cal A}_{0})$} 
  \kern -34.2pt\hbox{${\Omega}^{1}_{D}({\cal A}_0)$}\ }
\def\hbn{\hfill\break\noindent}
%%%%%%%%%%%%%%%%%%%%%%%%%%%%%%%%%%%%%%%%%%%%%%%%%%%%%%%%%%%%
%%%%%%%%%%%%%
{\nopagenumbers
\line{ETH-TH/96-45 \hfill HUTMP-98/B374}
\line{math-ph/9807006  \hfill IHES/P/98/46}
\sn
\bn\bn\bn%\bn\bn
\centerline{{\gross 
Supersymmetric Quantum Theory}}
\bn\sn
\centerline{{\gross and}}
\bn\sn
\centerline{{\gross Non-Commutative Geometry}}
\bn\bn
\vfill \centerline{by}\bn\bn \vfill
\centerline{{\bf J.\ Fr\"ohlich${}^1$, O.\ Grandjean${}^2$}\ and 
   {\bf A.\ Recknagel${}^{3}$}}
\bn\bn
\phtitle{${}^1\,$Institut f\"ur Theoretische Physik, ETH-H\"onggerberg}
\phtitle{\phantom{${}^1\,$}CH-8093 Z\"urich, Switzerland}
\mn
\phtitle{${}^2\,$Department of Mathematics, Harvard University}\hbn
\phtitle{\phantom{${}^2\,$}Cambridge,  MA 02138, USA}
\mn
\phtitle{${}^3\,$Institut des Hautes \'Etudes Scientifiques}
\phtitle{\phantom{${}^3\,$}35, route de Chartres, F-91440 Bures-sur-Yvette, 
                 France}
\bn\bn\vfill\vfill\vfill\bn
\centerline{\bf Abstract}
\bigskip\narrower
\noindent Classical differential geometry can be encoded in spectral 
data, such as Connes' spectral triples, involving supersymmetry 
algebras. In this paper, we formulate non-commutative geometry 
in terms of supersymmetric spectral data. This leads to generalizations  
of Connes' non-commutative spin geometry encompassing 
non-commutative Riemannian, symplectic, complex-Hermitian and (Hyper-)
K\"ahler geometry. A general framework for non-commutative geometry 
is developed from the point of view of supersymmetry and illustrated 
in terms of examples. In particular, the  non-commu\-tative  torus 
and the non-commutative 3-sphere are studied in some detail. 
\bn\bn
\bn\bn
\vfill\vfill\vfill
\smallskip
\medskip\noindent 
\ \ e-mail:\ $\,$juerg@itp.phys.ethz.ch, \ $\,$grandj@math.harvard.edu,\ $\,$
anderl@ihes.fr
\eject
\vfill $$\phantom{Leerseite}$$\vfill
\eject}
%%%%%%%%%%%%%%%%%%%%%%%%%%%%%%%%%%%%%%%%%%%%%%%%%%%%%%%%%%%
\pageno=-1\parindent=32pt
\phantom{xxx}\bn
\leftline{\bf Contents}
\bn\mn
1. Introduction\qdq \ 1
\medskip
\bn
2. Spectral data of non-commutative geometry\qdq \ 5
\medskip
\item{2.1} The $N=1$ formulation of non-commutative geometry\qdq \ 5
\item{2.2} The $N=(1,1)$ formulation of non-commutative
    geometry\qdq 21
\item{2.3} Hermitian and K\"ahler non-commutative geometry\qdq 31
\item{2.4} The $N=(4,4)$ spectral data\qdq 39
\item{2.5} Symplectic non-commutative geometry\qdq 39
\bn
3. The non-commutative 3-sphere\qdq 42
\medskip
\item{3.1} The $N=1$ data associated to the 3-sphere\qdq 42
\item{3.2} The topology of the non-commutative 3-sphere\qdq 45
\item{3.3} The geometry of the non-commutative 3-sphere\qdq 51
\item{3.4} Remarks on $N=(1,1)$\qdq 54
\bn
4. The non-commutative torus\qdq 58
\medskip
\item{4.1} The classical torus\qdq 58
\item{4.2} Spin geometry ($N=1$)\qdq 59
\item{4.3} Riemannian geometry ($N=(1,1)$)\qdq 63
\item{4.4} K\"ahler geometry ($N=(2,2)$)\qdq 66
\bn
5. Directions for future work\qdq 68
\bn
References \qdq 73
\bn
\vfill\eject 
{\nopagenumbers\phantom{xxx}\vfill\eject}
%%%%%%%%%%%%%%%%%%%%%%%%%%%%%%%%%%%%%%%%%%%%%%%%%%%%%%%%%%%
\pageno=1
\leftline{\bf 1. Introduction}
\bn
The study of highly singular geometrical spaces, such as the space of 
leaves of certain foliations, of discrete spaces, and the study of quantum 
theory have led A.\ Connes to develop a general theory of non-commutative 
geometry, involving non-commutative measure theory, cyclic cohomology, 
non-commutative differential topology and spectral calculus, \q{Co1\!-\!5}. 
A broad exposition of his theory and a rich variety of interesting 
examples can be found in his book \q{Co1}. Historically, the 
first examples of non-commutative spaces carrying geometrical 
structure emerged from non-relativistic quantum mechanics, as 
discovered by Heisenberg, Born, Jordan, Schr\"odinger and Dirac. 
Mathematically speaking, non-relativistic quantum mechanics is the 
theory of quantum phase spaces, which are non-commutative deformations 
of certain classical phase spaces (i.e., of certain symplectic 
manifolds), and it is the theory of dynamics on quantum phase spaces. 
Geometrical aspects of quantum phase spaces and supersymmetry entered the 
scene {\sl implicitly} in Pauli's theory of the non-relativistic, 
spinning electron and in the theory of non-relativistic positronium. 
Later on, the mathematicians discovered Pauli's and Dirac's theories 
of the electron as a powerful tool in algebraic topology and differential
geometry. 
\sn
In a companion paper \q{FGR1}, hereafter referred to as I, we have described a
formulation of classical differential geometry in terms of the spectral 
data of non-relativistic, supersymmetric quantum theory, in particular 
in terms of the quantum theory of the non-relativistic electron and 
of positronium propagating on a general (spin$^c$) Riemannian manifold. 
The work in I is inspired by Connes' fundamental work \q{Co1\!-\!5}, and 
by Witten's work on supersymmetric quantum theory and its applications in
algebraic topology \q{Wi1,2}; it attempts to merge these two threads 
of thought. Additional inspiration has come from the work in \q{AG,FW,AGF,HKLR} 
on the relation between index theory and supersymmetric quantum theory 
and on supersymmetric non-linear $\sigma$-models, as well as from the work 
by Jaffe and co-workers on connections between supersymmetry and cyclic 
cohomology \q{Ja1\!-\!3}. To elucidate the roots of some of these ideas in 
Pauli's non-relativistic quantum theory of the electron and of positronium 
has proven useful and suggestive of various generalizations. 
\sn
The work described in the present paper has its origins in an attempt 
to apply the methods of non-commutative geometry to exploring the geometry 
of string theory, in particular of superstring vacua; see \q{CF,FG}. 
In trying to combine quantum theory with the theory of gravitation, 
one observes that it is impossible to localize events in space-time 
arbitrarily precisely, and that, in a compact region of space-time, one can 
only resolve a finite number of distinct events \q{DFR}. One may then 
argue, heuristically, that space-time itself must have quantum-mechanical 
features at distance scales of the order of the Planck length, and that 
space-time and matter should be merged into a fundamental quantum theory of 
space-time-matter. Superstring theory \q{GSW} is a theoretical framework 
incorporating some of the features necessary for a unification of 
quantum theory and the theory of gravitation. Superstring vacua are 
described by certain superconformal field theories, see e.g.\ \q{GSW}. 
The intention of the program formulated in \q{CF,FG} is to reconstruct 
space-time geometry from algebraic data of superconformal field theory. 
In the study of concrete examples, one observes that, in general, the 
target spaces (space-times) of superconformal field theories are 
{\sl non-commutative} geometrical spaces, and the tools of Connes' 
non-commutative geometry become essential in describing their geometry. 
This observation has been confirmed more recently in the theory of 
D-branes \q{Pol,Wi4}. 
\sn
The purpose of this paper is to cast some of the tools of non-commutative
(differential) geometry into a form that makes connections to supersymmetric 
quantum theory manifest and that is particularly useful for applications 
to superconformal field theory. The methods and results of this paper are 
mathematically precise. Applications to physics are not treated here; but 
see e.g.\ \q{FGR2}. Instead, the general formalism developed in this paper 
is illustrated by an analysis of the geometry of the non-commutative torus 
and of the fuzzy 3-sphere; more details can be found in \q{Gr}. 
\mn 
Next, we sketch some of the key ideas underlying our approach to 
non-commutative geometry; for further background see also part I and 
\q{FGR2}. 
\sn
Connes has shown how to formulate classical geometry in terms 
of algebraic data, so-called {\sl spectral triples}, involving 
a commutative algebra $\a=C^{\infty}(M)$ of (smooth) functions on the smooth 
manifold 
$M$ under consideration, a Hilbert space $\h$ of spinors over $M$ 
on which the algebra \a\ acts by bounded operators, and a self-adjoint 
Dirac operator $D$ on \h\ satisfying certain properties with respect 
to \a. As explained in \q{Co1}, it is possible to extract complete 
geometrical information about $M$ from the spectral 
triple $(\a,\h,D)$. \hfill\break
\noindent The definition of spectral triples involves, in the classical 
case, a Clifford action on certain vector bundles 
over $M$, e.g.\  the spinor bundle or the bundle of differential 
forms. As was recalled in ref.\ I, the latter bundle actually 
carries two anti-commuting Clifford actions -- which can be used to 
define two Dirac-K\"ahler operators, $\d$ and $\bard$. It turns out that the  
algebraic relations between these operators are precisely 
those of the two supercharges of $N=(1,1)$ supersymmetric 
quantum mechanics (see part I, especially section 3, for the precise 
meaning of the terminology): These relations are 
$\{\,\d,\bard\,\}=0$ and $\d^2=\bard{}^2$. The commutators 
$\lb\,\d,a\,\rb$ and $\lb\,\bard,a\,\rb$, for arbitrary $a\in\a$, 
extend to bounded operators (anti-commuting sections of two Clifford 
bundles) acting on the Hilbert space \h\ of square-integrable differential 
forms. Furthermore, if the underlying manifold $M$ is compact, the 
operator $\exp(-\varepsilon \d^2)$ is trace-class for any 
$\varepsilon >0$. One may then introduce a nilpotent operator 
$$
d := \d -i \bard \ ,
$$
which turns out to correspond to exterior differentiation of 
differential forms. \hfill\break
\noindent
{}From the $N=(1,1)$ supersymmetric spectral data $(\a,\h,\d,\bard)$ 
just described, one can reconstruct the de Rham-Hodge theory and the 
Riemannian geometry of smooth (compact) Riemannian manifolds. 
\sn
$N=(1,1)$ supersymmetric spectral data are a variant of Connes' approach 
involving spectral triples. They are very natural from the point of view 
of supersymmetric quantum theory and encode the differential geometry 
of Riemannian manifolds (not required to be spin$^c$ manifolds). 
\sn
In a formulation of differential geometry in terms of spectral data 
$(\a,\h,\d,\bard,\ldots)$ with supersymmetry, additional geometrical structures, 
e.g.\ a symplectic or complex structure, appear in the form of {\sl global 
gauge symmetries} commuting with the elements of \a\  but acting 
non-trivially on the Dirac-K\"ahler operators $\d$ and $\bard$; see part I.  
For example, a global gauge symmetry group containing U(1)$\,\times\,$U(1) 
generates four  Dirac-K\"ahler operators -- the ``supercharges'' of 
$N=(2,2)$ supersymmetry -- from $\d$ and $\bard$ and identifies the 
underlying manifold $M$ as a {\sl K\"ahler manifold}. A global gauge 
symmetry group containing SU(2)$\,\times\,$SU(2) leads to eight 
supercharges generating an $N=(4,4)$ supersymmetry algebra and is 
characteristic of Hyperk\"ahler geometry; see also \q{AGF,HKLR}. 
Complex-Hermitian and symplectic geometry are encoded in $N=(2,2)$ 
supersymmetric spectral data with partially {\sl broken} supersymmetry. 
A systematic classification of different types of differential geometry 
in terms of supersymmetric spectral data extending the $N=(1,1)$ data 
of Riemannian geometry has been described in I (see section I$\,$3
for an overview, and \q{FGR2}). 
\mn
In this paper, we generalize these results from classical to 
non-commutative geometry, starting from the simple prescription to replace 
the commutative algebra of functions $C^{\infty}(M)$ over a classical 
manifold by a general, possibly non-commutative ${}^*$-algebra \a\  
satisfying certain properties. 
Section 2 contains general definitions and introduces 
various kinds of spectral data: We start with an exposition of 
Connes' non-commutative spin geometry; most of the material can 
be found in \q{Co1}, but we add some details on metric aspects 
ranging from connections over curvature and torsion to non-commutative
Cartan structure equations. In subsection 2.2, we introduce spectral 
data with $N=(1,1)$ supersymmetry that naturally lead to a non-commutative 
analogue of the de Rham complex of differential forms. 
Moreover, this ``Riemannian'' formulation of non-commutative 
geometry allows for immediate specializations to spectral 
data with extended supersymmetry -- which, in the classical case, 
correspond to manifolds carrying complex, K\"ahler, Hyperk\"ahler 
or symplectic structures. Spectral data with higher supersymmetry are 
treated in subsections 2.3 -- 2.5. In subsection 2.2.5, we discuss 
the relationship between spectral triples, as defined by Connes,  
and spectral data with $N=(1,1)$ supersymmetry: Whereas in the 
classical case, one can always pass from one description of a 
smooth manifold to the other, the situation is not quite as clear in 
the non-commutative framework. We propose a procedure how to construct 
$N=(1,1)$ data from a spectral triple -- heavily relying on Connes' notion 
of a real structure \q{Co4} --, but the construction is not complete for 
general spectral triples. Furthermore, subsection 2.2.6 contains 
proposals for definitions of non-commutative manifolds and 
non-commutative phase spaces, as suggested by the study of $N=(1,1)$ 
spectral data and by notions from quantum physics. 
\mn
In sections 3 and 4 we discuss two 
examples of non-commutative spaces, name\-ly the ``fuzzy 3-sphere'' 
and the non-commutative torus. The choice of the latter example 
does not require further explanation since 
it is one of the classic examples of a non-commutative space; see 
e.g.\ \q{Co1,Co5,Ri}. Here we add a description of the non-commutative 
2-torus in terms of spectral data with $N=(1,1)$ and $N=(2,2)$ 
supersymmetry, thus showing that this space can be endowed with a 
non-commutative Riemannian and a non-commutative K\"ahler 
structure. This is not too surprising, 
since the non-commutative torus can be regarded as a deformation of 
the classical flat torus. The calculations 
in section 4 also provide an example where the general ideas of 
subsection 2.2.5 on how to construct $N=(1,1)$ from $N=1$ spectral 
data can be carried out completely. \hfill\break
\noindent The other example, the non-commutative 3-sphere discussed 
in section 3 (see also \q{Gr}), represents a generalization of 
another prototype non-commutative geometrical space, 
name\-ly the fuzzy 2-sphere \q{Ber,Ho,Ma,GKP}. 
We choose to study the 3-sphere for the following reasons: First, 
in contrast to the fuzzy 2-sphere and the non-commutative torus, 
it cannot be viewed as a quantization of a classical phase 
space. Second, it is the simplest example of a series of quantized
spaces arising from so-called Wess-Zumino-Witten-models -- conformal 
field theories associated to non-linear $\sigma$-models with 
compact simple Lie groups as target manifolds, see \q{Wi3}. 
There is reason to expect that the spectral 
data arising from other WZW-models -- see \q{FG,FGR2} for a discussion 
-- can be treated essentially by the same methods as the fuzzy 
3-sphere associated to the group SU(2). \hfill\break
\noindent In view of the conformal field 
theory origin, one is led to conjecture that, as a non-commutative 
space, the non-commutative 3-sphere describes the non-commutative 
geometry of the quantum group 
$U_q(sl_2)$, for $q={\rm exp}(2\pi i/{k+2})$ where $k \in \Z_+$ is the 
level of the WZW-model. The parameter $k$ appears in the 
spectral data of the non-commutative 3-sphere in a natural way. 
One may expect that the fuzzy 3-sphere can actually be 
defined for arbitrary values of this parameter, since the same is true 
for the quantum group. As in the example of the non-commutative torus with 
rational deformation parameter, a truncation of the algebra of ``functions''
occurs for the special values $k\in\Z_+$, leading to  
the finite-dimensional matrix algebras used in section 3.
\mn
In section 5, we conclude with a list of open problems arising 
naturally from our discussion. In particular, 
we briefly comment on other, string theory motivated applications 
of non-commutative geometry; see also \q{FG,FGR2}. 
\mn
The present text is meant as a companion paper to I: Now and 
then, we will permit ourselves to refer to \q{FGR1} for technical 
details of proofs which proceed analogously to the classical case. 
More importantly, the study of classical geometry in part I provides 
the best justification -- besides the one of naturality -- of the 
expectation that our classification of (non-commutative) geometries 
according to the supersymmetry content of the spectral data leads to 
useful and fruitful definitions of non-commutative geometrical 
structure.  
\bn\vfill
\bn{\bf Acknowledgments}
\bn 
We thank A.\ Connes for useful criticism and suggestions that led to 
improvements of this paper. We thank O.\ Augenstein, A.H.\ Chamseddine, 
G.\ Felder and K.\ \gaw\ for previous collaborations that generated many 
of the ideas underlying our work. Useful discussions with A.\ Alekseev 
are gratefully acknowledged. We thank the referee for pointing 
out some imprecisions in Sect.\ 2.2.5.
\hbn
The work of O.G.\ is supported in part by the Department of Energy under 
Grant DE-FG02-94ER-25228 and by the National Science Foundation under 
Grant DMS-94-24334.
\hbn
The work of A.R.\ has been supported in part by the Swiss National 
Foundation. 
\eject
%%%%%%%%%%%%%%%%%%%%%%%%%%%%%%%%%%%%%%%%%%%%%%%%%%%%%%%%
\leftline{\bf 2. Spectral data of non-commutative geometry}
\bn
In the following, we  generalize the notions of part I {}from 
classical differential geometry to the non-commutative setting. 
The classification of geometrical structure according 
to the ``supersymmetry content'' of the relevant spectral data, 
which was uncovered in \q{FGR1}, 
will be our guiding principle. In the first part, we 
review Connes' formulation of non-commutative geometry 
using a single generalized Dirac        
operator, whereas, in the following subsections, spectral data 
with realizations of some genuine supersymmetry algebras will 
be introduced, allowing us to define non-commutative 
generalizations of Riemannian, complex, K\"ahler and 
Hyperk\"ahler, as well as of symplectic geometry. 
\bn\bn 
%%%%%%%%%%%%%%%%%%%%%%%%%%%%%%%%%%%%%%%%%%%%%%%%%%%%%%%%%%%%%%%
{\bf 2.1 The \Neqone formulation of non-commutative geometry}
\bn
This section is devoted to the non-commutative generalization 
of an algebraic description of spin geometry -- and, according
to the results of section I$\,$2, of general Riemannian geometry --
following the ideas of Connes \q{Co1}. The first two subsections 
contain the definition of abstract $N=1$ spectral data and of 
differential forms. In subsection 2.1.3, we describe a 
notion of integration which leads us to a definition of 
square integrable differential forms. After having introduced 
vector bundles and Hermitian structures in subsection 2.1.4, 
we show in subsection 2.1.5 that the module of square integrable 
forms always carries a generalized Hermitian structure. We then 
define connections, torsion, and Riemannian, Ricci 
and  scalar curvature in the next two subsections. Finally, in 
2.1.8, we derive non-commutative Cartan structure equations. 
Although much of the material in section 2.1 is contained (partly 
in much greater detail) in Connes' book \q{Co1}, it is reproduced 
here because it is basic for our analysis in later sections and
because we wish to make this paper accessible to non-experts. 
\bn\bn
%%%%%%%%%%%%%%%%%%%%%%%%%%%%%%%%%%%%%%%%%%%%%%%%%%%%%%%%%%%%%%%%%
{\bf 2.1.1 The \Neqone spectral data} 
\bn
{\bf Definition 2.1}\quad A quadruple $(\a,\h, D, \gamma)$ will 
be called a set of $N=1$ {\sl (even) spectral data} if 
\smallskip
\item {1)} \h\ is a separable Hilbert space; 
\smallskip
\item {2)} \a\ is a unital ${}^*$-algebra acting faithfully on \h\ 
by bounded operators;
\smallskip
\item {3)} $D$ is a self-adjoint operator on \h\  such that 
\itemitem {$i)$} for each $a\in\a$, the commutator $\lb\,D,a\,\rb$
defines a bounded operator on \h, 
\itemitem {$ii)$} the operator $\exp(-\varepsilon D^2)$ is trace class 
for all $\varepsilon >0\,$; 
\smallskip
\item {4)} $\gamma$ is a $\Z_2$-grading on \h, i.e., 
$\gamma = \gamma^* = \gamma^{-1}$, such that 
$$
\{\,\gamma,D\,\} = 0\ ,\quad 
\lb\, \gamma,a\,\rb = 0 \quad{\rm for\ all}\ \ a\in \a. 
$$ 
\mn
As mentioned before, in non-commutative geometry \a\  plays the role of 
the ``algebra of functions over a non-commutative space''. The 
existence of a unit in \a, together with property $3\,ii)$ above,  
reflects the fact that we are dealing with ``compact'' non-commutative 
spaces. Note that if the Hilbert space \h\ is infinite-dimensional, 
condition $3\,ii)$ implies that the operator $D$ is unbounded. By 
analogy with classical differential geometry, $D$ is interpreted as a 
(generalized) Dirac operator. 
\sn  
Also note that the fourth condition in Definition 2.1 does not impose 
any restriction on $N=1$ spectral data: In fact, given a triple 
$(\widetilde{\cal A}, \widetilde{\cal H}, \widetilde{D})$ satisfying 
the properties 1 - 3 {}from above, we can define a  set of $N=1$ 
{\sl even} spectral data $(\a,\h, D, \gamma)$ by setting   
$$\eqalign{
\h &= \widetilde{\cal H} \otimes \C^2\ ,\quad\quad 
\a = \widetilde{\cal A} \otimes \one_2\ ,
\cr
D &= \widetilde{D} \otimes \tau_1\ ,\quad\quad\  
\gamma\, = \one_{\tilde{\scriptscriptstyle \cal H}}\otimes \tau_3\ ,
\cr}$$
where $\tau_i$ are the Pauli matrices acting on $\C^2$. 
\bn\bn
%%%%%%%%%%%%%%%%%%%%%%%%%%%%%%%%%%%%%%%%%%%%%%%%%%%%%%%%%%%%%%%%
{\bf 2.1.2 Differential forms}
\bn
The construction of differential forms follows the same lines
as in classical differential geometry: We define the unital, 
graded, differential ${}^*$-algebra of universal forms, 
$\Omega^{\bullet}(\a)$, as in \q{Co1,CoK}: 
$$
\Omega^{\bullet}(\a) = \bigoplus_{k=0}^{\infty} \Omega^{k}(\a)\ ,
\quad\ \Omega^{k}(\a) := \{\,\sum_{i=1}^N a_0^i \delta a_1^i \cdots
\delta a_k^i \,|\, N\in \N,\; a_j^i \in \a\,\}\ ,\phantom{xx} 
\eqno(2.1{\rm a})$$ 
where $\delta$ is an abstract linear operator satisfying $\delta^2=0$ 
and the Leibniz rule. Note that,   
even in the classical case where $\a = C^{\infty}(M)$ for some smooth 
manifold $M$, no relations ensuring 
(graded) commutativity of $\Omega^{\bullet}(\a)$ are imposed. 
The complex conjugation of functions over $M$ 
is now to be replaced by the ${}^*$-operation of \a. We define  
$$
(\delta a)^* = - \delta(a^*)
\eqno(2.1{\rm b})$$
for all $a\in\a$. With the help of the (self-adjoint) 
generalized Dirac operator $D$, we 
introduce a ${}^*$-representation $\pi$ of $\Omega^{\bullet}(\a)$ 
on \h, 
$$
\pi(a) = a\ ,\quad\quad \pi(\delta a) = \lb\,D,a\,\rb \ ,  
$$
cf.\ \q{Co1} or eq.\ (I$\,$2.12). 
A graded ${}^*$-ideal $J$ of $\Omega^{\bullet}(\a)$ is defined by 
$$
J := \bigoplus_{k=0}^{\infty} J^k\ ,\quad\ \ 
J^k := {\rm ker}\,\pi\,|_{\Omega^k({\cal A})}\ .
\eqno(2.2)$$
Since $J$ is not a differential ideal, the graded quotient 
$\Omega^{\bullet}(\a)/J$ does not define a differential algebra 
and thus does not yield a satisfactory definition of the algebra 
of differential forms. This problem is solved as in the classical 
case.
\mn
{\bf Proposition 2.2}\ \q{Co1}\quad The graded sub-complex 
$$
J +\delta J=\bigoplus_{k=0}^{\infty}\,\bigl(J^k+\delta J^{k-1}\bigr)\ ,
$$
where $J^{-1} := 0$ and $\delta$ is the universal differential in 
$\Omega^{\bullet}(\a)$, is a two-sided graded differential ${}^*$-ideal 
of $\Omega^{\bullet}(\a)$. 
\mn
%{\kap Proof}: Since $\delta^2=0$, it is clear that $J+\delta J$ is 
%closed under $\delta$. We show that it is a two-sided ideal. 
%By linearity, it is sufficient to consider homogeneous elements. 
%Thus, let $\omega\in J^k$, $\eta\in J^{k-1}$, and 
%$\phi\in\Omega^p(\a)$; then the Leibniz rule gives 
%$$
%\phi\,(\omega+\delta\eta) = \phi\,\omega 
%- (-1)^p (\delta\phi)\eta + (-1)^p \delta(\phi\,\eta)\ .
%$$
%Since the first two terms are in $J^{p+k}$ and the last one is in 
%$\delta J^{p+k-1}$, we have $\phi(\omega+\delta\eta) \in J + \delta J$. 
%Analogously, we compute 
%$$
%(\omega+\delta\eta)\,\phi = \omega\,\phi - (-1)^{k-1} \eta\,\delta\phi 
%+\delta(\eta\,\phi)
%$$
%which is again  an element of $J + \delta J$. \hfill\qed
\mn
We define the unital graded differential ${}^*$-algebra of 
differential forms, $\Omega_D^{\bullet}(\a)$, as the graded 
quotient $\Omega^{\bullet}(\a)/(J+\delta J)$, i.e., 
$$
\Omega_D^{\bullet}(\a) := \bigoplus_{k=0}^{\infty}\, \Omega_D^k(\a)\ ,
\quad\ \ 
\Omega_D^k(\a) := \Omega^k(\a)/(J^k+\delta J^{k-1})\ .
\eqno(2.3)$$
\sn    Since 
$\Omega_D^{\bullet}(\a)$ is a graded algebra, each $\Omega^k_D(\a)$ 
is, in particular, a bi-module over $\a=\Omega_D^0(\a)$. 
\sn 
Note that $\pi$ does {\sl not} determine a 
representation of the algebra (or, for that matter, of the space) 
of differential forms $\Omega_D^{\bullet}(\a)$ on the Hilbert 
space \h: A differential $k$-form is an equivalence class 
$\lb\omega\rb\in\Omega_D^{k}(\a)$ with some representative 
$\omega\in\Omega^k(\a)$, and $\pi$ maps this class to a {\sl set} of 
bounded operators on \h, namely 
$$
\pi\bigl(\lb\omega\rb\bigr)=\pi(\omega)+\pi\bigl(\delta J^{k-1}\bigr)\ .
$$
In general, the only subspaces where we do not meet this complication 
are $\pi\bigl(\Omega^0_D(\a)\bigr) = \a$ and $\pi\bigl(
\Omega^1_D(\a)\bigr) \cong\pi\bigl(\Omega^1(\a)\bigr)$. However, the 
image of $\Omega_D^{\bullet}(\a)$ under $\pi$ is $\Z_2$-graded, 
$$
\pi\bigl(\Omega_D^{\bullet}(\a)\bigr) = 
\pi\Bigl(\,\bigoplus_{k=0}^{\infty} \Omega_D^{2k}(\a) \,\Bigr) \oplus
\pi\Bigl(\,\bigoplus_{k=0}^{\infty} \Omega_D^{2k+1}(\a) \,\Bigr)\ ,
$$
because of the (anti-)commutation properties of the 
$\Z_2$-grading $\gamma$ on \h, see Definition 2.1. 
\bn\bn
%%%%%%%%%%%%%%%%%%%%%%%%%%%%%%%%%%%%%%%%%%%%%%%%%%%%%%%%%%%%%%%%%%%%%%
{\bf 2.1.3 Integration}
\bn
Property $3 ii)$ of the Dirac operator in Definition 2.1 allows
us to define the notion of integration over a non-commutative 
space in the same way as in the classical case, see part I. Note that, 
for certain sets of $N=1$ spectral data, we could use the Dixmier 
trace, as Connes originally proposed; but the definition given 
below, first introduced in \q{CFF}, works in greater generality (cf.\ 
the remarks in section I$\,$2.1.3). Moreover, it is closer to notions 
coming up naturally in quantum field theory.    
\mn
{\bf Definition 2.3}\quad The {\sl integral} over the non-commutative
space described by the $N=1$ spectral data $(\a,\h,D,\gamma)$ is  
a state $\barint$ on $\pi\bigl(\Omega^{\bullet}(\a)\bigr)$
defined by 
$$
\Barint\,: \cases {&$\pi\bigl(\Omega^{\bullet}(\a)\bigr) \lra\, \C$\cr
&$\quad\quad \omega \phantom{xxx}\longmapsto\ \, 
{\displaystyle  \Barint\omega := \Lim_{\varepsilon\to 0^+} 
{ {\rm Tr}_{\scriptscriptstyle{\cal H}}\bigl( \omega e^{-\varepsilon
D^2}\bigr) 
\over {\rm Tr}_{\scriptscriptstyle{\cal H}}
\bigl( e^{-\varepsilon D^2}\bigr) } \ , }$\cr}
$$   
where $\Lim_{\varepsilon\to 0^+}$ denotes some limiting procedure 
making the functional $\barint$ linear and positive semi-definite;
the existence of such a procedure can be shown analogously to \q{Co1,3}, 
where the Dixmier trace is discussed. 
\mn
For this integral $\barint$ to be a useful tool, we need an 
additional property that must be checked in each example: 
\mn
{\bf Assumption 2.4}\quad The state $\barint$ on 
$\pi\bigl(\Omega^{\bullet}(\a)\bigr)$ is {\sl cyclic}, i.e.,
$$
\Barint \omega\,\eta^* = \Barint \eta^*\,\omega 
$$
for all $\omega,\eta \in \pi\bigl(\Omega^{\bullet}(\a)\bigr)$.
\mn
The state $\barint$ determines a positive semi-definite 
sesqui-linear form on $\Omega^{\bullet}(\a)$ by setting
$$
(\omega,\eta) := \Barint \pi(\omega)\,\pi(\eta)^*
\eqno(2.4)$$
for all $\omega,\eta \in \Omega^{\bullet}(\a)$. In the formulas below, 
we will often drop the representation symbol $\pi$ under the integral, 
as there is no danger of confusion.
\sn
Note that the commutation relations of the grading $\gamma$ 
with the Dirac operator imply that forms of odd degree are 
orthogonal to those of even degree with respect to $(\cdot,\cdot)$. 
\sn  
By $K^k$ we denote the kernel of this sesqui-linear form restricted 
to $\Omega^k(\a)$. More precisely we set 
$$
K := \bigoplus_{k=0}^{\infty} K^k \ ,\quad\ 
K^k  := \{\, \omega\in\Omega^k(\a)\,|\, 
(\,\omega,\omega\,)= 0\,\}\ .
\eqno(2.5)$$ 
Obviously, $K^k$ contains the ideal $J^k$ defined in eq.\ (2.2); in 
the classical case they coincide. Assumption 2.4 is needed to 
show that $K$ is a two-sided ideal of the algebra of universal
forms, so that we can pass to the quotient algebra. 
\mn
{\bf Proposition 2.5}\quad The set $K$ is a two-sided graded
${}^*$-ideal of $\Omega^{\bullet}(\a)$. 
\sn
{\kap Proof}: The Cauchy-Schwarz inequality for states implies
that $K$ is a vector space. If $\omega\in K^k$, then Assumption
2.4 gives  
$$
(\omega^*,\omega^*) = \Barint \pi(\omega)^* \pi(\omega)
= \Barint \pi(\omega) \pi(\omega)^* = 0\ ,
$$
i.e.\ that $K$ is closed under the involution *. With $\omega$ as 
above and $\eta\in\Omega^p(\a)$, we have that 
$$\eqalign{
(\eta\omega,\eta\omega) &= \Barint \pi(\eta)\pi(\omega)
\pi(\omega)^* \pi(\eta)^* 
= \Barint \pi(\omega)^* \pi(\eta)^* \pi(\eta)\pi(\omega)
\cr &\leq 
\Vert \pi(\eta)\Vert^2_{\scriptscriptstyle{\cal H}} 
\Barint \pi(\omega)^*\pi(\omega) = 0
\cr}$$
where $\Vert\cdot\Vert_{\scriptscriptstyle{\cal H}}$ is 
the operator norm on ${\cal B}(\h)$. 
On the other hand, we have that 
$$
(\omega\eta,\omega\eta) = 
 \Barint \pi(\omega) \pi(\eta) \pi(\eta)^*\pi(\omega)^* \leq 
\Vert \pi(\eta)\Vert^2_{\scriptscriptstyle{\cal H}} 
\Barint \pi(\omega)\pi(\omega)^* = 0\ ,
$$
and it follows that both $\omega\,\eta$ and $\eta\,\omega$ are 
elements of $K$, i.e., $K$ is a two-sided ideal.    \hfill\qed
\mn
We now define 
$$
\widetilde{\Omega}^{\bullet}(\a) 
:= \bigoplus_{k=0}^{\infty} \widetilde{\Omega}^k(\a)\ ,\quad
\widetilde{\Omega}^k(\a) := \Omega^k(\a)/K^k\ .
\eqno(2.6)$$
The sesqui-linear form $(\cdot,\cdot)$ descends to a positive 
definite scalar product on $\widetilde{\Omega}^k(\a)$, and we 
denote by $\widetilde{{\cal H}}^{k}$ the Hilbert space 
completion of this space with respect to the scalar product, 
$$
\widetilde{{\cal H}}^{\bullet} := 
\bigoplus_{k=0}^{\infty}\widetilde{{\cal H}}^k\ ,
\quad  \widetilde{{\cal H}}^k := 
\overline{\widetilde{\Omega}^k(\a)}^{\,{
\scriptscriptstyle (\cdot,\cdot)}}\ .
\eqno(2.7)$$
$\widetilde{{\cal H}}^k$ is to be interpreted as the {\sl space of 
square-integrable k-forms}. Note that $\widetilde{{\cal H}}^{\bullet}$ 
does not in general coincide with the Hilbert
space that would arise {}from a 
GNS construction using the state $\barint$ on $\widetilde{\Omega}^{
\bullet}(\a)$: Whereas in $\widetilde{{\cal H}}^{\bullet}$,   
orthogonality of forms of different degree is installed by definition, 
there may exist forms of even degree (or odd forms) in the GNS Hilbert
space that have different degrees but are not orthogonal. 
\mn  
{\bf Corollary 2.6}\quad The space $\widetilde{\Omega}^{\bullet}(\a)$
is a unital graded ${}^*$-algebra. For any 
$\omega\in\widetilde{\Omega}^k(\a)$, the left and right actions of 
$\omega$ on $\widetilde{\Omega}^p(\a)$ with
values in $\widetilde{\Omega}^{p+k}(\a)$, 
$$
m_L(\omega)\eta := \omega\eta\ ,\quad 
m_R(\omega)\eta := \eta\omega\ ,
$$
are continuous in the norm given by $(\cdot,\cdot)$. In particular, 
the Hilbert space $\widetilde{{\cal H}}^{\bullet}$ is a bi-module over 
$\widetilde{\Omega}^{\bullet}(\a)$ with continuous actions. 
\sn
{\kap Proof}: The claim follows immediately {}from the two estimates 
given in the proof of the previous proposition, applied to 
$\omega\in\widetilde{\Omega}^k(\a)$ and 
$\eta\in\widetilde{\Omega}^p(\a)$.   \hfill\qed 
\sn 
This remark shows that $\widetilde{\Omega}^{\bullet}(\a)$ and 
$\widetilde{{\cal H}}^{\bullet}$ are ``well-behaved'' with respect to the 
$\widetilde{\Omega}^{\bullet}(\a)$-action. Furthermore, Corollary 2.6 
will be useful for our discussion of curvature and torsion in 
sections 2.1.7 and 2.1.8. 
\mn
Since the algebra $\widetilde{\Omega}^{\bullet}(\a)$ may fail to be 
differential, we introduce the unital graded differential ${}^*$-algebra 
of square-integrable differential forms 
$\widetilde{\Omega}_D^{\bullet}(\a)$ as the graded quotient of 
$\Omega^{\bullet}(\a)$ by $K+\delta K$, 
$$
\widetilde{\Omega}_D^{\bullet}(\a) := \bigoplus_{k=0}^{\infty} 
\widetilde{\Omega}_D^k(\a)\ ,\quad
\widetilde{\Omega}_D^{k}(\a) := \Omega^k(\a)/(K^k+\delta K^{k-1}) 
\cong \widetilde{\Omega}^{k}(\a)/ \delta K^{k-1} \ .
\eqno(2.8)$$
In order to show  that $\widetilde{\Omega}_D^{\bullet}(\a)$ has the 
stated properties, one repeats the proof of Proposition 2.2. 
Note that we can regard the \a-bi-module 
$\widetilde{\Omega}_D^{\bullet}(\a)$ as a 
``smaller version'' of $\Omega_D^{\bullet}(\a)$ in the sense that 
there exists a projection {}from the latter onto the former; whenever 
one deals with a concrete set of $N=1$ spectral data that
satisfy Assumption 2.4, it will be advantageous to work with 
the ``smaller'' algebra of square-integrable differential forms. 
The algebra $\Omega_D^{\bullet}(\a)$, on the other hand, can be 
defined for arbitrary data. 
\sn
In the classical case, differential forms are identified 
with the orthogonal complement of $Cl^{(k-2)}$ within $Cl^{(k)}$, 
see \q{Co1} and 
the remarks in part I, after eq.\ (I$\,$2.15). Now, we use the 
scalar product $(\cdot,\cdot)$ on 
$\widetilde{{\cal H}}^{k}$ to introduce, for each 
$k\geq 1$,  the orthogonal projection 
$$
P_{\delta K^{k-1}}\,:\ 
\widetilde{{\cal H}}^k \lra\ \widetilde{{\cal H}}^k
\eqno(2.9)$$
onto the image of $\delta K^{k-1}$ in $\widetilde{{\cal H}}^k$, and 
we set 
$$
\omega^{\perp}:=(1-P_{\delta K^{k-1}})\,\omega\in\widetilde{{\cal H}}^k 
\eqno(2.10)$$
for each element $\lb\omega\rb\in\widetilde{\Omega}_D^{k}(\a)$. This 
allows us to define a positive definite scalar product on $\widetilde{
\Omega}_D^{k}(\a)$ via the representative $\omega^{\perp}\,$:
$$
(\,\lb\omega\rb, \lb\eta\rb\,) := (\,\omega^{\perp},\eta^{\perp}\,) 
\eqno(2.11)$$
for all $\lb\omega\rb, \lb\eta\rb \in \widetilde{\Omega}_D^{k}(\a)$. 
In the classical case, this is just 
the usual inner product on the space of square-integrable $k$-forms. 
\bn\bn
{\bf 2.1.4 Vector bundles and Hermitian structures}
\bn
Again, we simply follow the algebraic formulation of classical 
differential geometry in order to generalize the notion of a vector 
bundle to the non-commutative case:
\mn
{\bf Definition 2.7} \q{Co1}\quad A {\sl vector bundle} \e\ over the 
non-commutative space described by the $N=1$ spectral data 
$(\a,\h,D,\gamma)$ is a finitely generated projective left \a-module. 
\mn
Recall that a module \e\ is {\sl projective} if there exists another 
module ${\cal F}$ such that the direct sum ${\cal E} \oplus {\cal F}$
is {\sl free}, i.e., ${\cal E} \oplus {\cal F} \cong {\cal A}^n$ as 
left \a-modules,for some $n\in\N$. Since \a\ is an algebra, every 
\a-module is a vector space; therefore, left \a-modules are 
representations of the algebra \a, and \e\ is projective iff there 
exists a module ${\cal F}$ such that ${\cal E} \oplus {\cal F}$ is 
isomorphic to a multiple of the left-regular 
representation. 
\sn
By Swan's Lemma \q{Sw}, a finitely generated projective left module
corresponds, in  the commutative case, to the space of sections of 
a vector bundle. With this in mind, it is straightforward to define 
the notion of a Hermitian structure over a vector bundle:
\mn
{\bf Definition 2.8}\ \q{Co1}\quad A {\sl Hermitian structure} over a 
vector bundle \e\ is a sesqui-linear map (linear in the first argument) 
$$
\langle \cdot,\cdot \rangle\,:\ \e \times \e \lra\ \a 
$$
such that for all $a,b\in\a$ and all $s,t \in\e$ 
\smallskip
\item {1)}   $\langle\,as,bt\,\rangle = a\, \langle\,s,t\,\rangle\, 
   b^*\,$; 
\smallskip
\item {2)}   $\langle\,s,s\,\rangle \geq 0\,$;
\smallskip
\item {3)} the \a-linear map  
$$
g\,: \cases {&$\e\lra\, \ {\cal E}^*_R$\cr
&$\, s \longmapsto\ \,\vphantom{\sum^k}\langle\,s,\cdot\,\rangle  
   $\cr} \ ,
$$
where ${\cal E}^*_R := \{\, \phi\in{\rm Hom}(\e,\a)\,|\, 
\phi(as) = \phi(s)a^*\,\}$,\  
is an isomorphism of left \a-modules, i.e., $g$ can be regarded as 
a metric on \e.  
\bn\bn
%%%%%%%%%%%%%%%%%%%%%%%%%%%%%%%%%%%%%%%%%%%%%%%%%%%%%%%%%%%%%%%%%
%\eject
\leftline{\bf 2.1.5 Generalized Hermitian structure on \Omti}
\bn
In this section we show that the \a-bi-modules 
$\widetilde{\Omega}^k(\a)$ carry Hermitian structures in a slightly 
generalized sense. Let
$\overline{{\cal A}}$ be the weak closure of the algebra \a\ acting on 
$\widetilde{{\cal H}}^0$, i.e., $\overline{{\cal A}}$ is the von 
Neumann algebra generated by $\widetilde{\Omega}^0(\a)$ acting on the 
Hilbert space  $\widetilde{{\cal H}}^0$. 
\mn
{\bf Theorem 2.9}\quad There is a canonically defined sesqui-linear 
map 
$$
\langle\cdot,\cdot\rangle_{\scriptscriptstyle D}\,:\
\widetilde{\Omega}^k(\a)\times\widetilde{\Omega}^k(\a)
\lra\ \overline{\a}
$$                             such that for all 
$a,b\in\a$ and all $\omega,\eta\in\widetilde{\Omega}^k(\a)$ 
\smallskip
\item {1)} $\langle\,a\,\omega,b\,\eta\,\rangle_{\scriptscriptstyle D} 
= a\,\langle\,\omega,\eta\,\rangle_{\scriptscriptstyle D}\, b^*\,$; 
\smallskip
\item {2)} $\langle\,\omega,\omega\,\rangle_{\scriptscriptstyle D}
     \geq 0\,$;   
\smallskip
\item {3)} $\langle\,\omega\,a,\eta\,\rangle_{\scriptscriptstyle D} = 
\langle\,\omega ,\eta\,a^*\,\rangle_{\scriptscriptstyle D}\,$.
\sn
We call $\langle\cdot,\cdot\rangle_{\scriptscriptstyle D}$ a 
{\sl generalized Hermitian structure on} $\widetilde{\Omega}^k(\a)$. 
It is the non-commutative analogue of the Riemannian metric on the 
bundle of differential forms. Note that $\langle\cdot,\cdot\rangle_{
\scriptscriptstyle D}$  takes values in $\overline{\a}$ 
and thus property 3) of Definition 2.8 is not directly applicable.
\sn
{\kap Proof}: Let $\omega,\eta\in\widetilde{\Omega}^k(\a)$ and define 
the $\C$-linear map 
$$
\varphi_{\omega,\eta}(a) = \Barint a\,\eta\,\omega^*\ ,
$$
for all $a\in\widetilde{\Omega}^0(\a)$. Note that $a$ on the rhs 
actually is a representative in \a\ of the class  $a\in
\widetilde{\Omega}^0(\a)$, and analogously for $\omega$ and $\eta$ 
(and we have omitted the representation symbol $\pi$).  
The value of the integral is, however, independent of the choice of 
these representatives, which is why we used the same letters.  
The map $\varphi$ satisfies 
$$
|\varphi_{\omega,\eta}(a)| 
\leq \Big|\,\Barint aa^*\,\Big|^{1\over2} \Big|\,\Barint
\omega\eta^*\eta\omega^*\,
\Big|^{1\over2} \leq (a,a)^{1\over2} 
\Big|\,\Barint \omega\eta^*\eta\omega^*\,\Big|^{1\over2} \ .
$$
Therefore, $\varphi_{\omega,\eta}$ extends to a bounded linear 
functional on $\widetilde{{\cal H}}^0$, and there exists an element
$\langle\,\omega,\eta\,\rangle_{\scriptscriptstyle D} 
\in\widetilde{\cal H}^0$ such that 
$$
\varphi_{\omega,\eta}(x) = 
(x, \langle\,\omega,\eta\,\rangle_{\scriptscriptstyle D}\,)
$$
for all $x\in\widetilde{{\cal H}}^0$; since $(\cdot,\cdot)$ is 
non-degenerate, $\langle\,\omega,\eta\,\rangle_{\scriptscriptstyle D}$ 
is a well-defined element; but it remains to
show that it also acts as a bounded operator on this Hilbert space. 
To this end, choose a net $\{a_{\iota}\}\subset
\widetilde{\Omega}^0(\a)$ which converges to
$\langle\,\omega,\eta\,\rangle_{\scriptscriptstyle D}$. Then,  
for all $b,c\in\widetilde{\Omega}^0(\a)$,  
$$\eqalign{
(\,\langle\,\omega,\eta\,\rangle_{\scriptscriptstyle D} b,c\,) 
&= \lim_{\iota\to\infty}(\,a_{\iota}b,c\,) 
= \lim_{\iota\to\infty} \Barint a_{\iota}bc^* = \lim_{\iota\to\infty}
\Barint a_{\iota}(cb^*)^* 
\cr
&= \lim_{\iota\to\infty}(\,a_{\iota},cb^*\,) =  
(\,\langle\,\omega,\eta\,\rangle_{\scriptscriptstyle D}, cb^*\,) \ ,
\cr}$$
and it follows that 
$$\eqalign{
|(\,\langle\,\omega,\eta\,\rangle_{\scriptscriptstyle D} b,c\,)| &= 
|(\,\langle\,\omega,\eta\,\rangle_{\scriptscriptstyle D}, cb^*\,)| 
= |(\,cb^*, \langle\,\omega,\eta\,\rangle_{\scriptscriptstyle D}\,)| 
\cr
&= \Big|\, \Barint cb^* \eta\,\omega^*\, \Big| 
= \Big| \,\Barint\omega^* cb^* \eta\, \Big| 
= \Big| \,\Barint b^* \eta\,\omega^* c\, \Big|
\cr 
&\leq \Big|\,\Barint b^*b\,\Big|^{1\over2} \,
\Big|\,\Barint c^*\omega\,\eta^*\eta\omega^*c\,\Big|^{1\over2}
\leq \Vert\, \omega\eta^*\Vert_{\scriptscriptstyle{\cal H}}\, \Big|
\,\Barint b^*b\,\Big|^{1\over2}\Big|\,\Barint c^*c\,\Big|^{1\over2}
\cr
&\leq \Vert\,\omega\eta^*\Vert_{\scriptscriptstyle{\cal H}}\, 
(b,b)^{1\over2}  (c,c)^{1\over2} \ .
\cr}$$
In the third line, we first use the Cauchy-Schwarz inequality for the 
positive state $\barint$, and then an estimate which is true for 
all positive operators on a Hilbert space; the upper bound 
$\Vert\,\omega\eta^*\Vert_{\scriptscriptstyle{\cal H}}$ again involves 
representatives $\omega,\eta\in\pi\bigl(\Omega^k(\a)\bigr)$, which was 
not explicitly indicated above, since any two will do. \hfill\break
\noindent As $\widetilde{\Omega}^0(\a)$ is dense in 
$\widetilde{{\cal H}}^0$, we see that 
$\langle\,\omega,\eta\,\rangle_{\scriptscriptstyle D}$ 
indeed defines a bounded operator in $\widetilde{{\cal H}}^0$, which, 
by definition, is the weak limit of elements in 
$\widetilde{\Omega}^0(\a)$, i.e., it belongs to $\overline{\a}$.
Properties 1-3 of $\langle\cdot,\cdot\rangle_{\scriptscriptstyle D}$ 
are easy to verify.\hfill\qed
\mn
Note that the definition of the metric $\langle\cdot,
\cdot\rangle_{\scriptscriptstyle D}$ given here differs slightly {}from 
the one of refs.\ \q{CFF,CFG}. One can, however, show that in the 
$N=1$ case both definitions agree; moreover, the present one is 
better suited for the $N=(1,1)$ formulation to be introduced later.
\bn\bn
\eject
%%%%%%%%%%%%%%%%%%%%%%%%%%%%%%%%%%%%%%%%%%%%%%%%%%%%%%%%%%%%%
\leftline{\bf 2.1.6 Connections}
\bn
{\bf Definition 2.10}\quad A {\sl connection} $\nabla$ on a vector 
bundle \e\ over a non-commutative space is a $\C$-linear map 
$$
\nabla\,:\ \e \lra\  \widetilde{\Omega}^1_D(\a) 
\otimes_{\scriptscriptstyle {\cal A}}\e 
$$
such that $$
\nabla(as) = \delta a\otimes s + a \nabla s 
$$
for all $a\in\a$ and all $s \in \e$. 
\mn
Given a vector bundle \e, we define a space of \e-valued 
differential forms by 
$$
\widetilde{\Omega}^{\bullet}_D(\e) := \widetilde{\Omega}^{
\bullet}_D(\a)\otimes_{\scriptscriptstyle {\cal A}}\e \ ;
$$
if $\nabla$ is a connection on \e, then it extends uniquely to a 
$\C$-linear map, again denoted $\nabla$, 
$$
\nabla\,:\ \widetilde{\Omega}^{\bullet}_D(\e)\lra
\widetilde{\Omega}^{\bullet+1}_D(\e)
\eqno(2.12)$$
such that 
$$
\nabla(\omega s) = \delta\omega\,s + (-1)^k \omega\,\nabla s 
\eqno(2.13)$$
for all $\omega\in\widetilde{\Omega}^{k}_D(\a)$ and all 
$s \in  \widetilde{\Omega}^{\bullet}_D(\e)$. 
\mn
{\bf Definition 2.11}\quad The {\sl curvature} of a connection 
$\nabla$ on a vector bundle \e\ is given by 
$$
{\ttR}\,(\nabla) = - \nabla^2\,:\ \e \lra\ \widetilde{\Omega}^2_D(\a) 
\otimes_{\scriptscriptstyle {\cal A}}\e \ .
$$
Note that the curvature extends to a map 
$$ 
{\ttR}\,(\nabla)\,:\ \widetilde{\Omega}^{\bullet}_D(\e)\lra
\widetilde{\Omega}^{\bullet+2}_D(\e)
$$                                                   which is left 
\a-linear, as follows easily {}from eq.\ (2.12) and Definition 2.10.
\mn
{\bf Definition 2.12}\quad A connection $\nabla$ on a Hermitian vector 
bundle $(\e,\langle\cdot,\cdot\rangle)$ is called {\sl unitary} if 
$$
\delta\,\langle\,s,t\,\rangle = \langle\,\nabla s,t\,\rangle -
\langle\,s,\nabla t\,\rangle 
$$
for all $s,t\in\e$, where the rhs of this equation is defined by 
$$
\langle\,\omega\otimes s,t\,\rangle = \omega\,\langle\,s,t\,\rangle\ ,
\quad\quad
\langle\,s,\eta\otimes t\,\rangle = \langle\,s,t\,\rangle\,\eta^*
\eqno(2.14)$$
for all $\omega,\eta\in \widetilde{\Omega}^1_D(\a)$ and  all $s,t\in\e$. 
\bn\bn
%%%%%%%%%%%%%%%%%%%%%%%%%%%%%%%%%%%%%%%%%%%%%%%%
\eject
\leftline{\bf 2.1.7 Riemannian curvature and torsion}
\bn
Throughout this section, we  make three additional assumptions which 
limit the generality of our results, 
but turn out to be fulfilled in interesting examples. 
\mn
{\bf Assumption 2.13}\quad We assume that the $N=1$ spectral data under
consideration have the following additional properties:
\smallskip
\item {1)} $K^0= 0$. (This implies that $\widetilde{\Omega}_D^0(\a)=\a$ 
and $\widetilde{\Omega}_D^1(\a)=\widetilde{\Omega}^1(\a)$, 
thus $\widetilde{\Omega}_D^1(\a)$ carries a generalized Hermitian
structure.) 
\smallskip
\item {2)} $\widetilde{\Omega}_D^1(\a)$ is a vector bundle, called 
the {\sl cotangent bundle over} \a. ($\,\widetilde{\Omega}_D^1(\a)$ is 
always a left \a-module. Here, we assume, in addition, 
that it is {\sl finitely generated} and {\sl projective}.)
\smallskip
\item {3)} The generalized metric 
$\langle\cdot,\cdot\rangle_{\scriptscriptstyle D}$ on 
$\widetilde{\Omega}_D^1(\a)$ defines an isomorphism 
of left \a-modules between $\widetilde{\Omega}_D^1(\a)$ 
and the space of \a-anti-linear maps {}from $\widetilde{\Omega}_D^1(\a)$ 
to \a, i.e., for each \a-anti-linear map
$$
\phi\,:\ \widetilde{\Omega}^1_D(\a) \lra\ \a 
$$
satisfying $\phi(a\omega) = \phi(\omega)a^*$ for all
$\omega\in\widetilde{\Omega}^1_D(\a)$ and all $a\in\a$, there is a 
unique $\eta_{\phi}\in\widetilde{\Omega}^1_D(\a)$ with
$$
\phi(\omega) = \langle\,\eta_{\phi},\omega\,
\rangle_{\scriptscriptstyle D}\ .
$$
\mn
If $N=1$ spectral data $(\a,\h,D,\gamma)$ satisfy these assumptions, 
we are able to define non-commutative generalizations of  classical 
notions like torsion and curvature. Whereas torsion and Riemann 
curvature can be introduced whenever $\widetilde{\Omega}_D^1(\a)$ 
is a vector bundle, the last assumption in 2.13 will provide a 
substitute for the procedure of ``contracting indices'' leading to 
Ricci and scalar curvature. 
\mn
{\bf Definition 2.14}\quad Let $\nabla$ be a connection on the 
cotangent bundle $\widetilde{\Omega}^1_D(\a)$ over a non-commutative 
space $(\a,\h,D,\gamma)$ satisfying Assumption 2.13. The {\sl torsion} 
of $\nabla$ is the $\a$-linear map 
$$
{\ttT}(\nabla) := {\delta} - m \circ \nabla \,:\ 
\widetilde{\Omega}^1_D(\a)\lra\ \widetilde{\Omega}^2_D(\a)
$$ 
where $m\,:\ \widetilde{\Omega}^1_D(\a)\otimes_{\scriptscriptstyle 
{\cal A}}\widetilde{\Omega}^1_D(\a)
\lra\ \widetilde{\Omega}^2_D(\a)$ denotes the product of 1-forms in 
$\widetilde{\Omega}^{\bullet}_D(\a)$.  
\mn
Using the definition of a connection, \a-linearity of torsion is 
easy to verify. In analogy to the classical case, a unitary connection  
$\nabla$ with ${\ttT}(\nabla)=0$ is called a {\sl Levi-Civita connection}. 
In the classical case, there is exactly one Levi-Civita connection 
that, in addition, is a real operator on the complexified bundle of 
differential forms. 
In contrast, for a given set of non-commutative spectral data, there 
may be several (real) Levi-Civita connections -- or none at all. 
\mn
Since we assume that $\widetilde{\Omega}^1_D(\a)$ is a vector bundle, 
we can define the {\sl Riemannian curvature} of a connection $\nabla$ 
on the cotangent bundle as a specialization of Definition 2.11. To 
proceed further, we make use of part 2) of Assumption 2.13, which 
implies that there exists a finite set of generators $\{\, E^A\,\}$ 
of  $\widetilde{\Omega}^1_D(\a)$ and an associated ``dual basis'' 
$\{\,\varepsilon_A\,\}\subset \widetilde{\Omega}^1_D(\a)^*$, 
$$
\widetilde{\Omega}^1_D(\a)^* := \bigl\{\,\phi\,:\ 
\widetilde{\Omega}^1_D(\a) \lra\ \a\,\big\vert\, 
\phi(a\omega) = a\phi(\omega)\quad {\rm  for\  all}\  
a\in\a, \omega\in\widetilde{\Omega}^1_D(\a)\,\bigr\}\ ,
$$ 
such that each $\omega\in\widetilde{\Omega}^1_D(\a)$ can be written 
as $\omega = \varepsilon_A(\omega) E^A\,$, see e.g.\ \q{Jac}. 
Because the curvature is \a-linear, there is 
a family of elements $\{\, {\ttR}^A_{\phantom{A}B}\,\} \subset
\widetilde{\Omega}^2_D(\a)$ with 
$$
{\ttR}\,(\nabla) = 
\varepsilon_A \otimes {\ttR}^A_{\phantom{A}B} \otimes E^B \ ;
\eqno(2.15)$$
here and in the following the summation convention is used. 
Put differently, we have applied the canonical isomorphism of 
vector spaces 
$$
{\rm Hom}_{\scriptscriptstyle {\cal A}}\bigl(\widetilde{\Omega}^1_D(\a),
\  \widetilde{\Omega}^2_D(\a) \otimes_{\scriptscriptstyle {\cal A}} 
\widetilde{\Omega}^1_D(\a)\bigr) \ \cong \  
\widetilde{\Omega}^1_D(\a)^* \otimes_{\scriptscriptstyle {\cal A}} 
\widetilde{\Omega}^2_D(\a)   
\otimes_{\scriptscriptstyle {\cal A}}  \widetilde{\Omega}^1_D(\a) 
$$
-- which is valid because $\widetilde{\Omega}^1_D(\a)$ is projective 
-- and chosen explicit generators $E^A, \varepsilon_A$. Then we have  
that ${\ttR}\,(\nabla)\,\omega =
\varepsilon_A(\omega)\,{\ttR}^A_{\phantom{A}B} \otimes E^B$ for any 
1-form $\omega\in\widetilde{\Omega}^1_D(\a)$. 
\sn                         Note that although the components 
${\ttR}^A_{\phantom{A}B}$ need not be unique, the element on the rhs 
of eq.\ (2.15) is well-defined. Likewise, the Ricci and scalar 
curvature, to be introduced below, will be {\sl invariant} combinations  
of those components, as long as we make sure that all maps we use have 
the correct ``tensorial properties'' with respect to the \a-action. 
\sn 
The last part of Assumption 2.13 guarantees, furthermore, that to 
each $\varepsilon_A$ there exists a unique 1-form 
$e_A\in\widetilde{\Omega}^1_D(\a)$ such that 
$$
\varepsilon_A(\omega) = \langle\,\omega,e_A\,
\rangle_{\scriptscriptstyle D} 
$$
for all $\omega\in\widetilde{\Omega}^1_D(\a)$. By Corollary 2.6, every 
such $e_A$ determines a bounded operator $m_L(e_A)\,:\ 
\widetilde{{\cal H}}^1 \lra\widetilde{{\cal H}}^2$ acting on 
$\widetilde{{\cal H}}^1$ by left multiplication with $e_A$. The adjoint 
of this operator with respect to the scalar product $(\cdot,\cdot)$ on 
$\widetilde{\cal H}^{\bullet}$ is denoted by 
$$ 
e_A^{\rm ad}\,:\ \widetilde{{\cal H}}^2 \lra\ \widetilde{{\cal H}}^1 \ .
\eqno(2.16)$$
$e_A^{\rm ad}$ is a map of right \a-modules, and it is easy 
to see that also the correspondence $\varepsilon_A \mapsto 
e_A^{\rm ad}$ is right \a-linear: For all 
$b\in\a,\ \omega\in\widetilde{\Omega}^1_D(\a)$, we have 
$$
(\varepsilon_A\cdot b)(\omega) = \varepsilon_A(\omega)\cdot b = 
\langle\,\omega,e_A\,\rangle\,b=\langle\,\omega,b^* e_A\,\rangle\ ,
$$
and, furthermore, for all $\xi_1\in\widetilde{{\cal H}}^1,\ \xi_2\in
\widetilde{{\cal H}}^2$, 
$$
(\,b^*e_A(\xi_1),\xi_2\,) = (\,e_A(\xi_1),b \xi_2\,) = 
(\,\xi_1,e_A^{\rm ad}(b\xi_2)\,)\ ,
$$
where scalar products have to be taken in the appropriate spaces  
$\widetilde{{\cal H}}^k$. Altogether, the asserted right 
\a-linearity follows. Therefore, the map
$$
\varepsilon_A \otimes {\ttR}^A_{\phantom{A}B} \otimes E^B 
\longmapsto e_A^{\rm ad} \otimes {\ttR}^A_{\phantom{A}B} \otimes E^B
$$
is well-defined and has the desired tensorial properties. 
The definition of  Ricci curvature involves another operation which 
we require to be similarly well-behaved: 
\mn 
{\bf Lemma 2.15}\quad The orthogonal projections $P_{\delta K^{k-1}}$ 
on $\widetilde{{\cal H}}^k$, see eq.\ (2.9), satisfy 
$$
P_{\delta K^{k-1}}(axb) = a P_{\delta K^{k-1}}(x) b 
$$ 
for all $a,b\in\a$ and all $x\in\widetilde{{\cal H}}^k$. 
\sn
{\kap Proof}: Set $P := P_{\delta K^{k-1}}$, and let 
$y\in P\widetilde{{\cal H}}^k$. Then
$$
(\,P(axb), y\,) = (\,axb, P(y)\,) = (\,axb, y\,) = (\,x, a^*yb^*\,) 
= (\,x,P(a^*yb^*)\,) = (\,aP(x)b, y\,)\ , 
$$
where we have used that $P$ is self-adjoint with respect to $(\cdot,\cdot)$, 
that $Py=y$, and that the image of $P$ is an \a-bi-module.   \hfill\qed 
\mn
This lemma shows that projecting onto the ``2-form part'' of
${\ttR}^A_{\phantom{A}B}$ is an \a-bi-module map, i.e., we may apply  
$$
e_A^{\rm ad} \otimes {\ttR}^A_{\phantom{A}B} \otimes E^B
\longmapsto e_A^{\rm ad} 
\otimes \bigl({\ttR}^A_{\phantom{A}B}\bigr)^{\perp}\otimes E^B
$$ 
with $\bigl({\ttR}^A_{\phantom{A}B}\bigr)^{\perp}=(1-P_{\delta K^1})\,
{\ttR}^A_{\phantom{A}B}$ as in\ eq.\ (2.10).  Altogether, we arrive 
at the following definition of the {\sl Ricci curvature}, 
$$
{\ttRic}(\nabla) = e^{\rm ad}_A \Bigl(\bigl(
{\ttR}^A_{\phantom{A}B}\bigr)^{\perp}
\Bigr)\otimes E^B \in \widetilde{{\cal H}}^1 
\otimes_{{\cal A}} \widetilde{\Omega}^1_D(\a) \ ,
$$
which is in fact independent of any choices. In the following, 
we will also use the abbreviation 
$$
{\ttRic}_B := e^{\rm ad}_A \Bigl(\bigl({\ttR}^A_{\phantom{A}B}
\bigr)^{\perp}\Bigr) 
$$ 
for the components (which, again, are not uniquely defined).   
\sn     {}From  
the components ${\ttRic}_B$ we can pass to  scalar curvature. 
Again, we have to make sure that all maps occurring in this process are 
\a-covariant so as to obtain an invariant definition.  For any 1-form 
$\omega\in\widetilde{\Omega}^1_D(\a)$, right multiplication on 
$\widetilde{{\cal H}}^0$ with $\omega$ defines a bounded operator 
$m_R(\omega)\,:\ \widetilde{{\cal H}}^0 \lra\ \widetilde{{\cal H}}^1$, 
and we denote by  
$$
\omega_R^{\rm ad}\, :\ 
\widetilde{{\cal H}}^1 \lra\  \widetilde{{\cal H}}^0 
\eqno(2.17)$$
the adjoint of this operator. In a similar fashion as above, one 
establishes that 
$$
(\omega a)_R^{\rm ad}\,(x) = \omega_R^{\rm ad}\,(xa^*)
$$
for all $x\in\widetilde{{\cal H}}^1$ and $a\in\a$. This makes it 
possible to define the {\sl scalar curvature} ${\ttr}\,(\nabla)$ of 
a connection $\nabla$ as 
$$
{\ttr}\,(\nabla) = \bigl( E^{B*} \bigr)_R^{\rm ad} ({\ttRic}_B) 
\in \widetilde{{\cal H}}^0\ .
$$
As was the case for the Ricci tensor, acting with the adjoint of 
$m_R\bigl( E^{B*} \bigr)$ serves as an analogue for ``contraction of
indices''. We summarize our results in the following 
\mn
{\bf Definition 2.16}\quad Let  $\nabla$ be a connection on the 
cotangent bundle $\widetilde{\Omega}^1_D(\a)$ over a non-commutative 
space $(\a,\h,D,\gamma)$ satisfying Assumption 2.13. The {\sl 
Riemannian curvature} ${\ttR}\,(\nabla)$ is the left \a-linear map 
$$
{\ttR}\,(\nabla) = - \nabla^2\,:\ \widetilde{\Omega}^1_D(\a)
 \lra\ \widetilde{\Omega}^2_D(\a) 
\otimes_{\scriptscriptstyle {\cal A}} \widetilde{\Omega}^1_D(\a)\ .
$$
Choosing a set of generators $E^A$ of $\widetilde{\Omega}^1_D(\a)$  
and dual generators $\varepsilon_A$ of $\widetilde{\Omega}^1_D(\a)^*$, 
and writing  ${\ttR}\,(\nabla) = 
\varepsilon_A \otimes {\ttR}^A_{\phantom{A}B} \otimes E^B$
as above, the {\sl Ricci tensor} ${\ttRic}\,(\nabla)$ is given by 
$$
{\ttRic}(\nabla) =
{\ttRic}_B \otimes E^B \in \widetilde{{\cal H}}^1 
\otimes_{{\cal A}} \Omega^1_D(\a) \ ,
$$
where ${\ttRic}_B := e^{\rm ad}_A \Bigl(\bigl({\tt
R}^A_{\phantom{A}B}\bigr)^{\perp}\Bigr)$, see eqs.\ (2.10) and (2.16). 
Finally, the {\sl scalar curvature} ${\ttr}\,(\nabla)$ of the 
connection $\nabla$ is defined as 
$$
{\ttr}\,(\nabla) = \bigl( E^{B*} \bigr)_R^{\rm ad} ({\ttRic}_B) 
\in \widetilde{{\cal H}}^0\ , 
$$
with the notation of eq.\ (2.17). (Note that, in the classical case, 
our definition of the scalar curvature differs from the usual one 
by a sign.) Both ${\ttRic}(\nabla)$ and ${\ttr}\,(\nabla)$ do not 
depend on the choice of generators.  
\bn\bn
%%%%%%%%%%%%%%%%%%%%%%%%%%%%%%%%%%%%%%%%%%%%%%%%%%%%%%%%%%%%%%%%%%%%
{\bf 2.1.8 Non-commutative Cartan structure equations}
\bn
The classical Cartan structure equations are an important tool for
explicit calculations in differential geometry. Non-commutative 
analogues of those equations were obtained in \q{CFF,CFG}. Since  
proofs were only sketched in these references, we will give a rather 
detailed account of their results in the following. Throughout 
this section, we  assume that the space $\widetilde{\Omega}^1_D(\a)$ 
is a vector bundle over \a. In fact, no other properties of this 
space are used. Therefore all the statements on the non-commutative 
Cartan structure equations for the curvature will hold for any 
finitely generated projective module \e\ over \a; the torsion tensor, 
on the other hand, is defined only on the cotangent bundle over a 
non-commutative space. 
\mn                           Let $\nabla$ be a 
connection on the vector bundle $\widetilde{\Omega}^1_D(\a)$, 
then the curvature and the torsion of $\nabla$ are 
the left \a-linear maps given in Definitions 2.16 and 2.14, 
$$\eqalign{
\ttR\,(\nabla)\,:\ \ &\widetilde{\Omega}^1_D(\a) \lra\ 
\widetilde{\Omega}^2_D(\a) \otimes_{\scriptscriptstyle {\cal A}} 
\widetilde{\Omega}^1_D(\a)\ ,
\cr
\ttT\,(\nabla)\,:\ \ &\widetilde{\Omega}^1_D(\a) \lra\ 
\widetilde{\Omega}^2_D(\a)\ . \cr}$$
Since the left \a-module $\widetilde{\Omega}^1_D(\a)$ is finitely 
generated, we can choose a finite set of generators 
$\{\,E^A\,\}_{A=1,\ldots,N}\subset 
\widetilde{\Omega}^1_D(\a)$, and define the components
$\Omega^A_{\phantom{A}B}\in\widetilde{\Omega}^1_D(\a)$, 
${\ttR}^A_{\phantom{A}B}\in\widetilde{\Omega}^2_D(\a)$ and 
${\ttT}^A \in\widetilde{\Omega}^2_D(\a)$ of connection, curvature 
and torsion, resp., by setting
$$\eqalignno{
&\nabla\, E^A = - \,\Omega^A_{\phantom{A}B} \otimes E^B  \ ,
&(2.18)\cr
&\ttR\,(\nabla)\, E^A = {\ttR}^A_{\phantom{A}B} \otimes E^B\ , 
&(2.19)\cr
&\ttT\,(\nabla)\, E^A = {\ttT}^A \ .
&(2.20)\cr}$$
Note that the components $\Omega^A_{\phantom{A}B}$ and
${\ttR}^A_{\phantom{A}B}$ 
are {\sl not} uniquely defined if  $\widetilde{\Omega}^1_D(\a)$ is not 
a free module. Using Definitions 2.16 and 2.14, the components of the 
curvature and torsion tensors can be expressed in terms of the 
connection components: 
$$\eqalignno{
{\ttR}^A_{\phantom{A}B} &= \delta\,\Omega^A_{\phantom{A}B} + 
\Omega^A_{\phantom{A}C}\,\Omega^C_{\phantom{C}B}  \ ,
&(2.21)\cr
{\ttT}^A\   &= \delta E^A + \Omega^A_{\phantom{A}B} E^B\ .   
&(2.22)\cr}$$
As they stand, eqs.\ (2.21) and (2.22) cannot be applied for solving 
typical problems like finding a connection without torsion, because 
the connection components $\Omega^A_{\phantom{A}B}$ cannot be 
chosen at will unless $\widetilde{\Omega}^1_D(\a)$ is free.
We obtain more useful Cartan structure equations if we can 
relate the components $\Omega^A_{\phantom{A}B}$ to those of a 
connection $\widetilde{\nabla}$ on a free module $\a^N$. To this 
end, we employ some general constructions valid for any 
finitely generated projective left \a-module \e. 
\mn
Let $\{\,\widetilde{E}{}^A\,\}_{A=1,\ldots,N}$ be the canonical basis 
of the standard module $\a^N$, and define a left \a-module homomorphism 
$$
p\,:\ \cases{ &$\ \ \a^N\   \lra\ \widetilde{\Omega}^1_D(\a)$ \cr
&$a_A \widetilde{E}
\vphantom{M^{M^M}}{}^A \longmapsto\  a_A E^A$ \cr} 
\eqno(2.23)$$ 
for all $a_A\in\a$. Since $\widetilde{\Omega}^1_D(\a)$ is projective 
there exists a left \a-module ${\cal F}$ such that 
$$
\widetilde{\Omega}^1_D(\a) \oplus {\cal F} \cong \a^N \ .
\eqno(2.24)$$
Denote by  $i\,:\ \widetilde{\Omega}^1_D(\a) \lra\ \a^N$  the 
inclusion map determined by the isomorphism (2.24), which satisfies  
$p\circ i = {\rm id}$ on $\widetilde{\Omega}^1_D(\a)$. 
For each $A=1,\ldots,N$, we define a left \a-linear map 
$$
\widetilde{\varepsilon}_A\,:\ \cases{ 
&$\ \ \a^N\ \lra\ \a$\cr  
&$a_B\widetilde{E}\vphantom{M^{M^M}}{}^B \longmapsto\ a_A$\cr}\ .
\eqno(2.25)$$
It is clear that $\widetilde{\varepsilon}_A(\omega)\widetilde{E}{}^A  
= \omega$ for all $\omega\in\a^N$. With the help of the inclusion 
$i\,$, we can introduce the  left \a-linear maps  
$$
\varepsilon_A\,:\ \cases{ 
&$\widetilde{\Omega}^1_D(\a) \lra\ \a$\cr  
&$\quad \ \omega\ \ \ \longmapsto\ \widetilde{\varepsilon}_A
\vphantom{M^{M^M}}\bigl(i(\omega)\bigr) $\cr} 
\eqno(2.26)$$
for all $A=1,\ldots,N$. With these, 
$\omega\in\widetilde{\Omega}^1_D(\a)$ can be written as 
$$
\omega = p\bigl(i(\omega)\bigr) = 
p\bigl(\widetilde{\varepsilon}_A(i(\omega))\widetilde{E}{}^A \bigr)
= \varepsilon_A(\omega) E^A\ ,
\eqno(2.27)$$
and we see that $\{\,\varepsilon_A\,\}$ is the dual basis already 
used in section 2.1.7. The first step towards the non-commutative 
Cartan structure equations is the following result; see also \q{Kar}. 
\mn
{\bf Proposition 2.17}\quad Every connection 
$\widetilde{\nabla}$ on $\a^N$
$$
\widetilde{\nabla}\,:\ \a^N \lra\ \widetilde{\Omega}^1_D(\a)
\otimes_{\scriptscriptstyle {\cal A}} \a^N
$$
determines a connection $\nabla$ on $\widetilde{\Omega}^1_D(\a)$ by 
$$
\nabla = ({\rm id}\,\otimes p) \circ \widetilde{\nabla} \circ i  ,
\eqno(2.28)$$
and every connection on $\widetilde{\Omega}^1_D(\a)$ is of this form. 
\sn
{\kap Proof}: Let $\widetilde{\nabla}$ be a connection on $\a^N$ -- 
which always exists (see the remarks after the proof).  
Clearly, $\nabla = ({\rm id}\,\otimes p) \circ \widetilde{\nabla} 
\circ i $ is a well-defined map, and it satisfies 
$$
\nabla (a\,\omega) =({\rm id}\,\otimes p)\bigl(\widetilde{\nabla}(a\,
i(\omega))\bigr)=({\rm id}\,\otimes p)\bigl(\delta a\otimes i(\omega) 
+a \widetilde{\nabla} i(\omega) \bigr) 
= \delta a \otimes \omega + a \nabla \omega 
$$
for all $a\in\a$ and all $\omega\in \widetilde{\Omega}^1_D(\a)$. 
This  proves that $\nabla$ is a connection on 
$\widetilde{\Omega}^1_D(\a)$.  \hfill\break 
\noindent If $\nabla'$ is any other connection on 
$\widetilde{\Omega}^1_D(\a)$, then 
$$
\nabla' - \nabla \in {\rm Hom}_{\scriptscriptstyle {\cal A}}
\bigl(\widetilde{\Omega}^1_D(\a),\ 
\widetilde{\Omega}^1_D(\a)\otimes_{\scriptscriptstyle {\cal A}} 
\widetilde{\Omega}^1_D(\a)\bigr)\ ,
$$
where Hom$_{\scriptscriptstyle {\cal A}}$ denotes the space of 
homomorphisms of left \a-modules. Since 
$$
{\rm id}\,\otimes p\,:\  \widetilde{\Omega}^1_D(\a) 
\otimes_{\scriptscriptstyle {\cal A}} \a^N \lra\ 
\widetilde{\Omega}^1_D(\a)\otimes_{\scriptscriptstyle {\cal A}} 
\widetilde{\Omega}^1_D(\a)
$$
is surjective and $\widetilde{\Omega}^1_D(\a)$ is a projective 
module, there exists a module map 
$$
\varphi\,:\ \widetilde{\Omega}^1_D(\a) \lra
\widetilde{\Omega}^1_D(\a) \otimes_{\scriptscriptstyle{\cal A}}\a^N 
$$
with 
$$\nabla' - \nabla = ({\rm id}\,\otimes p) \circ \varphi\ .
$$
Then $\widetilde{\varphi} := \varphi\circ p \in 
{\rm Hom}_{\scriptscriptstyle {\cal A}}\bigl( \a^N,\,\widetilde{
\Omega}^1_D(\a)\otimes_{\scriptscriptstyle {\cal A}}\a^N\,\bigr)$, 
and $\widetilde{\nabla} + \widetilde{\varphi}$ is a connection on 
$\a^N$ whose associated connection on $\widetilde{\Omega}^1_D(\a)$ is 
given by $\nabla'$:  
$$
({\rm id}\,\otimes p) \circ (\widetilde{\nabla} + \widetilde{\varphi}) 
\circ i = \nabla + ({\rm id}\,\otimes p) \circ \varphi = \nabla'\ .
$$
This proves that every connection on $\widetilde{\Omega}^1_D(\a)$  
comes {}from  a connection on $\a^N$.   \hfill\qed
\mn 
The importance of this proposition lies in the fact that an 
arbitrary collection of 1-forms 
$\bigl\{\,\widetilde{\Omega}^A_{\phantom{A}B}\,\bigr\}_{A,B=1,\ldots,N} 
\subset \widetilde{\Omega}^1_D(\a)$ defines a connection 
$\widetilde{\nabla}$ on $\a^N$ by the formula 
$$
\widetilde{\nabla}\,\bigl(a_A\widetilde{E}{}^A\bigr) 
= \delta a_A \otimes \widetilde{E}{}^A - 
a_A \widetilde{\Omega}^A_{\phantom{A}B} \otimes \widetilde{E}{}^B\ , 
$$ 
and conversely. Thus, not only the existence of connections on $\a^N$ 
and $\widetilde{\Omega}^1_D(\a)$ is guaranteed, but eq.\ (2.28) allows 
us to compute the components $\Omega^A_{\phantom{A}B}$ of the induced 
connection $\nabla$ on $\widetilde{\Omega}^1_D(\a)$. The action of 
$\nabla$ on the generators is 
$$\eqalign{
\nabla E^A&=({\rm id}\otimes p)\bigl(\widetilde{\nabla}\,i(E^A)\bigr) 
= ({\rm id}\otimes p)\bigl(\widetilde{\nabla}
\,\widetilde{\varepsilon}_B(i(E^A))\widetilde{E}{}^B\bigr) 
= ({\rm id}\otimes p) \bigl(
\widetilde{\nabla}\,\varepsilon_B(E^A) \widetilde{E}{}^B \bigr) 
\cr
&= ({\rm id}\otimes p)\bigl(\delta \varepsilon_B(E^A) \otimes
\widetilde{E}{}^B - \varepsilon_B(E^A)\, 
\widetilde{\Omega}^B_{\phantom{B}C} \otimes\widetilde{E}{}^C\bigr) 
\cr
&= \delta \varepsilon_B(E^A) \otimes E^B-  \varepsilon_C(E^A)\, 
\widetilde{\Omega}^C_{\phantom{C}B} \otimes E^B\ ,  
\cr}$$
where we have used some of the general properties listed before. 
In short, we get the relation 
$$
\Omega ^A_{\phantom{A}B} = \varepsilon_C(E^A)\, 
\widetilde{\Omega}^C_{\phantom{C}B} - \delta \varepsilon_B(E^A)
\eqno(2.29)$$
expressing the components of the connection $\nabla$ on 
$\widetilde{\Omega}^1_D(\a)$ in terms of the  components 
of the connection $\widetilde{\nabla}$ on $\a^N$. 
\sn
Upon inserting (2.29) into (2.21,22), one arrives at Cartan structure 
equations which express torsion and curvature in terms of these 
unrestricted components. We can, however, obtain equations of a 
simpler form if we exploit the fact that the map $\widetilde{\nabla}
\mapsto \nabla$ is many-to-one; this allows us to impose some extra 
symmetry relations on the components of the connection $\widetilde{\nabla}$. 
\mn
{\bf Proposition 2.18}\quad Let $\widetilde{\Omega}^A_{\phantom{A}B}$ 
be the coefficients of a connection $\widetilde{\nabla}$ on $\a^N$, 
and denote by $\nadt$ the connection on $\a^N$ whose components 
are given by 
$\omdt{}^A_{\phantom{A}B} := \varepsilon_C(E^A)\,
\widetilde{\Omega}^C_{\phantom{C}D}\,\varepsilon_B(E^D)\,$.
Then, these components enjoy the symmetry relations 
$$
\varepsilon_C(E^A)\,\omdt{}^C_{\phantom{C}B}=\omdt{}^A_{\phantom{A}B}\ 
\quad\quad    \omdt{}^A_{\phantom{A}C}\,
\varepsilon_B(E^C)=\omdt{}^A_{\phantom{A}B}\ ,
\eqno(2.30)$$
and $\nadt$ and $\widetilde{\nabla}$ induce the same 
connection on $\widetilde{\Omega}_D^1(\a)$.  In particular, every 
connection on $\widetilde{\Omega}_D^1(\a)$ is induced by a connection 
on $\a^N$ that satisfies (2.30). 
\sn
{\kap Proof}: We explicitly compute the action of the connection $\nabla$
induced by $\nadt$ on a generator, using eqs.\ (2.27,28) and the fact 
that all maps and the tensor product are \a-linear: 
$$\eqalign{
\nabla\,E^A &= -\,\Omega^A_{\phantom{A}B} \otimes E^B = 
\delta \varepsilon_B(E^A) \otimes E^B
-  \varepsilon_C(E^A)\, \omdt{}^C_{\phantom{C}B} \otimes E^B
\cr
&=\delta\varepsilon_B(E^A)\otimes E^B-\varepsilon_C(E^A)
\varepsilon_D(E^C)\,\widetilde{\Omega}^D_{\phantom{D}F} 
\varepsilon_B(E^F) \otimes E^B 
\cr
&= \delta \varepsilon_B(E^A) \otimes E^B  - \varepsilon_D
\bigl(\varepsilon_C(E^A) E^C\bigr)\,\widetilde{\Omega}^D_{\phantom{D}F} 
\otimes \varepsilon_B(E^F) E^B 
\cr
&= \delta \varepsilon_B(E^A) \otimes E^B  - \varepsilon_D(E^A)
\,\widetilde{\Omega}^D_{\phantom{D}F} \otimes E^F 
\cr}$$
This shows that $\nabla$ is identical to the connection induced by 
$\widetilde{\nabla}$. The symmetry relations (2.30) follow directly 
{}from \a-linearity and (2.27).   \hfill\qed 
\mn
We are now in a position to state the {\sl Cartan structure equations} 
in a simple form.
\mn
{\bf Theorem 2.19}\quad Let $\widetilde{\Omega}^A_{\phantom{A}B}$ and 
$\omdt{}^A_{\phantom{A}B}$ be as in Proposition 2.18. Then the 
curvature and torsion components of the induced connection on  
$\widetilde{\Omega}_D^1(\a)$ are given by 
$$\eqalign{
&{\ttR}{}^A_{\phantom{A}B} = 
  \varepsilon_C(E^A)\,\delta\,\omdt{}^C_{\phantom{C}B} 
+ \omdt{}^A_{\phantom{A}C}\,\omdt{}^C_{\phantom{C}B}
+ \delta \varepsilon_C(E^A)\,\delta\,\varepsilon_B(E^C) \ ,
\cr
&{\ttT}{}^A\ \,= \varepsilon_B(E^A)\,\delta E^B 
  + \omdt{}^A_{\phantom{A}B} E^B\ .
\cr}$$
\sn
{\kap Proof}: With eqs.\ (2.21,29,30) and the Leibniz rule, we get 
$$\eqalign{
{\ttR}{}^A_{\phantom{A}B} &= \delta\,\omdt{}^A_{\phantom{A}B} 
+ \bigl(\,\omdt{}^A_{\phantom{A}C}-\delta\varepsilon_C(E^A) \bigr) 
\bigl(\,\omdt{}^C_{\phantom{C}B} - \delta\varepsilon_B(E^C) \bigr)
\cr
&= \delta\,\bigl(\varepsilon_C(E^A)\,\omdt{}^C_{\phantom{C}B}\bigr) 
+ \bigl(\,  \omdt{}^A_{\phantom{A}C} - \delta\varepsilon_C(E^A) \bigr) 
\bigl(\, \omdt{}^C_{\phantom{C}B} - \delta\varepsilon_B(E^C) \bigr)
\cr
&= \varepsilon_C(E^A)\delta\,\omdt{}^C_{\phantom{C}B} 
+ \omdt{}^A_{\phantom{A}C}\,\omdt{}^C_{\phantom{C}B}
+ \delta \varepsilon_C(E^A)\,\delta\,\varepsilon_B(E^C)
- \omdt{}^A_{\phantom{A}C} \delta\varepsilon_B(E^C)\ . 
\cr}$$
The last term does in fact not contribute to the curvature, as can be 
seen after tensoring with $E^B$:
$$
\omdt{}^A_{\phantom{A}C} \delta\varepsilon_B(E^C) \otimes E^B 
= - \delta\bigl(\,\omdt{}^A_{\phantom{A}B}\bigr) \otimes E^B 
+\bigl(\delta\,\omdt{}^A_{\phantom{A}C}\bigr)\otimes \varepsilon_B(E^C) 
E^B = 0\ ,
$$
where we have used the Leibniz rule, the relations (2.30) and 
\a-linearity of the tensor product.    \hfill\break
\noindent To compute the components of the torsion, we use 
eqs.\ (2.22,29) analogously, 
$$
{\ttT}{}^A = \delta E^A + \omdt{}^A_{\phantom{A}B}E^B - \delta 
\varepsilon_B(E^A) E^B 
=  \delta E^A + \omdt{}^A_{\phantom{A}B}E^B - \delta E^A 
+ \varepsilon_B(E^A)\delta E^B \ ,
$$ 
which gives the result.  \hfill\qed 
\mn
The Cartan structure equations of Theorem 2.19 are considerably simpler 
than those one would get directly {}from (2.29) and (2.21,22). The price 
to be paid is that the components $\omdt{}^A_{\phantom{A}B}$ are not 
quite independent {}from each other, but of course they can easily be 
expressed in terms of the arbitrary components 
$\widetilde{\Omega}^A_{\phantom{A}B}$ according to Proposition 2.18. 
Therefore, the equations of Theorem 2.19 are useful e.g.\ for 
determining connections on $\widetilde{\Omega}_D^1(\a)$ with special 
properties. We refer the reader to \q{CFG} for an explicit application 
of the Cartan structure equations. 
\bn\bn
%%%%%%%%%%%%%%%%%%%%%%%%%%%%%%%%%%%%%%%%%%%%%%%%%%%%%%%%%%%%%%%%%%%%%%%%
{\bf 2.2 The \Noneone formulation of non-commutative geometry}
\bn
In this section, we introduce the non-commutative generalization 
of the description of Riemannian geometry by a set of $N=(1,1)$ 
spectral data, which was presented, for the classical case, in 
section 2.2 of part I. The advantage over the $N=1$ formulation is that now 
the algebra of differential forms is naturally represented on the 
Hilbert space \h. Therefore, calculations in concrete 
examples and also the study of cohomology rings will become much 
easier. There is the drawback that the algebra of 
differential forms is no longer closed under the ${}^*$-operation 
on \h, but we will introduce an alternative involution below and 
add further remarks in section 5. \hfill\break
\noindent 
The $N=(1,1)$ framework explained in the following will  
also provide the basis for  the definition of various 
types of complex non-commutative geometries in sections 2.3 and 2.4. 
\bn\bn
%%%%%%%%%%%%%%%%%%%%%%%%%%%%%%%%%%%%%%%%%%%%%%%%%%%%%%%%%%%%%%%%
{\bf 2.2.1 The \Noneone spectral data}
\bn
{\bf Definition 2.20}\quad A quintuple $(\a,\h, \dd, \gamma, *)$
is called a {\sl set of $N=(1,1)$ spectral data} if 
\smallskip
\item {1)} \h\ is a separable Hilbert space;
\smallskip
\item {2)} \a\ is a unital ${}^*$-algebra acting faithfully on \h\ 
by bounded operators;
\smallskip
\item {3)} \dd\ is a densely defined closed operator on \h\ such that 
\itemitem{$i)$} $\dd^2=0\,$,
\itemitem{$ii)$} for each $a\in\a$, the commutator $\lb\,\dd,a\,\rb$ 
extends uniquely to a bounded operator on \h,
\itemitem{$iii)$} the operator $\exp(-\varepsilon\triangle)$ with 
$\triangle=\dd\dd^*+\dd^*\dd$ is trace class for all $\varepsilon>0\,$; 
\smallskip
\item {4)} $\gamma$ is a $\Z_2$-grading on \h, i.e.,
$\gamma=\gamma^*=\gamma^{-1}$, such that 
\itemitem{$i)$} $\lb\,\gamma,a\,\rb=0$ for all $a\in\a\,$,
\itemitem{$ii)$} $\{\,\gamma,\dd\,\}=0\,$;
\smallskip
\item {5)} $*$ is a unitary operator on \h\ such that 
\itemitem{$i)$} $*\,\dd = \zeta\, \dd^*\,*\,$ for some $\zeta\in\C$ 
  with $|\zeta|=1\,$, 
\itemitem{$ii)$} $\lb\,*,a\,\rb=0$ for all $a\in\a\,$.
\mn
Several remarks are in order. First of all, note that we can 
introduce the two operators 
$$
\d = \dd + \dd^*\ ,\quad \bard = i\,(\dd- \dd^*\,) 
$$
on \h\ which satisfy the relations 
$$
\d^2 = \bard{}^2\ ,\quad\ \{\,\d,\bard\,\}=0\ ,
$$
cf.\ Definition I$\,$2.6. Thus, our 
notion of $N=(1,1)$ spectral data is an immediate generalization 
of a classical $N=(1,1)$ Dirac bundle -- except for the boundedness 
conditions to be required on infinite-dimensional Hilbert spaces, and 
the existence of the additional operator $*$ (see the comments below).   
\sn
As in the $N=1$ case, the $\Z_2$-grading $\gamma$ may always be 
introduced if not given {}from the start, simply by ``doubling'' the 
Hilbert space -- see the remarks following Definition 2.1.
\sn
Moreover, if $(\widetilde{\a}, \widetilde{{\cal H}}, \widetilde{\dd},
\tilde\gamma)$ is a quadruple satisfying conditions 1-4 of Definition
2.20, we obtain a full set of $N=(1,1)$ spectral data by setting 
$$\eqalign{
\h &= \widetilde{{\cal H}} \otimes \C^2\ ,\quad\quad\   
\a = \widetilde{{\cal A}} \otimes \one_2\ ,
\cr
\dd&= \widetilde{\dd}\otimes {1\over2}\,(\one_2+\tau_3)- 
\widetilde{\dd}^*  \otimes {1\over2}\,(\one_2-\tau_3)\ ,
\cr
\gamma &= \tilde\gamma\otimes\one_2\ , \quad\quad\quad 
*\, = \one_{\tilde{\scriptscriptstyle {\cal H}}} \otimes \tau_1 
\cr}$$
with the Pauli matrices $\tau_i$ as usual. Note that, in this example, 
$\zeta=-1$, and the $*$-operator additionally 
satisfies $*^2=1$ as well as  $\lb\,\gamma,*\,\rb=0\,$. 
\sn
The unitary operator $*$ was not present in our algebraic formulation 
of classical Riemannian geometry. But for a compact oriented manifold, 
the usual Hodge $*$-operator acting on differential forms satisfies 
all the properties listed above, after appropriate rescaling in each 
degree. (Moreover, one can always achieve $*^2=1$ {\sl or} $\zeta=-1$.)
For a non-orientable manifold, we can apply the construction of the 
previous paragraph to obtain a description of the differential forms 
in terms of $N=(1,1)$ spectral data including a Hodge operator. In 
our approach to the non-commutative case, we will make essential use  
of the existence of $*$, which we will also call {\sl Hodge operator},
in analogy to the classical case.   
\bn\bn
%%%%%%%%%%%%%%%%%%%%%%%%%%%%%%%%%%%%%%%%%%%%%%%%%%%%%%%%%%%%%%%%%%%%
\eject
\leftline{\bf 2.2.2 Differential forms}
\bn
We first introduce an involution, $\natural$, called {\sl complex
conjugation}, on the algebra of universal forms: 
$$
\natural\,:\ \Omega^{\bullet}(\a) \lra\ \Omega^{\bullet}(\a) 
$$
is the unique $\C$-anti-linear anti-automorphism such that
$$
\natural(a) \equiv a^{\natural} := a^*\ ,\quad\quad
\natural(\delta a) \equiv (\delta a)^{\natural} := \delta(a^*\,) 
\eqno(2.31)$$
for all $a\in\a$. Here we choose a sign convention that 
differs {}from the $N=1$ case, eq.\ (2.1). 
If we write $\hat\gamma$ for the mod 2 reduction of the 
canonical $\Z$-grading on $\Omega^{\bullet}(\a)$, we have 
$$
\delta\natural\hat\gamma = \natural\delta\ .
\eqno(2.32) 
$$
We define a representation of $\Omega^{\bullet}(\a)$ on \h, 
again denoted by $\pi$, by
$$
\pi(a) := a\ ,\quad\quad 
\pi(\delta a) := \lb\,\dd,a\,\rb 
\eqno(2.33)$$
for all $a\in \a$. The map $\pi$ is a $\Z_2$-graded representation 
in the sense that 
$$
\pi(\hat\gamma \omega\hat\gamma) = \gamma\pi(\omega)\gamma
\eqno(2.34)$$
for all $\omega\in\Omega^{\bullet}(\a)$. 
\sn
Although the abstract algebra of universal forms is the 
same as in the $N=1$ setting, the interpretation of the universal
differential $\delta$ has changed: In the $N=(1,1)$ framework, it is 
represented on \h\ by the nilpotent operator \dd, instead of the 
self-adjoint Dirac operator $D$, as before. In particular, we now have 
$$
\pi(\delta\omega) = \lb\,\dd,\pi(\omega)\,\rb_g
\eqno(2.35)$$
for all $\omega\in\Omega^{\bullet}(\a)$, where $\lb\cdot,\cdot\rb_g$
denotes the graded commutator (defined with the  canonical 
$\Z_2$-grading on $\pi(\Omega^{\bullet}(\a))$, see (2.34)). 
The validity of eq.\ (2.35) is the main difference between the
$N=(1,1)$ and the $N=1$ formalism. It ensures that there do not exist 
any forms  $\omega\in\Omega^p(\a)$ with $\pi(\omega)=0$ but
$\pi(\delta\omega) \neq 0$, in other words:
\mn
{\bf Proposition 2.21}\quad The graded vector space 
$$
J=\bigoplus_{k=0}^{\infty} J^k\ ,\quad  
J^k := {\rm ker}\,\pi\,|_{\Omega^k({\cal A})} 
$$
with $\pi$ defined in (2.33) is a two-sided graded {\sl differential} 
${}^{\natural}$-ideal of $\Omega^{\bullet}(\a)$.
\sn
{\kap Proof}: The first two properties are obvious, the third one is 
the content of eq.\ (2.35). Using (2.31) and the relations satisfied 
by the Hodge $*$-operator according to part 5) of Definition 2.20, 
we find that  
$$\eqalign{
\pi\bigl((\delta a)^{\natural}\bigr) &= \pi(\delta( a^*\,)) = 
\lb\, \dd,a^*\,\rb = \lb\,a,\dd^*\,\rb^* 
= \zeta\, \lb\, a, *\,\dd\, *^{-1}\,\rb^* 
\cr
&= \zeta\, *\,\lb\,a,\dd\,\rb^*\,*^{-1}= 
-\zeta\,*\,\pi(\delta a)^*\,*^{-1}\ ,\cr}$$
which implies 
$$
\pi\bigl(\omega^{\natural}\bigr)=(-\zeta)^k\,*\,\pi(\omega)^*\,*^{-1}
\eqno(2.36)$$
for all  $\omega\in\Omega^k(\a)$. In particular, $J = {\rm ker}\,\pi$ 
is a ${}^{\natural}$-ideal.   \hfill\qed
\mn
As a consequence of this proposition, the algebra of differential forms
$$
\Omega_{\rm d}^{\bullet}(\a) := \bigoplus_{k=0}^{\infty}\, 
\Omega_{\rm d}^k(\a)\ ,\quad\ \ 
\Omega_{\rm d}^k(\a) := \Omega^k(\a)/J^k\ , 
\eqno(2.37)$$
is represented on the Hilbert space \h\ via $\pi$. For later purposes, 
we will also need an involution on $\Omega_{\rm d}^{\bullet}(\a)$, and 
according to Proposition 2.21, this is given by the anti-linear map
$\natural$ of (2.31). Note that the ``natural'' involution  
$\omega \mapsto \omega^*$, see eq.\ (2.1), which is inherited {}from \h\ 
and was used in the $N=1$ case, is no longer available here:  The 
space $\pi(\Omega^k(\a))$ is not closed under taking adjoints,
because \dd\ is not self-adjoint.   \hfill\break
In summary, the space $\Omega_{\rm d}^{\bullet}(\a)$ is a unital 
graded differential ${}^{\natural}$-algebra and the representation 
$\pi$ of $\Omega^{\bullet}(\a)$ determines a representation of 
$\Omega_{\rm d}^{\bullet}(\a)$ on \h\ as a unital differential algebra.
\bn\bn
%%%%%%%%%%%%%%%%%%%%%%%%%%%%%%%%%%%%%%%%%%%%%%%%%%%%%%%%%%%%%%
%\vfil\eject
\leftline{\bf 2.2.3 Integration}
\bn
The integration theory follows the same lines as in the $N=1$ case:
The state $\barint$ is given as in Definition 2.3 with $\d^2$
written as $\triangle= \dd\dd^*+\dd^*\dd$. Again, we make Assumption 
2.4 about the cyclicity of the integral. This yields a sesqui-linear
form on $\Omega_{\rm d}^{\bullet}(\a)$ as before:
$$
(\omega,\eta) = \Barint \omega\,\eta^* 
\eqno(2.38)$$
for all $\omega,\eta\in\Omega_{\rm d}^{\bullet}(\a)$, where we have 
dropped the representation symbols $\pi$ under the integral. 
\sn
Because of the presence of the Hodge $*$-operator, the form 
$(\cdot,\cdot)$ has an additional feature in the $N=(1,1)$ setting:
\mn
{\bf Proposition 2.22}\quad If the phase in part 5) of Definition 2.20 
is $\zeta=\pm 1$, then the  inner product defined in eq.\ (2.38)
behaves like a real functional with respect to the involution 
$\natural$, i.e., for $\omega,\eta\in\Omega_{\rm d}^{\bullet}(\a)$
we have 
$$
(\,\omega^{\natural},\eta^{\natural}\,) = \overline{(\omega,\eta)}
$$
where the bar denotes ordinary complex conjugation. 
\sn
{\kap Proof}: First, observe that the Hodge operator commutes with the 
Laplacian, which is verified e.g.\ by taking the adjoint of the relation  
$*\,\dd = \zeta\, \dd^*\,*\,$. Then the claim 
follows immediately using eq.\ (2.36), unitarity of 
the Hodge operator, and cyclicity of the trace on \h: Let
$\omega\in\Omega^p_{\rm d}(\a),\ \eta\in \Omega^q_{\rm d}(\a)$, then 
$$\eqalign{
(\,\omega^{\natural},\eta^{\natural}\,) 
&= \Barint \omega^{\natural} \bigl(\eta^{\natural}\bigr)^*
= (-\zeta)^p (-\bar\zeta)^q \Barint *\, \omega^* *^{-1}* \eta\, *^{-1} 
= (-\zeta)^{p-q} \Barint \omega^* \eta \cr
&= (-\zeta)^{p-q} \Barint \eta\, \omega^* = 
(-\zeta)^{p-q}\, \overline{(\omega,\eta)}\ ;\cr}
$$ 
again, we have suppressed the representation symbol $\pi$. The claim 
follows since the $\Z_2$-grading implies $(\omega,\eta)=0$ unless 
$p-q \equiv 0\ ({\rm mod}\,2)$.       \hfill\qed
\sn
Note that, in examples, $p$- and $q$-forms for $p\neq q$ are often
orthogonal with respect to the inner product $(\cdot,\cdot)$; then Proposition 
2.22 holds independently of the value of $\zeta$. 
\mn
Since $\Omega_{\rm d}^{\bullet}(\a)$ is a ${}^{\natural}$- and 
{\sl not} a ${}^*$-algebra, Proposition 2.5 is to be replaced by 
\mn 
{\bf Proposition 2.23}\quad The graded kernel $K$, see eq.\ (2.5),  
of the sesqui-linear form $(\cdot,\cdot)$ is a two-sided graded 
${}^{\natural}$-ideal of $\Omega_{\rm d}^{\bullet}(\a)$. 
\sn
{\kap Proof}: The proof that $K$ is a two-sided graded ideal is 
identical to the one of Proposition 2.5. That $K$ is closed under 
$\natural$ follows immediately {}from the proof of Proposition 
2.22.\hfill\qed
\mn
The remainder of section 2.1.3 carries over to the $N=(1,1)$ case, with 
the only differences that $\widetilde{\Omega}^{\bullet}(\a)$ is a 
${}^{\natural}$-algebra and that the quotients 
$\Omega^k(\a)/\bigl(K^k+\delta K^{k-1}\bigr) \cong
\widetilde{\Omega}^k(\a)/\delta K^{k-1}$ are denoted by 
$\widetilde{\Omega}_{\rm d}^k(\a)$. 
\sn
While $\Omega_{\rm d}^{\bullet}(\a)$ is a differential algebra 
(by construction), $\widetilde{\Omega}^{\bullet}(\a)$ is {\sl not},
in general, a differential algebra, because the ideal $K$ may not 
be a differential ideal (i.e.\ there may exist
$\omega\in K^{k-1}$ with $\delta\omega\notin K^k\,$). However, $K$ 
is trivial in many interesting examples. If $K$ is trivial then 
the algebra $\widetilde{\Omega}^{\bullet}(\a)$ of square-integrable
forms is a differential algebra which is faithfully represented
on $\widetilde{{\cal H}}^{\bullet}$.
%Apart {}from that, a sufficient condition ensuring that $K$ is a 
%differential ideal is given by  
%$$
%\Barint \lb\,\dd,\omega\,\rb_g = 0 
%$$     for all $\omega$ 
%in the algebraic span of $\pi\bigl(\Omega_{\rm d}^{\bullet}
%(\a)\bigr)+\pi\bigl(\Omega_{\rm d}^{\bullet}(\a)\bigr)^*$: If this 
%is true, the Cauchy-Schwarz inequality leads to the estimate 
%$$
%\Big\vert\, \Barint \lb\,\dd,\omega\,\rb_g\eta^* \,\Big\vert 
%= \Big\vert\, \Barint \omega\,\lb\,\dd,\eta^*\,\rb_g \Big\vert 
%\leq \Bigl( \Barint \omega\,\omega^* \Bigr)^{1\over2} 
%\Bigl( \Barint \lb\,\dd,\eta^*\,\rb_g\lb\,\dd,\eta^*\,\rb^*_g 
%\Bigr)^{1\over2} \ .
%$$
%Applying this to $\eta= \lb\,\dd,\omega\,\rb_g$ for 
%$\omega\in K^{k-1}$, we obtain that $\delta\omega\in K^k$. 
\bn\bn
%%%%%%%%%%%%%%%%%%%%%%%%%%%%%%%%%%%%%%%%%%%%%%%%%%
{\bf 2.2.4 Unitary connections and scalar curvature}
\bn
Except for the notions of unitary connections and scalar curvature, 
all definitions and results of sections 2.1.4-8
literally apply to the $N=(1,1)$ case as well. The two exceptions 
explicitly involve the ${}^*$-involution on the algebra 
of differential forms, which is no longer available now. Therefore, 
we have to modify the definitions for $N=(1,1)$ non-commutative 
geometry as follows:
\mn
{\bf Definition 2.24}\quad A connection $\nabla$ on a Hermitian 
vector bundle $\bigl(\e,\langle\cdot,\cdot\rangle\bigr)$ over 
an \break \noindent \hbox{$N=(1,1)$} non-commutative space is called 
{\sl unitary} if 
$$
\dd\,\langle\,s,t\,\rangle = 
\langle\,\nabla s,t\,\rangle + \langle\,s,\nabla t\,\rangle
$$
for all $s,t\in\e$; the Hermitian structure on the rhs is extended 
to  
\e-valued differential forms by
$$
\langle\,\omega\otimes s,t\,\rangle = \omega\,\langle\,s,t\,\rangle\ ,
\quad\quad
\langle\, s,\eta \otimes t\,\rangle  
= \langle\,s,t\,\rangle\,\eta^{\natural}
$$
for all $\omega,\eta \in \widetilde{\Omega}^{\bullet}_{{\rm d}}(\a)$ 
and $s,t\in\e$. 
\mn
{\bf Definition 2.25}\quad The {\sl scalar curvature} of a connection
$\nabla$ on $\widetilde{\Omega}^{1}_{{\rm d}}(\a)$ is defined by 
$$
{\ttr}\,(\nabla) = \bigl( E^{B\,\natural}\bigr)^{\rm ad}_R ({\ttRic}_B)
\in \widetilde{{\cal H}}_0\ .
$$
\bn\bn
%%%%%%%%%%%%%%%%%%%%%%%%%%%%%%%%%%%%%%%%%%%%%%%%%%%%%%%%%%%%%%%
%%%%%%%%%%%%%%%%%%%%%%%%%%%%%%%%%%%%%%%%%%%%%%%%%%%%%%%%%%%%%%%
\def\hnull{\hskip-3pt \hbox{ $ {\cal H}\kern-5pt \raise7pt
\hbox{$\scriptscriptstyle {\rm o}$}$}\hskip3pt}
%%%%%%%%%%%%%%%%%%%%%%%%%%%%%%%%%%%%%%%%%%%%%%%%%%%%%%%%%%%%%%%
%%%%%%%%%%%%%%%%%%%%%%%%%%%%%%%%%%%%%%%%%%%%%%%%%%%%%%%%%%%%%%%
{\bf 2.2.5 Remarks on the relation of \Neqone and \Noneone 
spectral data}
\bn
The definitions of $N=1$ and $N=(1,1)$ non-commutative spectral
data provide two different generalizations of classical Riemannian 
differential geometry. In the latter context, one can always find 
an $N=(1,1)$ description of a manifold originally given by an 
$N=1$ set of data (see part I), whereas a non-commutative set of 
$N=(1,1)$ spectral data seems to require a different mathematical 
structure than a spectral triple, because of 
the additional generalized Dirac operator which must be given on the 
Hilbert space. Thus, it is a natural and important question under which 
conditions on an $N=1$ spectral triple $(\a, \h, D)$ there exists an 
associated $N=(1,1)$ set of data $(\a, \widetilde{\cal H}, {\tt d}, *)$ 
over the same non-commutative space \a. 
%As was mentioned in section 4, 
%we believe that the existence of an $N=(1,1)$ description is part of  
%a convenient characterization of {\sl non-commutative manifolds}. 
\sn
We have not been able yet to answer the question of how to pass {}from 
$N=1$ to $N=(1,1)$ data in a general way. But in the following  we 
present a procedure that might lead to a solution. Our 
guideline is the classical case, where the main step in passing {}from
$N=1$ to $N=(1,1)$ data is to replace the Hilbert space $\h=L^2(S)$ by 
$\widetilde{\cal H} = L^2(S\otimes\overline S)$ carrying 
two actions of the Clifford algebra and therefore two anti-commuting 
Dirac operators $\d$ and $\bard$ -- which yield a description 
equivalent to the one involving the nilpotent differential {\tt d}, 
see the remark after Definition 2.20. \hfill\break 
\noindent It is plausible that there are other approaches 
to this question, in particular approaches of a more operator 
algebraic nature, e.g.\ using a ``Kasparov product of spectral 
triples'', but we will not enter these matters here. 
\sn                               The first problem 
one meets when trying to copy the classical step {}from $N=1$ to 
$N=(1,1)$ is that \h\ should be an \a-bi-module.  To ensure this, we 
require that the set of $N=1$ (even) spectral data $(\a,\h,D,\gamma)$
is endowed with a {\sl real structure} \q{Co4}, i.e.\ 
that there exists an anti-unitary operator $J$ on \h\ such that 
$$
J^2 = \epsilon\,\one\ ,\quad\quad J\gamma= \epsilon' \gamma J\ ,
\quad\quad      JD = DJ 
$$
for some (independent) signs $\epsilon, \epsilon' = \pm 1$, and 
such that, in addition,
$$
J a J^*\quad \hbox{commutes with $b$ and $\lb\,D,b\,\rb$ for all}\ 
a,b\in\a\ .
$$
\sn
This definition of a real structure was introduced by Connes in 
\q{Co4}; $J$ is of course a variant of Tomita's modular conjugation 
(cf.\ the next subsection).
\hfill\break
\noindent In the present context, $J$ provides a canonical 
right \a-module structure on \h\ by defining 
$$
\xi\cdot a := Ja^* J^* \xi 
$$
for all $a\in\a$, $\xi\in\h$, see \q{Co4}. We can extend this to a 
right action of $\Omega^1_D(\a)$ on \h\ if we set 
$$
\xi\cdot\omega := J\omega^* J^* \xi 
$$ 
for all $\omega\in\Omega^1_D(\a)$ and $\xi\in\h$; for simplicity, 
the representation symbol $\pi$ has been omitted. Note that by the 
assumptions on $J$, the right action commutes with the left action of 
\a. Thus \h\ is an \a-bi-module, and we can form tensor products 
of bi-modules {\sl over the algebra} \a\ just as in 
the classical case. 
{}From now on, we assume that \h\ contains a dense projective left
${\cal A}$-module $\hnull$ which is stable under $J$ and $\gamma$.
In particular, $\hnull$ is itself an ${\cal A}$-bi-module.
Since $\hnull$ is projective, it carries a Hermitian structure,
see Definition 2.8, that induces a scalar product on
$\hnull\otimes_{\cal A}\hnull$ (see also Sect.\ 4.3). We shall denote by
$\widetilde{\cal H}$ the Hilbert space completion of 
$\hnull\otimes_{\cal A}\hnull$ with respect to this scalar product.
%If \h\ carries a Hermitian structure, see 
%Definition 2.8, then $\h\otimes_{\cal A}\h$ can be endowed 
%with a natural inner product. 
\sn
The real structure $J$ allows us to define the anti-linear
``flip'' operator 
\def\vps{\vphantom{m_{{\displaystyle \sum}}}}
$$
\Psi\,:\ \cases{&${\displaystyle \Omega^1_D(\a)\vps\otimes_{\cal A}\hnull
\lra \hnull\otimes_{\cal A}\Omega^1_D(\a)}$\cr
&$\phantom{xxxxxx}\omega\otimes\xi \longmapsto J\xi \otimes 
\omega^*$\cr} \ .
$$             It is straightforward 
to verify that $\Psi$ is well-defined and that it satisfies 
$$
\Psi(a\,s) = \Psi(s)\,a^*
$$
for all $a\in\a$, $s\in\Omega^1_D(\a)\otimes_{\cal A}\hnull$. 
\sn    
%{}From now on, we assume furthermore that \h\ is a 
%{\sl projective} left \a-module. Then it admits connections 
Since $\hnull$ is projective, it admits connections
$$
\nabla\,:\ \hnull \lra \Omega^1_D(\a)\otimes_{\cal A}\hnull\ ,
$$
i.e.\ $\C$-linear maps such that 
$$
\nabla(a\xi) = \delta a\otimes\xi + a \nabla\xi 
$$ 
for all $a\in\a$ and $\xi\in\hnull$. 
We assume that $\nabla$ commutes with the grading $\gamma$ on 
$\hnull$, i.e.\ $\nabla\,\gamma\,\xi= 
(\one\otimes\gamma)\,\nabla\xi$ for all $\xi\in\hnull$.  
For each connection $\nabla$ on 
\hnull, there is an ``associated right-connection'' $\overline{\nabla}$ 
defined with the help of the flip $\Psi$: 
$$
\overline{\nabla}\,:\ \cases{&${\displaystyle \hnull\lra 
\hnull\vps\otimes_{\cal A}\Omega^1_D(\a)}$\cr
&$\phantom{l}\xi \longmapsto -\Psi(\nabla J^* \xi)$\cr}
$$
$\overline{\nabla}$ is again $\C$-linear and satisfies 
$$
\overline{\nabla}(\xi a)=\xi\otimes\delta a+(\overline{\nabla}\xi)a\ .
$$
A connection $\nabla$ on \hnull, together with its associated right 
connection $\overline{\nabla}$, induces a $\C$-linear ``tensor product 
connection'' $\widetilde{\nabla}$ on $\hnull\otimes_{\cal A}\hnull$ 
of the form 
$$
\widetilde{\nabla}\,:\ \cases{&${\displaystyle \hnull\otimes_{\cal A}\hnull \lra 
\hnull\vps\otimes_{\cal A}\Omega^1_D(\a)\otimes_{\cal A}\hnull}$\cr
&\phantom{w}$\xi_1 \otimes \xi_2 \longmapsto \overline{\nabla}\xi_1 
\otimes \xi_2 + \xi_1 \otimes \nabla\xi_2$\cr}\ .
$$
Because of the position of the factor $\Omega^1_D(\a)$, 
$\widetilde{\nabla}$ is not quite a connection in the usual sense. 
In the classical case, the last ingredient needed for the definition of  
the two Dirac operators of an $N=(1,1)$ Dirac bundle are the two 
anti-commuting Clifford actions on $\widetilde{\h}$. Their obvious 
generalizations to the non-commutative case are the $\C$-linear maps 
$$
{\c}\,:\ \cases{&${\displaystyle \hnull\otimes_{\cal A}\Omega^1_D(\a)
\otimes_{\cal A}\hnull\lra \hnull\vps\otimes_{\cal A}\hnull}$\cr
&$\phantom{xxxxww}\xi_1\otimes\omega\otimes\xi_2 \longmapsto
\xi_1\otimes\omega\,\xi_2$\cr}
$$
and \def\c{\hbox{{\tpwrt c}}}
$$
\overline{{\c}}\,:\ \cases{&${\displaystyle \hnull\otimes_{\cal A}
\Omega^1_D(\a)\otimes_{\cal A}\hnull\lra \hnull\vps\otimes_{\cal A}\hnull}$\cr
&\phantom{xxxxxxw}$\xi_1\otimes\omega\otimes\xi_2 \longmapsto
\xi_1\,\omega\otimes\gamma\xi_2$\cr}\ .
$$
With these, we may introduce two operators $\d$ and $\bard$ on 
$\hnull\otimes_{\cal A}\hnull$ in analogy to the classical case:
$$
\d := \c \circ \widetilde{\nabla}\ ,\quad\quad
\bard := \overline{\c} \circ \widetilde{\nabla}\ .
$$
In order to obtain a set of $N=(1,1)$ spectral data, one has to find a 
connection $\nabla$ on \hnull\ which makes the operators $\d$ and $\bard$ 
essentially self-adjoint and ensures that the relations 
$\d^2 = \bard{}^2$ and $\{\,\d,\bard\,\}=0$ of 
Definition 2.20 are satisfied. The $\Z_2$-grading on 
%$\h\otimes_{\cal A}\h$ 
$\widetilde{\h}$ is simply the tensor product grading, and the 
Hodge operator can be taken to be $* = \gamma\otimes\one$. 
\sn      
In section 4 below, we will verify these conditions in the 
example of the non-commutative torus. In the general case, we have, 
up to now,  not been able to prove the existence of a connection 
$\nabla$ on \hnull\ which supplies $\d$ and $\bard$ with the correct 
algebraic properties, but the naturality of the construction presented 
above as well as the similarity with the procedure of section I$\,$2.2.2 
lead us to expect that this problem can be solved in many cases of 
interest. \hfill\break
\noindent More precisely, we expect that the relation 
$\{\,\d,\bard\,\}=0$ can be satisfied under rather general assumptions, 
whereas it may often be appropriate to deal with a non-vanishing 
operator $\d^2-\bard{}^2$ that generates an $S^1$-action. 
\bn\bn
%%%%%%%%%%%%%%%%%%%%%%%%%%%%%%%%%%%%%%%%%%%%%%%%%%%%%%%%%%%%%%%%%%%%
{\bf 2.2.6 Riemannian and  Spin\CC\ ``manifolds'' in non-commutative 
geometry}
\bn
In this section, we address the following question: What is 
the additional structure that makes an $N=(1,1)$ non-commutative 
space into a {\sl non-commutative ``manifold''}, into a {\sl Spin${}^c$ 
``manifold''}, or into a {\sl quantized phase space}? There is a 
definition of non-commutative manifolds in terms of $K$-homology, see
e.g.\ \q{Co1}. In our
search for the characteristic features of non-commutative manifolds
we will, as before, be guided by the classical case and by the 
principle that they should be natural {}from the point of view of 
quantum physics. 
\sn                                         Extrapolating {}from 
classical geometry, we are e.g.\ led to the following 
requirement an $N=(1,1)$ space $(\a, \h, {\tt d}, \gamma, *)$ should 
satisfy in order to describe a ``manifold'': The data must extend to a 
set of $N=2$ {\sl spectral data} $(\a, \h, {\tt d}, T, *)$ where $T$ 
is a self-adjoint operator on \h\ such that 
\smallskip
\item{$i)$} $\lb\,T,a\,\rb=0$ for all $a\in\a\,$;
\smallskip
\item{$ii)$} $\lb\,T,{\tt d}\,\rb={\tt d}\,$; 
\smallskip
\item{$iii)$} $T$ has integral spectrum, and $\gamma$ is the mod 2 
reduction of $T$, i.e.\ $\gamma=\pm1$ on $\h_{\pm}$, where 
$$
\h_{\pm} = {\rm span}\,\bigl\{\, \xi\in\h\,|\, T\xi=n\,\xi 
\hbox{ for some } n\in\Z, (-1)^n = \pm 1 \,\bigr\}\ .
$$
\sn
Such $N=2$ spectral data have been used in section I$\,$1.2 already, and 
have also been briefly discussed in section I$\,$3.  
\bn
Before we can formulate further properties that we suppose to 
characterize non-commu\-tative manifolds, we recall some  
basic facts about {\sl Tomita-Takesaki theory}. Let \m\  be a 
von Neumann algebra acting on a separable Hilbert space 
\h, and assume that $\xi_0\in\h$ is a cyclic and separating vector 
for \m, i.e.\ 
$$
\overline{\m\,\xi_0} = \h 
$$
and 
$$
a\,\xi_0 = 0 \quad \Longrightarrow\quad a=0 
$$
for any $a\in\m$, respectively. Then we may define an anti-linear 
operator $S_0$ on \h\ by setting 
$$
S_0\, a\, \xi_0 = a^* \xi_0 
$$
for all $a\in\m$. One can show that $S_0$ is closable, and we denote 
its closure by $S$. The polar decomposition of $S$ is written as 
$$
S= J \Delta^{1\over2}
$$
where $J$ is an anti-unitary involutive operator, referred  to as 
{\sl modular conjugation}, and the so-called {\sl modular operator} 
$\Delta$ is a positive self-adjoint operator on \h. The fundamental 
result of Tomita-Takesaki theory is the following theorem: 
$$
J \m J = \m'\ ,\quad \Delta^{it} \m \Delta^{-it} = \m
$$
for all $t\in\R$; here, $\m'$ denotes the commutant of \m\ on \h. 
Furthermore, the vector state $\omega_0(\cdot):=(\xi_0,\cdot\,\xi_0)$
is a {\sl KMS-state} for the automorphism $\sigma_t := 
{\rm Ad}_{\Delta^{it}}$ of \m, i.e. 
$$
\omega_0(\sigma_t(a)\,b) = \omega_0(b\,\sigma_{t-i}(a))
$$
for all $a,b\in\m$ and all real $t$. 
\bn
Let  $(\a, \h, {\tt d}, T, *)$ be a set of $N=2$ spectral data 
coming {}from an $N=(1,1)$ space as above. We define the analogue 
$Cl_{\cald}(\a)$ of the space of sections of the Clifford bundle, 
$$
Cl_{\cald}(\a)=\bigl\{\,a_0\,\lb\,\d,a_1\,\rb\ldots \lb\,\d,a_k\,\rb
   \,|\, k\in\Z_+,\ a_i\in\a\,\bigr\}\ ,
$$
where $\d = \dd + \dd^*$, and, corresponding to the second 
generalized Dirac operator $\bard = i(\dd - \dd^*)\,$, 
$$
Cl_{\calbd}(\a) = \bigl\{\, a_0\,\lb\,\bard,a_1\,\rb \ldots 
\lb\,\bard,a_k\,\rb   \,|\, k\in\Z_+,\ a_i\in\a\,\bigr\}\ .
$$
In the classical setting, the sections $Cl_{\cald}(\a)$ and 
$Cl_{\calbd}(\a)$ operate on \h\ by the two actions $c$ and 
$\overline{c}$, respectively, see Definition I$\,$2.6. In the 
general case, we notice that, in contrast to the algebra
$\Omega^{\bullet}_{\rm d}(\a)$ introduced before,
$Cl_{\cald}(\a)$ and $Cl_{\calbd}(\a)$ form ${}^*$-algebras of 
operators on \h, but are neither $\Z$-graded nor differential. 
\sn
We want to apply Tomita-Takesaki theory to the von Neumann algebra 
$\m := \bigl(Cl_{\cald}(\a)\bigr)''\,$. Suppose there exists a vector 
$\xi_0\in\h$ which is cyclic and separating for $\m$, and let $J$ be 
the anti-unitary conjugation associated to $\m$ and $\xi_0$. Suppose, 
moreover, that for all $a\in {}^J\!\a := J\a J$ the operator 
$\lb\,\bard,a\,\rb$ uniquely extends to a bounded operator on \h. 
Then we can form the algebra of bounded operators $Cl_{\calbd}\,({}^J\!\a)$ 
on \h\ as above. The properties $J\a J \subset \a'$ and 
$\{\,\d,\bard\,\}=0$ imply that $Cl_{\cald}(\a)$ and 
$Cl_{\calbd}\,({}^J\!\a)$ commute in the graded sense; to arrive at 
truly commuting algebras, we first 
decompose $Cl_{\calbd}\,({}^J\!\a)$ into a direct sum 
$$
Cl_{\calbd}\,({}^J\!\a)=Cl_{\calbd}^+\,({}^J\!\a)\oplus 
Cl_{\calbd}^-\,({}^J\!\a)
$$                                      with 
$$
Cl_{\calbd}^{\pm}\,({}^J\!\a)=\bigl\{\,\omega\in Cl_{\calbd}\,({}^J\!\a)
\,|\, \gamma\,\omega = \pm\omega\,\gamma\bigr\}\ .
$$
Then we define the ``twisted algebra'' $\widetilde{Cl}_{\calbd}
\,({}^J\!\a) := Cl_{\calbd}^+\,({}^J\!\a) \oplus
\gamma\,Cl_{\calbd}^-\,({}^J\!\a)$. This algebra  
%$\widetilde{Cl}_{\calbd}\,({}^J\!\a)$ 
commutes with $Cl_{\cald}(\a)$.  
\sn
We propose the following definitions:  The $N=2$ spectral data 
$(\a, \h, {\tt d}, T, *)$ describe a {\sl non-commutative manifold} if 
$$
\widetilde{Cl}_{\calbd}\,({}^J\!\a) = J\, Cl_{\cald}(\a)\, J \ .
$$
Furthermore, inspired by classical geometry, we say that a 
non-commutative manifold $(\a, \h, {\tt d}, T, *, \xi_0)$ is 
{\sl spin${}^c$} if the Hilbert space factorizes as a 
$Cl_{\cald}(\a)\otimes \widetilde{Cl}_{\calbd}\,({}^J\!\a)$ module 
in the form 
$$
\h = \h_{\cald} \otimes_{\cal Z} \h_{\calbd} 
$$
where ${\cal Z}$ denotes the center of $\m$. 
\mn
Next, we introduce a notion of ``quantized phase space''. We consider a 
set of $N=(1,1)$ spectral data $(\a, \h, \dd, \gamma, *)$, where we now 
think of \a\ as the algebra of phase space ``functions'' (i.e.\ of
pseudo-differential operators, in the Schr\"odinger picture of quantum 
mechanics) rather than functions over configuration space. 
We are, therefore, not postulating the existence of a cyclic and 
separating  vector for the algebra $Cl_{\cald}(\a)$. 
Instead, we define for each $\beta>0$ the {\sl temperature} or 
{\sl KMS state} 
$$
\Barintbeta\,: \cases {&$Cl_{\cald}(\a) \lra\, \C$\cr
&$\quad\quad \omega \phantom{w}\longmapsto\ \, 
{\displaystyle  \Barintbeta\omega := { {\rm Tr}_{
\scriptscriptstyle{\cal H}}\bigl(\omega e^{-\beta\cald^2}\bigr) 
\over {\rm Tr}_{\scriptscriptstyle{\cal H}}
\bigl( e^{-\beta\cald^2}\bigr) } \ , }$\cr}
$$
with no limit $\beta\to 0$ taken, in contrast to Definition 2.3. The 
$\beta$-integral $\barintbeta$ clearly is a faithful state, and 
through the GNS-construction we obtain a faithful representation of 
$Cl_{\cald}(\a)$ on a Hilbert space $\h_{\beta}$ with a cyclic and 
separating vector $\xi_{\beta}\in \h_{\beta}$ for $\m$. Each bounded 
operator $A\in{\cal B}(\h)$ on \h\ induces a bounded operator 
$A_{\beta}$ on $\h_{\beta}$; this is 
easily seen by computing matrix elements of $A_{\beta}$,  
$$
\langle\,A_{\beta} x, y\,\rangle = \Barintbeta A x y^* 
$$ 
for all $x,y\in\m \subset \h_{\beta}$, and using the explicit 
form of the $\beta$-integral. We denote the modular conjugation 
and the modular operator on $\h_{\beta}$ by $J_{\beta}$ and
$\triangle_{\beta}$, respectively, and we assume that for 
each $a\in\m$ the commutator 
$$
\lb\,\bard, J_{\beta}aJ_{\beta}\,\rb = {1\over i}\,{d\over dt}\,
\left( \left( e^{it\calbd}\right)_{\beta} J_{\beta}a J_{\beta}
\left( e^{-it\calbd}\right)_{\beta} \right) \bigg\vert_{t=0} 
$$
defines a bounded operator on $\h_{\beta}$.   \hfill\break  
\noindent Then we can define an algebra of bounded operators 
$\widetilde{Cl}_{\calbd}\,({}^{J_{\beta}}\a)$ on $\h_{\beta}$, 
which is contained in the commutant of $Cl_{\cald}\,(\a)$, 
and we say that 
the $N=(1,1)$ spectral data $(\a, \h, \dd, \gamma, *)$ describe a  
{\sl quantized phase space} if the following equation holds:  
$$
J_{\beta}\,Cl_{\cald}\,(\a)\, J_{\beta} 
= \widetilde{Cl}_{\calbd}\,({}^{J_{\beta}}\a)
$$
\bn\bn
%%%%%%%%%%%%%%%%%%%%%%%%%%%%%%%%%%%%%%%%%%%%%%%%%%%%%%%%%%%%%%%%%%%%%
%\vfil\eject
\leftline{\bf 2.3 Hermitian and K\"ahler non-commutative geometry}
\bn                     In this section, we  
introduce the spectral data describing complex non-commutative 
spaces, more specifically spaces that carry a Hermitian or a K\"ahler 
structure; the terminology is of course carried over from 
the classical case, see part I. Since these structures are more 
restrictive than the data of 
Riemannian non-commutative geometry, we will be able to derive some 
appealing properties of the space of differential forms. We also find a  
necessary condition for a set of $N=(1,1)$ spectral data to extend to 
Hermitian data. A different approach to complex non-commutative 
geometry has been proposed in \q{BC}. 
\bn\bn
{\bf 2.3.1 Hermitian and \Ntwotwo spectral data}
\bn
{\bf Definition 2.26}\quad A set of data $(\a,\h, \partial, \overline{\partial}, 
T, \overline{T}, \gamma, *)$ is called a set of {\sl Hermitian spectral data}
if 
\smallskip
\item {1)} the quintuple $(\a,\h, \partial+ \overline{\partial}, \gamma, *)$
forms a set of $N=(1,1)$ spectral data;
\smallskip
\item {2)} $T$ and $\overline{T}$ are self-adjoint bounded operators on \h, 
$\partial$ and $\overline{\partial}$ are densely defined, closed operators on \h\  
such that the following (anti-)commutation relations hold:
$$\eqalign{
&\partial^2 = \overline{\partial}{}^2 = 0\ , \quad\,
\{\,\partial, \overline{\partial}\,\} = 0\ , 
\cr
&\lb\, T, \partial\,\rb = \partial\ ,\quad\quad \lb\,
T,\overline{\partial}\,\rb = 0\ , 
\cr
&\lb\, \overline{T}, \partial\,\rb = 0\ ,\quad\quad 
\lb\, \overline{T},\overline{\partial}\,\rb = \overline{\partial}\ , 
\cr
&\lb\,T,\overline{T}\,\rb = 0 \ ;
\cr}$$ 
\smallskip
\item {3)} for any $a\in\a$, $\lb\,T,a\,\rb = \lb\,\overline{T},a\,\rb = 0\,$ 
and each of the operators $\lb\,\partial,a\,\rb$,  
$\lb\,\overline{\partial},a\,\rb$ and 
$\{\,\partial, \lb\,\overline{\partial},a\,\rb\,\}$
extends uniquely to a bounded operator on \h; 
\smallskip
\item {4)} the $\Z_2$-grading $\gamma$ satisfies 
$$\eqalign{
\{\,\gamma,\partial\,\} &= \{\,\gamma,\overline{\partial}\,\} = 0\ ,
\cr
\lb\,\gamma, T\,\rb &=\, \lb\,\gamma, \overline{T}\,\rb\, = 0\ ;
\cr}$$
\smallskip
\item{ 5)} the Hodge $*$-operator satisfies 
$$
*\,\partial = \zeta\,\overline{\partial}{}^*\,*\ ,\quad\quad 
*\,\overline{\partial} = \zeta\,\partial^*\,*  
$$
for some phase $\zeta\in\C$. 
\mn
Some remarks on this definition may be useful: The Jacobi identity and the  
equation $\{\,\partial,\overline{\partial}\,\}=0$ show that condition 3 above 
is in fact symmetric in $\partial$ and $\overline{\partial}$. 
\sn
As in section 2.2.1, a set $(\a,\h,\partial,\overline{\partial},T,\overline{T})$ 
that satisfies the first three conditions but does not involve $\gamma$ or $*$, 
can be made into a complete set of Hermitian spectral data. 
\sn
In classical Hermitian geometry, the $*$-operator can always be taken to be 
the usual Hodge $*$-operator -- up to a multiplicative redefinition in each 
degree -- since complex manifolds are orientable. 
\mn
Next, we describe conditions sufficient to equip a set of 
$N=(1,1)$ spectral data with a Hermitian structure. In subsection 2.3.2, 
Corollary 2.34, a necessary criterion is given as well. 
\mn
{\bf Proposition 2.27}\quad Let $(\a,\h,{\tt d}, \gamma, *)$ be a set of 
$N=(1,1)$ spectral data with $\lb\,\gamma,*\,\rb=0$, and let $T$ be 
a self-adjoint bounded operator on \h\ such that 
\smallskip
\item {a)} the operator $\partial := \lb\,T,{\tt d}\,\rb$ is nilpotent: 
$\partial^2 =0$; 
\smallskip
\item {b)} $\lb\,T,\partial\,\rb = \partial\,$; 
\smallskip
\item {c)} $\lb\,T, a\,\rb = 0$ for all $a\in\a$; 
\smallskip
\item {d)} $\lb\,T,\omega\,\rb \in \pi(\Omega^1(\a))$ for all 
$\omega\in\pi(\Omega^1(\a))$;
\smallskip
\item {e)} the operator $\overline{\partial} := {\tt d} - \partial$ satisfies 
$*\,\partial = \zeta\,\overline{\partial}{}^*\,*\,$, where $\zeta$ is the
phase appearing in the relations of $*$ in the $N=(1,1)$ data; 
\smallskip
\item {f)} $\lb\, T,\gamma\,\rb = 0$ and $\lb\,T,\overline{T}\,\rb = 0$,  
where $\overline{T} := - *\,T\,*^{-1}\,$.  
\sn
Then $(\a,\h,\partial,\overline{\partial},T,\overline{T}, \gamma,*)$ forms a set 
of Hermitian spectral data. 
\sn
Notice that the conditions a - d) are identical to those in Definition I$\,$2.20 of
section I$\,$2.4.1. Requirement e) will turn out to correspond to part e) of that 
definition. The relations in f) ensure compatibility of the 
operators $T$, $\gamma$ and $*$ and were not needed in the classical setting. 
\sn
{\kap Proof}: We check each of the conditions in Definition 2.26: The first
one is satisfied by assumption, 
since ${\tt d} = \partial + \overline{\partial}$ is the differential 
of $N=(1,1)$ spectral data. \hfill\break
\noindent The equalities $\partial^2 = \overline{\partial}{}^2 =
\{\,\partial,\overline{\partial}\,\}= \lb\,T,\overline{\partial}\,\rb =0$
follow {}from a) and b), as in the proof of Lemma I$\,$2.21. With this, we compute 
$$
\lb\,\overline{T},\overline{\partial}\,\rb = - \lb\,*\,T\,*^{-1},
\overline{\partial}\,\rb
= - \zeta\,*\,\lb\,T,\partial^*\,\rb\,*^{-1} = \overline{\partial} \ , 
$$ 
and since
$$
\lb\,\overline{T},\dd\,\rb = \lb\,*\,T\,*^{-1},\dd^*\,\rb^* 
= \zeta\,*\,\lb\,T,\dd\,\rb^*\,*^{-1} =  \overline{\partial}\ ,
$$ 
we obtain $\lb\,\overline{T},\partial\,\rb =0$. 
The relation $\lb\,T,\overline{T}\,\rb=0$ and self-adjointness of $T$ 
were part of the assumptions, and $\overline{T}{}^*=\overline{T}$ is 
clear {}from the unitarity of the Hodge $*$-operator. \hfill\break
\noindent That $\lb\,\partial,a\,\rb$ and  $\lb\,\overline{\partial},a\,\rb$ 
are bounded for all $a\in\a$ follows {}from the corresponding property of {\tt d} 
and {}from the assumption that 
$T$ is bounded. As in the proof of Proposition I$\,$2.22, one shows 
that $\{\,\partial, \lb\,\overline{\partial},a\,\rb\}\in\pi(\Omega_{\rm d}^2(\a))$, 
and therefore $\{\,\partial, \lb\,\overline{\partial},a\,\rb\}$ 
is a bounded operator. $T$ and $*$ commute with all
$a\in\a$ by assumption, and thus the same is true for
$\overline{T}$. \hfill\break
\noindent Using f) and the Jacobi identity, we get 
$$
\{\,\gamma,\partial\,\} = \{\,\gamma, \lb\,T,{\tt d}\,\rb\,\} = 
\lb\,T,\{\,{\tt d},\gamma\,\}\,\rb + \{\,{\tt d},\lb\,\gamma,T\,\rb\,\} =0 
$$ 
and 
$$
\{\,\gamma,\overline{\partial}\,\} = \{\,\gamma, {\tt d}- \partial\,\} = 0 \ .
$$
By assumption, $\gamma$ commutes with $T$ and $*$, therefore also with
$\overline{T}$.    \hfill\break
\noindent Finally, the relations of condition 5 in Definition 2.26 between
the $*$-operator and $\partial$, $\overline{\partial}$ follow directly {}from e) 
and $*\,{\tt d} = \zeta\,{\tt d}^*\,*\,$. \hfill\qed 
\mn
As  in classical differential geometry,  K\"ahler spaces arise 
as a special case of Hermitian geometry. In particular, 
K\"ahler spectral data provide a realization of the $N=(2,2)$ 
supersymmetry algebra:
\mn
{\bf Definition 2.28}\quad Hermitian spectral data 
$(\a,\h, \partial, \overline{\partial}, T, \overline{T}, \gamma, *)$ are called
$N=(2,2)$ or {\sl K\"ahler spectral data} if 
$$\eqalign{
&\{\,\partial,\overline{\partial}{}^*\,\} = \{\,\overline{\partial},\partial^*
\,\} = 0\ ,
\cr 
&\{\,\partial,\partial^*\,\} = \{\,\overline{\partial},\overline{\partial}{}^*
\,\} \ .
\cr}$$
\sn
Note that the first line is a consequence of the second one in classical 
complex geometry, but has to be imposed as a separate condition in the 
non-commutative setting. 
\bn
%In our discussion of Hermitian and K\"ahler geometry, we have, {}from the 
%start, worked with data whose $N=2$ supersymmetry algebra splits into two 
%``chiral halves'' -- in the terminology of section I$\,$3. This is 
%advantageous when we treat differential forms, which now form a bi-graded 
%complex. Alternatively, o
One can also define K\"ahler spectral data, as in 
section I$\,$1.2, as containing a nilpotent differential {\tt d} -- 
together with its adjoint ${\tt d}^*$ -- and two commuting
U(1) generators $L^3$ 
and $J_0$, say, which satisfy the relations (I$\,$1.49-51). This approach has 
the virtue that the complex structure familiar {}from classical differential 
geometry is already present in the algebraic formulation; see eq.\ (I$\,$1.54) 
for the precise relationship with $J_0$. Moreover, this way of 
introducing non-commutative complex geometry makes the role of Lie group 
symmetries of the spectral data explicit, which is somewhat hidden in 
the formulation of Definitions 2.26 and 2.28 and in Proposition 2.27: 
The presence of the U(1)$\,\times\,$U(1) symmetry, acting in an 
appropriate way, ensures that a set of  $N=(1,1)$ spectral data 
acquires an $N=(2,2)$ structure.  \hfill\break
\noindent               Because of the 
advantages in the treatment of differential forms, we will 
stick to the setting using $\partial$ and $\overline{\partial}$ 
for the time being, but the data with generators $L^3$ and $J_0$ will 
appear naturally in the context of symplectic geometry in section 2.5. 
\bn\bn 
%%%%%%%%%%%%%%%%%%%%%%%%%%%%%%%%%%%%%%%%%%%%%%%%%%%%%%%%%%%%%%%%%%
%\eject
\leftline{\bf 2.3.2 Differential forms}
\bn
In the context of Hermitian non-commutative geometry, we have two 
differential operators $\partial$ and $\overline{\partial}$ at our disposal. 
We begin this section with the definition of an abstract
algebra of universal forms which is appropriate for this situation. 
\mn
{\bf Definition 2.29}\quad A {\sl bi-differential algebra} \b\ is 
a unital algebra together with two {\sl anti-commuting} nilpotent 
derivations $\delta, \overline{\delta}\,:\ \b \lra\ \b\,$.   \hfill\break
\noindent A {\sl homomorphism of bi-differential algebras} 
$\varphi : \b \lra\ \b'$ is a unital algebra homomorphism  
which intertwines the derivations. 
\mn
{\bf Definition 2.30}\quad The {\sl algebra of complex universal forms}
$\Omega^{\bullet,\bullet}(\a)$ over a unital algebra \a\ is the 
(up to isomorphism) unique pair $(\iota, \Omega^{\bullet,\bullet}(\a))$
consisting of a unital bi-differential algebra $\Omega^{\bullet,\bullet}(\a)$ 
and an injective unital algebra homomorphism 
$\iota : \a \lra\ \Omega^{\bullet,\bullet}(\a)$
such that the following universal property holds: For any bi-differential 
algebra \b\ and any unital algebra homomorphism $\varphi : \a \lra\ \b\,$,  
there is a unique homomorphism $\widetilde{\varphi} : 
\Omega^{\bullet,\bullet}(\a) \lra\ \b$ of bi-differential algebras such 
that $\varphi = \widetilde{\varphi}\circ\iota$. 
\mn
The description of $\Omega^{\bullet,\bullet}(\a)$ in terms of generators 
and relations is analogous to the case of $\Omega^{\bullet}(\a)$, 
and it shows that $\Omega^{\bullet,\bullet}(\a)$ is a {\sl bi-graded
bi-differential algebra}  
$$
\Omega^{\bullet,\bullet}(\a) = \bigoplus_{r,s=0}^{\infty} 
\Omega^{r,s}(\a)
\eqno(2.39)$$
by declaring the generators $a, \delta a, \overline{\delta} a$ and 
$\delta\overline{\delta} a$, $a\in\a$, to have bi-degrees 
(0,0), (1,0), (0,1) and (1,1), respectively. 
\sn
As in the $N=(1,1)$ framework, we introduce an involution $\natural$, 
called {\sl complex conjugation}, on the algebra of complex universal 
forms, provided \a\ is a ${}^*$-algebra: 
$$
\natural\,:\ \Omega^{\bullet,\bullet}(\a) \lra\ \Omega^{\bullet,\bullet}(\a)
$$
is the unique anti-linear anti-automorphism acting on generators by 
$$\eqalignno{
&\natural(a) \equiv a^{\natural} :=  a^*\ , 
&\cr
&\natural(\delta a) \equiv (\delta a)^{\natural} := \overline{\delta} ( a^*\,)\ ,
\quad\quad\quad\quad  
\natural(\overline{\delta} a) \equiv (\overline{\delta} a)^{\natural} 
:= \delta ( a^*\,)\ , 
\phantom{XXX}&(2.40)\cr
&\natural(\delta\overline{\delta} a) \equiv (\delta\overline{\delta} a)^{\natural} 
:=  \delta\overline{\delta}(a^*\,)\ . 
&\cr}$$
Let $\tilde\gamma$ be the $\Z_2$-reduction of the total 
grading on $\Omega^{\bullet,\bullet}(\a)$, i.e., $\tilde\gamma = (-1)^{r+s}$ 
on $\Omega^{r,s}(\a)$. Then it is easy to verify that 
$$
\overline{\delta} \natural \tilde\gamma = \natural \delta \ .
\eqno(2.41)$$
This makes $\Omega^{\bullet,\bullet}(\a)$ into a unital bi-graded
bi-differential ${}^{\natural}$-algebra.
\mn
Let $(\a,\h,\partial,\overline{\partial},T,\overline{T},\gamma,*)$ be a set of
Hermitian spectral data. Then we define a $\Z_2$-graded representation
$\pi$ of $\Omega^{\bullet,\bullet}(\a)$ as a unital bi-differential 
algebra on \h\ by setting
$$\eqalignno{
&\pi(a) = a\ , 
&\cr
&\pi(\delta a) = \lb\,\partial,a\,\rb\ ,\quad\quad\quad 
\pi(\overline{\delta} a) = \lb\,\overline{\partial},a\,\rb\ ,\phantom{XXXXX}
&(2.42)\cr
&\pi(\delta\overline{\delta} a)=\{\,\partial,\lb\,\overline{\partial},a\,\rb\,\}\ .
&\cr}$$
Note that, by the Jacobi identity, the last equation is compatible with 
the anti-commutati\-vi\-ty of $\delta$ and $\overline{\delta}$.  \hfill\break
\noindent As in the case of $N=(1,1)$ geometry, we have that
$$
\pi(\delta\omega) = \lb\,\partial, \pi(\omega)\,\rb_g\ ,\quad\quad\quad\quad  
\pi(\overline{\delta}\omega) = \lb\,\overline{\partial}, \pi(\omega)\,\rb_g\ ,
\eqno(2.43)$$
for any $\omega\in\Omega^{\bullet,\bullet}(\a)$,  
and therefore the graded kernel of the representation $\pi$ has good 
properties: We define 
$$
J^{\bullet,\bullet} := \bigoplus_{r,s=0}^{\infty} J^{r,s}\ ,\quad\quad
J^{r,s} := \{\, \omega\in\Omega^{r,s}(\a)\,|\, \pi(\omega) = 0\,\}\ , 
\eqno(2.44)$$
and we prove the following statement in the same way as 
Proposition 2.21: 
\mn
{\bf Proposition 2.31}\quad The set $J$ is a two-sided, bi-graded, 
bi-differential ${}^{\natural}$-ideal of $\Omega^{\bullet,\bullet}(\a)$. 
\mn
We introduce the space of complex differential forms as 
$$
\Omega_{\partial,\bar\partial}^{\bullet,\bullet}(\a) := \bigoplus_{r,s=0}^{\infty}
\Omega_{\partial,\bar\partial}^{r,s}(\a)\ , \quad\quad
\Omega_{\partial,\bar\partial}^{r,s}(\a):= \Omega^{r,s}(\a)/ J^{r,s}\ . 
\eqno(2.45)$$     The algebra 
$\Omega_{\partial,\bar\partial}^{\bullet,\bullet}(\a)$ is a unital bi-graded
bi-differential ${}^{\natural}$-algebra, too, and the representation $\pi$ 
determines a representation, still denoted $\pi$,  of this algebra on \h. 
%The differentials induced on $\Omega_{\partial,\bar\partial}^{\bullet,\bullet}(\a)$ 
%by $\delta$ and $\overline{\delta}$ will be denoted $\partial$ and 
%$\overline{\partial}$, respectively. 
\mn
Due to the presence of the operators $T$ and $\overline{T}$ among the Hermitian 
spectral data, the image of $\Omega_{\partial,\bar\partial}^{\bullet,\bullet}(\a)$  
under $\pi$ enjoys a property not  present in the $N=(1,1)$ case:
\mn
{\bf Proposition 2.32}\quad The representation of the algebra of complex 
differential forms satisfies 
$$
\pi\bigl(\Omega_{\partial,\bar\partial}^{\bullet,\bullet}(\a) \bigr) = 
\bigoplus_{r,s=0}^{\infty}\pi\bigl(\Omega_{\partial,\bar\partial}^{r,s}(\a)\bigr)\ .
\eqno(2.46)$$
In particular, $\pi$ is a representation of
$\Omega_{\partial,\bar\partial}^{\bullet,\bullet}(\a)$ as a unital, bi-graded, 
bi-differential 
${}^{\natural}$-algebra. The ${}^\natural$-operation is implemented on 
$\pi\bigl(\Omega_{\partial,\bar\partial}^{\bullet,\bullet}(\a) \bigr)$ with
the help of the Hodge $*$-operator and the ${}^*$-operation on ${\cal B}(\h)$:
$$
\natural\,:\ \cases{
&${\displaystyle \pi\bigl(\Omega_{\partial,\bar\partial}^{r,s}(\a)\bigr) 
\lra\ \pi\bigl(\Omega_{\partial,\bar\partial}^{r,s}(\a)\bigr) }$ 
\cr
&$\phantom{MMM}\vphantom{\sum^{M}}\omega\quad \longmapsto \omega^{\natural} 
:= (-\zeta)^{r+s}\,*\,\omega^*\,*^{-1} $\cr}\ . 
$$
\sn
{\kap Proof}: Let
$\omega\in\pi\bigl(\Omega_{\partial,\bar\partial}^{r,s}(\a)\bigr)$. 
Then part 2) of Definition 2.26 implies that 
$$
\lb\,T, \omega\,\rb = r\,\omega\ ,\quad\quad 
\lb\,\overline{T}, \omega\,\rb = s\,\omega\ ,
$$
which gives the direct sum decomposition (2.46). It remains to 
show that the ${}^\natural$-operation is implemented on the space 
$\pi\bigl(\Omega_{\partial,\bar\partial}^{\bullet,\bullet}(\a) \bigr)$: 
For $a\in\a$, we have that  
$$\eqalign{
\pi\bigl((\delta a)^{\natural}\bigr) 
&=\pi\bigl(\overline{\delta}(a^*\,)\bigr) = \lb\,\overline{\partial},a^*\,\rb 
= - \lb\,\overline{\partial}{}^*, a\,\rb^* 
= - \lb\,\bar\zeta\,*\,\partial\,*^{-1}, a\,\rb^* 
= -\zeta\,*\,\lb\,\partial, a\,\rb^*\,*^{-1} 
\cr
&= -\zeta\,*\,\pi(\delta a)^*\,*^{-1}\ ,
\cr}$$ 
and, similarly, using (2.40) and the properties of the Hodge
$*$-operator,   
$$
\pi\bigl((\overline{\delta} a)^{\natural}\bigr) 
= -\zeta\,*\,\pi(\overline{\delta} a)^*\,*^{-1}\ ,
\quad\quad\ \ \pi\bigl((\delta\overline{\delta} a)^{\natural}\bigr) 
= \zeta^2\,*\,\pi(\delta\overline{\delta} a)^*\,*^{-1}\ .
$$ 
This proves that $\pi(\omega^{\natural}) = \pi(\omega)^{\natural}$. \hfill\qed
\sn
As an aside, we mention that the implementation of $\natural$ on 
$\pi\bigl(\Omega_{\partial,\bar\partial}^{\bullet,\bullet}(\a) \bigr)$ 
via the Hodge $*$-operator shows that the conditions e) of the 
``classical'' Definition I$\,$2.20 and of Proposition 2.27 are related; 
more precisely, the former is a consequence of the latter.   
\mn
Hermitian spectral data carry, in particular, an $N=(1,1)$ structure, and 
thus we have two notions of differential forms available. Their relation 
is described in our next proposition.  
\mn
{\bf Proposition 2.33}\quad The space of $N=(1,1)$ differential forms is 
included in the space of Hermitian forms, i.e., 
$$
\pi\bigl(\Omega_{\rm d}^{p}(\a) \bigr) \subset \bigoplus_{r+s=p} 
\pi\bigl(\Omega_{\partial,\bar\partial}^{r,s}(\a) \bigr) \ ,
\eqno(2.47)$$ 
and the spaces coincide if and only if 
$$
\lb\,T, \omega\,\rb \in \pi\bigl(\Omega_{\rm d}^{1}(\a) \bigr)
\quad\ {\rm for\ all}\quad \omega\in\Omega_{\rm d}^{1}(\a)\ .
\eqno(2.48)$$
\sn
{\kap Proof}: The inclusion (2.47) follows simply {}from ${\tt d} = 
\partial + \overline{\partial}$. If the spaces are equal then the equation   
$$
\lb\,T, \omega\,\rb = r\,\omega\ , 
$$ 
for all $\omega\in \pi\bigl(\Omega_{\partial,\bar\partial}^{r,s}(\a)
\bigr)\,$,   
implies (2.48). The converse is shown as in the proof of Proposition 
I$\,$2.22 in section 2.4.1 of part I, concerning classical Hermitian 
geometry.    \hfill\qed
\mn
Note that even if the spaces of differential forms do not coincide, 
the algebra of complex forms contains a graded differential algebra 
$\bigl( \Omega_{\partial,\bar\partial}^{\ \bullet}(\a), {\tt d}\bigr)$
with ${\tt d} = \partial + \overline{\partial}$ and
$$
\Omega_{\partial,\bar\partial}^{\ \bullet}(\a) 
   := \bigoplus_p \Omega_{\partial,\bar\partial}^{\ p}(\a)\ ,  \quad\quad
\Omega_{\partial,\bar\partial}^{\ p}(\a) := \bigoplus_{r+s=p} 
\Omega_{\partial,\bar\partial}^{r,s}(\a)\ .
\eqno(2.49)$$
By Proposition 2.32, we know that 
$$
\pi\bigl( \Omega_{\partial,\bar\partial}^{\bullet,\bullet}(\a)\bigr)
= \bigoplus_p \pi\bigl(\Omega_{\partial,\bar\partial}^{\ p}(\a)\bigr) \ ,
$$
and hence we obtain a necessary condition for $N=(1,1)$ spectral data 
to extend to Hermitian spectral data:
\mn
{\bf Corollary 2.34}\quad If a set of $N=(1,1)$ spectral data extends to  
a set of Hermitian spectral data then
$$
\pi\bigl( \Omega_{\rm d}^{\bullet}(\a)\bigr)
= \bigoplus_p \pi\bigl(\Omega_{\rm d}^{p}(\a)\bigr) \ .
$$
\mn
This condition is clearly not sufficient since it is always satisfied 
in classical differential geometry. 
\mn
Beyond the complexes (2.45) and (2.49), one can of course also consider 
the analogue of the {\sl Dolbeault complex} using only the differential 
$\overline{\partial}$ acting on
$\Omega^{\bullet,\bullet}_{\partial,\bar\partial}(\a)$. 
The details are straightforward. 
\bn           We conclude this  
subsection with some remarks concerning possible variations of our 
Definition 2.26 of Hermitian spectral data. For example, one may wish to 
drop the boundedness condition on the operators $T$ and $\overline{T}$, 
in order to include infinite-dimensional spaces into the theory.   
This is possible, but then one has to make some stronger assumptions 
in Proposition 2.27. 
\sn
Another relaxation of the requirements in Hermitian spectral data  
is to avoid introducing $T$ and $\overline{T}$ altogether, and to 
replace them by a decomposition of the $\Z_2$-grading 
$$
\gamma = \gamma_{\partial} + \gamma_{\bar\partial}
$$
such that 
$$\eqalign{
&\{\,\gamma_{\partial}, \partial\,\} = 0\ ,\quad\quad       
\lb\,\gamma_{\partial}, \overline{\partial}\,\rb = 0\ , 
\cr
&\{\,\gamma_{\bar\partial}, \overline{\partial}\,\} = 0\ ,\quad\quad 
\lb\,\gamma_{\bar\partial}, \partial\,\rb = 0\ . 
\cr}$$
Then the space of differential forms may be defined as above, but 
Propositions 2.32 and 2.33, as well as the good properties of the 
integral established in the next subsection, will not hold in general.
\bn\bn
%%%%%%%%%%%%%%%%%%%%%%%%%%%%%%%%%%%%%%%%%%%%%%%%%%%%%%%%%%%%%%%%%%
%\vfil\eject
\leftline{\bf 2.3.3 Integration in complex non-commutative geometry}
\bn
The definition of the integral is completely analogous to the $N=(1,1)$ 
setting: Again we use the operator $\triangle = {\tt d}\,{\tt d}^* + 
{\tt d}^*\,{\tt d}$, where now ${\tt d} = \partial + \overline{\partial}$. 
Due to the larger set of data, the space of square-integrable,  
complex differential forms, now obtained after quotienting by the
two-sided {\sl bi-graded} ${}^{\natural}$-ideal $K$, has better properties 
than the corresponding space of forms in Riemannian non-commutative 
geometry. There, two elements 
$\omega\in\Omega^p_{\rm d}(\a)$ and $\eta\in\Omega^q_{\rm d}(\a)$
with $p\neq q$ were not necessarily orthogonal with respect to the 
sesqui-linear form 
$(\cdot,\cdot)$ induced by the integral. For Hermitian and K\"ahler
non-commutative geometry, however, we can prove the following orthogonality 
statements: 
\mn
{\bf Proposition 2.35}\quad Let $\omega_i \in
\pi\bigl(\Omega_{\partial,\bar\partial}^{r_i,s_i}(\a)\bigr)$, $i=1,2$. 
Then 
$$
(\omega_1, \omega_2) = 0  
\eqno(2.50)$$
if $r_1+s_1 \neq r_2+s_2$ in the Hermitian case; if the spectral data also 
carry an $N=(2,2)$ structure, then eq.\  (2.50) holds as soon as $r_1\neq
r_2$ or $s_1\neq s_2$. 
\sn
{\kap Proof}: In the case of Hermitian spectral data, the assertion follows
immediately {}from cyclicity of the trace, {}from the commutation relations 
$$
\lb\, T, \omega_i\,\rb = r_i\,\omega_i\ , \quad\quad 
\lb\, \overline{T}, \omega_i\,\rb = s_i\,\omega_i\ , 
$$
which means that $T+\overline{T}$ counts the total degree of a differential 
form, and {}from the equation  
$$
\lb\, T+\overline{T}, \triangle\,\rb = 0\ .
$$
In the K\"ahler case, Definition 2.28 implies the stronger relations 
$$
\lb\, T, \triangle\,\rb = \lb\, \overline{T}, \triangle\,\rb = 0\ .
\eqno{\qed} 
$$
\bn\bn
{\bf 2.3.4 Generalized metric on \Omtipar} 
\bn
The notions of vector bundles, Hermitian structure, torsion, etc.\ 
are defined just as for $N=(1,1)$ spectral data in section 2.2.  
The definitions of holomorphic vector bundles and connections 
can be carried over {}from the classical case; see section I$\,$2.4.4.
Again, we pass {}from $\Omega^{\ 1}_{\partial,\bar\partial}$, 
see eq.\ (2.49), to the space of all square-integrable 1-forms  
$\widetilde{\Omega}^{\ 1}_{\partial,\bar\partial}$, which is equipped with 
a generalized Hermitian structure $\langle\cdot,\cdot\rangle_{\scpbp}\,$ 
according to the construction in Theorem 2.9. Starting {}from here, we 
can define an analogue 
$$
\langle\!\langle\cdot,\cdot\rangle\!\rangle\,:\ 
\widetilde{\Omega}^{\ 1}_{\partial,\bar\partial}(\a)
\times \widetilde{\Omega}^{\ 1}_{\partial,\bar\partial}(\a)  \lra\ \C 
$$
of the $\C$-bi-linear metric in classical complex geometry by 
$$
\langle\!\langle\,\omega,\eta\,\rangle\!\rangle := 
\langle\,\omega,\eta^{\natural}\,\rangle_{\scpbp}
\ .
$$
\mn
{\bf Proposition 2.36}\quad  The generalized metric 
$\langle\!\langle\cdot,\cdot\rangle\!\rangle$ on 
$\widetilde{\Omega}^{\ 1}_{\partial,\bar\partial}(\a)$
has the following properties: 
\smallskip
\item {1)} $\langle\!\langle\,a\omega,\eta\,b\,\rangle\!\rangle
= a\,\langle\!\langle\,\omega,\eta\,\rangle\!\rangle\,b\,$;
\smallskip
\item {2)} $\langle\!\langle\,\omega\,a,\eta\,\rangle\!\rangle
= \langle\!\langle\,\omega,a\ \eta\,\rangle\!\rangle\,$;
\smallskip
\item {3)} $\langle\!\langle\,\omega,\omega^{\natural}\,\rangle\!\rangle
\geq 0\ $; 
\sn
here $\omega,\,\eta\in\widetilde{\Omega}^{\ 1}_{\partial,\bar\partial}(\a)$ 
and $a,b\in\a$. If the underlying spectral data are K\"ahlerian, one has that 
$$
\langle\!\langle\,\omega,\eta\,\rangle\!\rangle= 0
$$ 
if $\,\omega,\eta\in\widetilde{\Omega}^{0,1}_{\partial,\bar\partial}(\a)\,$
or $\,\omega,\eta\in\widetilde{\Omega}^{1,0}_{\partial,\bar\partial}(\a)\,$.
\sn
{\kap Proof}: The first three statements follow directly {}from the 
definition of $\langle\!\langle\cdot,\cdot\rangle\!\rangle$ and the 
corresponding properties of $\langle\cdot,\cdot\rangle_{\scpbp}$ 
listed in Theorem 2.9. The last 
assertion is a consequence of Proposition 2.35, using the 
fact that the spaces $\widetilde{\Omega}^{r,s}_{\partial,\bar\partial}(\a)$
are \a-bi-modules. Note that this property of the metric $\langle\!\langle
\cdot,\cdot\rangle\!\rangle$ corresponds to the property 
$g_{\mu\nu}=g_{\bar\mu\bar\nu}=0$ (in complex coordinates) in the 
classical case.  \hfill\qed
\bn\bn
%%%%%%%%%%%%%%%%%%%%%%%%%%%%%%%%%%%%%%%%%%%%%%%%%%%%%%%%%
{\bf 2.4 The \Nfourfour spectral data}
\bn
We just present the definition of spectral data 
describing non-commutative Hyperk\"ahler spaces.
Obviously, it is chosen in analogy to the discussion of the classical 
case in section 2.5 of part I.
\mn
{\bf Definition 2.37}\quad A set of data $(\a,\h,G^{a\pm}, 
\overline{G}{}^{a\pm}, T^i, \overline{T}{}^i, \gamma,*)$ with 
$a=1,2$, $i=1,2,3$, is called a set of $N=(4,4)$ or {\sl 
Hyperk\"ahler spectral data} if 
\smallskip
\item {1)} the subset $(\a,\h,G^{1+}, \overline{G}{}^{1+}, 
T^3, \overline{T}{}^3, \gamma,*)$ forms a set of $N=(2,2)$ 
spectral data; 
\smallskip
\item {2)} $G^{a\pm}$, $a=1,2$  are closed, densely defined operators   
on \h, and $T^i$, $i=1,2,3$,  are bounded operators on \h\ which 
satisfy $\bigl(G^{a\pm}\bigr)^*=  G^{a\mp}\,,\ \bigl(T^i\bigr)^* = T^i$ 
and the following (anti-)commutation relations ($a,b = 1,2,\ i,j=1,2,3$, 
and $\tau^i$ are the Pauli matrices): 
$$\eqalign{
&\{\, G^{a+}, G^{b+} \} = 0\ ,\quad\  \ \ \  \ \,
\{\, G^{a-}, G^{b+} \} = \delta^{ab}\,\square\,\ ,\   
\cr
&\lb\, \square\,, G^{a+}\, \rb = 0\ , \quad\quad \; \ \  \ \ \,\ \,
\lb\, \square\,, T^i\, \rb = 0\ , 
\cr
&\lb \,T^i, T^j \,\rb = i \epsilon^{ijk}\, T^k\ ,\quad\ \ \ 
\lb \,T^i, G^{a+} \,\rb = {1\over2}\,\overline{\tau^i_{ab}}\, G^{b+}\ ,
\phantom{xxxxxxx}\cr}$$
for some self-adjoint operator\ $\,\square\,$\ on $\h$, which, in the classical 
case, is the holomorphic part of the Laplace operator;
%%%%%%%%%%%%%%%%%%%%%%%%
%$$\eqalignno{
%&\lb\, \square\,, G^{a+}\, \rb = 0\ , \quad a=1,2\ , 
%&(3.83)\cr
%&\lb\, \square\,, T^i\, \rb = 0\ , \quad i=1,2,3\ , 
%&(3.84)\cr
%&\{\, G^{a+}, G^{b+} \} = 0\ ,\quad a,b = 1,2\ ,  
%&(3.85)\cr
%&\{\, G^{a-}, G^{b+} \} = \delta^{ab}\,\square\,\ ,\quad a,b = 1,2\ ,  
%&(3.86)\cr
%&\lb \,T^i, G^{a+} \,\rb = {1\over2}\,\overline{\tau^i_{ab}}\, G^{b+}\ ,
%\quad i=1,2,3\,,\ a=1,2\ ,  
%&(3.87)\cr
%&\lb \,T^i, T^j \,\rb = i \epsilon^{ijk}\, T^k\ ,\quad i,j = 1,2,3\ ;  
%&(3.88)\cr}$$
%$\tau^i$ are the Pauli matrices defined in the first section. 
%In addition, the following Hermiticity conditions hold: 
%$$
%\square^* = \square\ ,\quad 
%\bigl(G^{a\pm}\bigr)^*=  G^{a\mp}\ ,\quad  
%\bigl(T^i\bigr)^* = T^i \ .
%\eqno(3.89)$$
%In particular, these operators generate a finite-dimensional
%$\Z_2$-graded (or super) Lie algebra. 
%%%%%
\smallskip
\item {3)} the operators $\overline{G}{}^{a\pm}$, $a=1,2$, and   
$\overline{T}{}^i$, $i=1,2,3$, also satisfy the conditions in 
2) and (anti-)commute with $G^{a\pm}$ and  $T^i$. 
\mn
The construction of non-commutative differential forms and the 
integration theory 
is precisely the same as for $N=(2,2)$ spectral data. We therefore  
refrain {}from giving more details. It might, however, be interesting to see  
whether the additional information encoded in $N=(4,4)$ 
spectral data gives rise to special properties,  beyond 
the ones found for K\"ahler data in subsection 2.3.3. 
\bn\bn
\vfill\break
%%%%%%%%%%%%%%%%%%%%%%%%%%%%%%%%%%%%%%%%%%%%%%%%%%%%%%%%%
\noindent
{\bf 2.5 Symplectic non-commutative geometry}
\bn             
Once more, our description in the non-commutative context 
follows the  algebraic characterization 
of classical symplectic manifolds given in section 2.6 of part I. 
The difference between our approaches to the classical
and to the non-commutative case is that, in the former, we 
could derive most of the algebraic relations -- including  
the SU(2) structure showing up on symplectic manifolds -- {}from 
the specific properties of the symplectic 2-form, whereas now 
we will instead include those relations into the defining data, as 
a ``substitute'' for the symplectic form. 
\def\tdst{\widetilde{{\tt d}}^*}\def\td{\widetilde{{\tt d}}}
\mn
{\bf Definition 2.38}\quad The set of data $(\a,\h, {\tt d}, L^3, 
L^+, L^-, \gamma,*)$ is called a set of {\sl symplectic  
spectral data} if 
\smallskip
\item {1)} $(\a,\h, {\tt d}, \gamma,*)$ is a set of $N=(1,1)$ 
spectral data; 
\smallskip
\item {2)} $L^3$, $L^+$ and $L^-$ are bounded operators on \h\ which  
commute with all $a\in\a$ and satisfy the sl$_2$ commutation relations 
$$ 
\lb\, L^3, L^{\pm}\,\rb = \pm 2 L^{\pm}\ ,\quad 
\lb\, L^+, L^-\,\rb = L^3\ 
$$                   as well as 
the Hermiticity properties $(L^3)^*=L^3$, $(L^{\pm})^*= L^{\mp}$;
furthermore, they commute with the grading $\gamma$ on \h; 
\item {3)} the operator $\tdst := \lb\,L^-, {\tt d}\,\rb$ 
is densely defined and closed, and together with {\tt d} it forms 
an SU(2) doublet, i.e., the following commutation relations hold:  
$$\eqalign{
&\lb\, L^3, {\tt d} \,\rb = {\tt d}\ ,\quad\ 
\lb\, L^3, {\tdst} \,\rb = - \tdst\ ,
\cr
&\lb\, L^+, {\tt d} \,\rb = 0 \ ,\quad\  
\lb\, L^+, {\tdst} \,\rb = {\tt d}\ ,\quad
\cr 
&\lb\, L^-, {\tt d} \,\rb = \tdst\ ,\quad
\lb\, L^-, {\tdst} \,\rb = 0\ .
\cr}$$
\mn
As in the classical case, there is a second SU(2) doublet spanned 
by the adjoints ${\tt d}^*$ and $\td$. The Jacobi identity 
shows that $\tdst$ is nilpotent and that it anti-commutes with {\tt d}. 
\sn
Differential forms and integration theory are formulated just as for 
$N=(1,1)$ spectral data, but the presence of SU(2) generators among 
the symplectic spectral data leads to additional interesting features, 
such as the following: Let $\omega\in\Omega_{\rm d}^k(\a)$ 
and $\eta\in\Omega_{\rm d}^l(\a)$ be two differential forms. 
Then their scalar product, see eq.\ (2.38), vanishes unless $k=l$:
$$
(\omega,\eta) = 0 \quad\hbox{if $k\neq l$}\ .
\eqno(2.51)$$ 
This is true because, by the SU(2) commutation relations listed above, 
the operator $L^3$ induces a $\Z$-grading on differential forms, and 
because $L^3$ commutes with the Laplacian $\triangle = {\tt d}^*{\tt d}
+{\tt d}{\tt d}^*$. One consequence of (2.51) is that 
the reality property of $(\cdot,\cdot)$ stated in   
Proposition 2.22 is valid independently of the phase occurring in the 
Hodge relations. 
\mn
The following proposition shows that we can introduce an $N=(2,2)$ 
structure on a set of symplectic spectral data if certain 
additional properties are satisfied. As was the case for 
Definition 2.38, the extra requirements are slightly stronger 
than in the classical situation, where some structural elements like 
the almost-complex structure are given automatically. In the K\"ahler 
case, the latter allows for a separate counting of holomorphic 
resp.\ anti-holomorphic degrees of differential forms, which in turn 
ensures that the symmetry group of the symplectic data associated 
to a classical K\"ahler manifold  
is in fact SU(2)$\,\times\,$U(1) -- see also section 3 of part I. 
Without this enlarged symmetry group, it is impossible to re-interpret 
the $N=4$ as an $N=(2,2)$ supersymmetry algebra. Therefore, we explicitly 
postulate the existence of an additional U(1) generator  in the 
non-commutative context --  which coincides with the U(1) generator $J_0$ in  
eq.\ (I$\,$1.49) of section I$\,$1.2 and is intimately related to the complex 
structure.  
\sn
{\bf Proposition 2.39}\quad Suppose that the SU(2) generators of a  
set of symplectic spectral data satisfy the following relations with 
the Hodge operator:
$$
*\,L^3  = - L^3\,*\ ,\quad\quad *\,L^+ = - \zeta^2\,L^-\,*\ ,
$$
where $\zeta$ is the phase appearing in the Hodge relations of the $N=(1,1)$ 
subset of the symplectic data. Assume, furthermore, that there exists a 
bounded self-adjoint operator $J_0$ on \h\ which commutes with all $a\in\a$, 
with the grading $\gamma$, and with $L^3$, whereas it acts like  
$$
\lb\,J_0, {\tt d}\,\rb = -i\,\td\ ,\quad\quad
\lb\,J_0, \td\,\rb = i\,{\tt d}
$$
between the SU(2) doublets. 
Then the set of symplectic data carries an $N=(2,2)$ K\"ahler structure 
with 
$$\eqalign{
\partial &= {1\over2}\,({\tt d} - i\,\td)\ ,\quad\quad\quad 
\overline{\partial} =  {1\over2}\,({\tt d} + i\,\td)\ ,
\cr
T &= {1\over2}\,(L^3 + J_0)\ ,\quad\quad\, 
\overline{T} = {1\over2}\,(L^3 - J_0)\ .
\cr}$$
\sn
{\kap Proof}: All the conditions listed in Definition 2.26 of Hermitian
spectral data can be verified easily: Nilpotency of $\partial$ and 
$\overline{\partial}$ follows from ${\tt d}^2= \td^2=0$ and 
$$\{\,{\tt d},\td\,\}=0\ ,\eqno(2.52)$$ 
and the action of the Hodge operator on the SU(2) generators 
ensures that $*$ intertwines $\partial$ and $\overline{\partial}$ in the 
right way. As for the extra conditions in Definition 2.28 of K\"ahler 
spectral data, one sees that the first one is always true for 
symplectic spectral data, whereas the second one, namely the equality of 
the ``holomorphic'' and ``anti-holomorphic'' Laplacians, is again a 
consequence of relation (2.52).      \hfill\qed 
\bn\bn
\vfill\eject 
%\bye
%%%%%%%%%%%%%%%%%%%%%%%%%%%%%%%%%%%%%%%%%%%%%%%%%%%%%%%%%%%%%%%%%%%%%%%%
%%%%%%%%%%%%%%%%%%%%%%%%%%%%%%%%%%%%%%%%%%%%%%%%%%%%%%%%%%%%%%%%%%%%%%%
%%%%%%%%%%%%%%%%%%%%%%%%%%%%%%%%%%%%%%%%%%%%%%%%%%%%%%%%%%%%%%%%%%%%%%%%%
%\pageno=42
\leftline{\bf 3. The non-commutative  3-sphere}
\bn
Here and in the next section, we present two examples of 
non-commutative spaces and show how the general methods developed 
above can be applied. We first discuss the ``quantized'' or ``fuzzy'' 
3-sphere. We draw some inspiration from the conformal field theory 
associated to a non-linear $\sigma$-model with target being a
3-sphere, the so-called SU(2)-WZW model, see \q{Wi3} and also 
\q{FGK,PS}. But while the ideas on a 
non-commutative interpretation of conformal field theory models 
proposed in \q{FG} are essential for placing non-commutative geometry 
into a string theory context, the following calculations are self-contained; 
the results of subsections 3.2 and 3.3 are taken from \q{Gr}. 
Although there is no doubt that the methods used in \q{Gr} and below 
can be extended to arbitrary compact, connected and simply connected 
Lie groups, we will, for simplicity, restrict ourselves to the case of SU(2). 
\sn
We first introduce a set of $N=1$ spectral data describing 
the non-commutative 3-sphere, then discuss the de Rham complex
and its cohomology, and finally turn towards geometrical aspects 
of this non-commutative space. Subsection 3.4 briefly describes the 
$N=(1,1)$ formalism.
\bn\bn
%%%%%%%%%%%%%%%%%%%%%%%%%%%%%%%%%%%%%%%%%%%%%%%%%%%%%%%
\leftline{\bf 3.1 The $\Neqone$ data associated to the 3-sphere}
\bn
In this subsection, we introduce $N=1$ data describing the non-commutative
3-sphere.
\sn
Since the 3-sphere is diffeomorphic to the Lie group $G=\rm{SU}(2)$, we are 
looking for data describing a Lie group $G\,$. Let $\{T_A\}$ be a basis of 
{\tt g}$\;= T_eG\,$, the Lie algebra of $G\,$. By $\vartheta_A$ and 
$\overline{\vartheta}_A$
we denote the left- and right-invariant vector fields associated to the basis
elements $T_A\,$, and by $\theta^A$ and $\overline{\theta}{}^A$ the corresponding
dual basis of 1-forms. The structure constants $f_{AB}^C$ are defined, 
as usual, by
$$
	\lb\vartheta_A,\,\vartheta_B\rb=f_{AB}^C\vartheta_C\ .
\eqno(3.1)	
$$
The Killing form on {\tt g} induces a canonical Riemannian
metric on $TG$ given by
$$
	g_{AB}\; \equiv \; g(\vartheta_A,\,\vartheta_B)=
	-{\rm{Tr}}\,({\rm{ad}}_{T_A}\circ{\rm{ad}}_{T_B})=
	-f_{AC}^Df_{BD}^C\ ,
\eqno(3.2)	
$$
and the Levi-Civita connection reads
$$
	\nabla_{\!A}\vartheta_B\,\equiv\,\nabla_{\vartheta_A}\vartheta_B=
	{1\over2}f_{AB}^C\vartheta_C\ .
\eqno(3.3)	
$$
The left-invariant vector fields $\vartheta_A$ define a trivialization 
of the (co-)tangent bundle. We denote by $\nabla^L$  the 
flat connection associated to that trivialization, 
$$
\nabla^L \theta^A = 0 
$$
for all $A$. 
We introduce the operators 
$$
a^{A\,*} = \theta^A\, \wedge\ \,,\quad a^A = g^{AB}\,\vartheta_B \;\llcorner 
$$
on the space of differential forms, as well as the usual gamma matrices 
$$
\gamma^A = a^{A\,*} - a^A\ , \quad 
\overline{\gamma}{}^A = i\,(\, a^{A\,*} + a^A\,)\ .
\eqno(3.4)$$
It is easy to verify that $\gamma^A$ and $\overline{\gamma}{}^A$ 
generate two anti-commuting copies of the Clifford algebra, 
$$
	\{\,\gamma^A,\,\gamma^B\}=
	\{\,\overline{\gamma}{}^A,\,\overline{\gamma}{}^B\}=-2g^{AB}\ ,\quad
	\{\,\gamma^A,\,\overline{\gamma}{}^B\}=0\ .
\eqno(3.5)
$$
Following the notations of section I$\,$2.2, we shall denote by $\S$ the 
bundle of differential forms endowed with the above structures. 
\sn
We define two connections $\nabla^{\S}$ and $\overline{\nabla}{}^{\S}$ 
on $\S$ by setting 
$$\eqalign{
\nabla^{\S}&=\theta^A\otimes(\nabla^L_{\vartheta_A} 
    + {1\over12}\,f_{ABC}\gamma^B\gamma^C)\ ,
\cr
\overline{\nabla}{}^{\S}&=
	\overline{\theta}^A\otimes(\nabla^L_{\overline{\vartheta}_A} 
     -	{1\over12}\,f_{ABC}\overline{\gamma}{}^B\overline{\gamma}{}^C) \ ,
\cr}\eqno(3.6)
$$
where $f_{ABC}=f_{AB}^Dg_{DC}\,$, and we put 
$$
J_A:=i\,\nabla^L_{\vartheta_A}\ ,\quad \psi^A:=-i\,\gamma^A\ ,\quad
\overline{J}_A:=-i\,\nabla^L_{\overline{\vartheta}_A}\ ,\quad 
\overline{\psi}{}^A:=i\,\overline{\gamma}{}^A\ .
\eqno(3.7)
$$
These objects satisfy the commutation relations
$$
	\lb\,J_A,J_B\,\rb=if_{AB}^C\,J_C\ ,\quad
	\{\,\psi^A,\psi^B\,\}=2g^{AB}\ , 
\eqno(3.8)
$$
with analogous relations for $\overline{J}_A$ and $\overline{\psi}{}^A$; 
barred and unbarred operators (anti-)commute. 
\sn
The two anti-commuting Dirac operators ${\cal D}$ and 
$\overline{\cal D}$ on $\S$ read \q{FG} 
$$\eqalign{
{\cal D}&=\psi^A J_A-{i\over12}\,f_{ABC}\psi^A\psi^B\psi^C
\cr
\overline{\cal D}&=
	\overline{\psi}^A\overline{J}_A-{i\over12}\,f_{ABC}
	\overline{\psi}{}^A\overline{\psi}{}^B\overline{\psi}{}^C
\cr}\eqno(3.9)
$$
where $\c$ and $\overline{\c}$ are the Clifford actions defined by 
the gamma matrices of eq.\ (3.4). The $\Z_2$-grading operator $\gamma$ 
on $\S\,$, anti-commuting with ${\cal D}$ and 
$\overline{\cal D}$, is given by
$$
\gamma={1\over i\,(3!)^2}\,g\;\varepsilon_{ABC}\,\varepsilon_{DEF}\;
\psi^A\psi^B\psi^C\,\overline{\psi}{}^D\overline{\psi}{}^E
\overline{\psi}{}^F\ ,
\eqno(3.10)$$
where $g=\det g_{AB}\,$. By $ {L}^2(\S)\simeq {L}^2(G)\otimes W$, 
where $W$ is the irreducible representation of the Clifford algebra 
of eqs.\ (3.4,5), 
we denote the Hilbert space of square integrable sections of the 
bundle $\S\,$, with respect to the normalized Haar measure on $G\,$. In
the language of Connes' spectral triples, the classical 3-sphere is 
described by the $N=1$ data $( {L}^2(\S),\,\cinftyg,\,D,\,\gamma)\,$,
with $D\, \equiv\,{\cal D}$.
\sn
The Hilbert space $ {L}^2(\S)$ carries a unitary representation $\pi$ of
$G\times G$ given by
$$
	\bigl(\pi(g_1,\,g_2)f\bigr)(h)=f(g_1^{-1}hg_2)\ ,
\eqno(3.11)$$
for all $g_i,\,h\in G$ and $f\in {L}^2(G)\,$. For each
$j\in{1\over2}\Z_+$ we denote by $(\pi^j,\,V_j)$ the irreducible unitary
$(2j+1)$-dimensional (spin $j$) representation of $G\,$, and to each vector
$\xi^*\otimes\eta\in V_j^*\otimes V_j$ we associate a smooth function 
$f_{\xi^*\otimes\eta}\in\cinftyg$ by setting
$$
f_{\xi^*\otimes\,\eta}(g)={1\over \sqrt{2j+1}}\,
\langle\,\xi^*, \pi^j(g)\eta\,\rangle\ .
\eqno(3.12)$$
This defines a linear isometry
$$
	\varphi\,:\ \bigoplus_{j\in{1\over2}\Z_+}V_j^*\otimes V_j
	\longrightarrow {L}^2(G)\ ,
\eqno(3.13)$$
and the Peter-Weyl theorem states that the image of $\varphi$ is dense in
$ {L}^2(G)$ and also in $C(G)$ in the supremum norm topology. It is easy
to verify that the operators $J_A$ and $\overline{J}_A$ act 
on $\bigoplus_{j\in{1\over2}\Z_+}V_j^*\otimes V_j\,$ as 
$d\pi(T_A,\,1)$ and $d\pi(1,\,T_A)\,$,
respectively. For each positive integer $k\,$, we denote by $P_{(k)}$ the
orthogonal projection
$$
	P_{(k)}\,:\ {L}^2(G)\lra \h_0 :=
	\bigoplus_{j=0,{1\over2},\,\ldots}^{k/2}V_j^*\otimes V_j\ .
\eqno(3.14)$$
The Dirac operator $D$ and the $\Z_2$-grading $\gamma$ clearly leave the
{\sl finite-dimensional} Hilbert space $\h_0\otimes W$ invariant. We define
$\a_0$ to be the unital subalgebra of ${\rm End}(\h_0)$ generated by
operators of the form $P_{(k)}f_{\xi^*\otimes\eta}\,$, where
$\xi^*\otimes\eta\in\h_0\,$. The following theorem is proven in \q{Gr}:
\sn
{\bf Theorem 3.1}\quad 
	The algebra $\a_0$ coincides with the algebra of endomorphisms
   of $\h_0\,$,	i.e., 
$$\a_0={\rm End}(\h_0)\ .$$
\sn
The proof in \q{Gr} shows that $\a_0$ is a full matrix algebra for any
compact, connected and simply connected group. That $\a_0$ equals the
endomorphism ring of $\h_0$ was only proved for SU(2), but a slight
generalization of the proof for SU(2) should yield the result for all
groups of the above type.
\mn
We define the non-commutative 3-sphere by the $N=1$ data
$(\a_0,\,\h_0\otimes W,\,D,\,\gamma)\,$. Notice that this definition of the
non-commutative 3-sphere is very close to that of the non-commutative 2-sphere
\q{Ber,Ho,Ma,GKP}. For an alternative derivation of this definition, 
the reader is referred to \q{FG} where it is shown how this
space arises as the quantum target of the WZW model based on
$SU(2)\,$. 
\sn
We note that $1/k$ plays the role of Planck's constant $\hbar$ in the 
quantization of symplectic manifolds, i.e., it is a deformation parameter. 
Formally, the classical 3-sphere emerges as the limit of non-commutative 
3-spheres as the deformation parameter $1/k$ tends to zero. 
\bn\bn
\eject
%%%%%%%%%%%%%%%%%%%%%%%%%%%%%%%%%%%%%%%%%%%%%%%%%%%%%%%%
\leftline{\bf 3.2 The topology of the non-commutative 3-sphere}
\bn
In this subsection, we shall apply the tools of subsection 2.1 to 
the non-commutative space $(\a_0,\,\h_0\otimes W,\,D,\,\gamma)$ describing
the non-commutative 3-sphere; we follow the presentation in \q{Gr}. 
For convenience, we shall choose the basis $\{T_A\}$ of $T_eG$ in such a way
that $g_{AB}=2\delta_{AB}\,$. The structure constants are then given by the
Levi-Civita tensor, $f_{AB}^C=\varepsilon_{ABC}\,$.
\bn\bn
%%%%%%%%%%%%%%%%%%%%%%%%%%%%%%%%%%%%%%%%%%%%%%%%%%%%%%%%%%%%%%%%%%%%%%%%%
%\eject
\leftline{\bf 3.2.1 The de Rham complex}
\bn
First, we determine the structure of the spaces of differential
forms \df{n} and the action of the exterior differentiation
$\delta:\df{\bullet}\longrightarrow\df{\bullet}\,$. We use the same notations
as in subsection 2.1.2.
\sn
The space of $1$-forms is
$$
	\df{1}\,\simeq\,\pi(\uf{1})\,=\,
	\Bigl\{\,\sum_i a_0^i\,\lb\, J^A,a_1^i\,\rb\otimes\psi^A\;
	\Bigl|\;a_j^i\in\a_0 \,\Bigr\}\ .
\eqno(3.15)
$$
Since $\a_0$ is a full matrix algebra, see Theorem 3.1, it follows that
$$
	\df{1}\simeq\{\,a_A\otimes\psi^A\,|\,a_A\in\a_0\,\}\ .
\eqno(3.16)
$$
Using the fact that any element of $\pi(\uf{2})$ can be written as a linear
combination of products of pairs of elements in $\pi(\uf{1})$, we get
$$
	\pi(\uf{2})=\{\,a_{AB}\otimes\psi^A\psi^B\,|\,a_{AB}\in\a_0\,\}\ .
\eqno(3.17)
$$
Our next task is to determine the space $\pi(\delta J^1)$ of so-called 
``auxiliary $2$-forms'', see eq.~(2.2). To this end, let 
$\omega=\sum_i a_i\delta b_i\in\uf{1}$ be such that
$$
	\pi(\omega)=\sum_i a_i\lb D\,,\,b_i\rb=0\ .
\eqno(3.18)
$$
Using eqs.~(3.8) and (3.18), we see that the coefficient of 
$\lb\,\psi^A,\psi^B\,\rb$ in $\pi(\delta\omega)$ is proportional to
$$\eqalign{
\varepsilon^{AB}\sum_i
	{}[\, &J^A,a_i{}\,]{}[\, J^B,b_i{}\,]=
	-\varepsilon^{AB}\sum_i 
	a_i\bigl[\,J^A,{}[\, J^B,b_i{}\,]\bigr]\quad\quad\cr
	&=-{1\over2}\varepsilon^{AB}\sum_i
	a_i\bigl[\,{}[\, J^A,J^B{}\,],b_i\,\bigr]=
	-{i\over2}\varepsilon^{AB}\varepsilon^{ABC}\sum_i
	a_i{}[\, J^C,b_i{}\,]=0\ ,
\cr}$$
where $\varepsilon^{AB}$ denotes the Levi-Civita antisymmetric tensor.
This shows that $\pi(\delta J^1)$ is included in $\a_0\,$, and since $\a_0$ is a
full matrix algebra, this implies that $\pi(\delta J^1)$ is either $0$ or 
equal to $\a_0\,$. We construct a non-vanishing element of $\pi(\delta J^1)$
explicitly. Let $P_j$ be the orthogonal projection onto $V_j^*\otimes V_j$. We
define $a\,,\,b\in\a_0$ by
$$
	a=P_0\,a\,P_{1/2}\ ,\quad b=P_{1/2}\,b\,P_0
$$
and
$$\eqalign{
	&a\,:\;V_{1/2}^*\otimes V_{1/2}\ni
	%\left(\begin{array}{c} \alpha\\ \beta \end{array} \right)
{\alpha \choose \beta} \otimes {\gamma \choose \delta} 	
%\otimes  \left(\begin{array}{c} \gamma\\ \delta \end{array} \right)
	\ \ \longmapsto\ \ 
	\alpha-2\beta+2\gamma+\delta\ \ 
	\in \quad V_0^*\otimes V_0\cr
	&b\,:\ V_0^*\otimes V_0\quad\ni\quad \alpha
	\ \ \longmapsto	\ \ 
\alpha\cdot{1 \choose -2} \otimes {2 \choose 1} 
%	\left(\begin{array}{c} 1\\-2 \end{array} \right)
%	\otimes
%	\left(\begin{array}{c} 2\\ 1 \end{array} \right)\quad
	\in \quad V_{1/2}^*\otimes V_{1/2}\ .
\cr}$$
It is straightforward to verify that $\omega:=a\delta b$ satisfies 
$\pi(\omega)=0$ and $\pi(\delta\omega)\neq0\,$. 
\hbn
This proves that $\pi(\delta J^1)=\a_0$, and we get
$$
\df{2}\simeq\{\,a_{AB}\otimes\psi^A\psi^B\,|\,a_{AB}=-a_{BA}\in\a_0\,\}\ .
\eqno(3.19)$$
In order to determine the space of $3$-forms, we first notice that
$$
	\pi(\uf{3})=\{\,a_{ABC}\otimes\psi^A\psi^B\psi^C\,\}\ ,
\eqno(3.20)$$
and we compute the space $\pi(\delta J^2)\,$. Let 
$a_i,b_i,c_i\,\in\a_0$ be such that 
$\omega=\sum_i a_i\delta b_i\delta c_i$ satisfies
$$
	\pi(\omega)=\sum_i a_i\lb\, D,b_i\,\rb\lb\,D,c_i\,\rb=0\ .
\eqno(3.21)$$
The coefficient of $\psi^1\psi^2\psi^3$ in $\pi(\delta\omega)$ is proportional
to
$$\eqalign{
	\varepsilon^{ABC}&\sum_i\, 
	{}[\, J^A,a_i{}\,]{}[\, J^B,b_i{}\,]{}[\, J^C,c_i{}\,]
=	-\varepsilon^{ABC}\sum_i\,
	a_i\left\lb\, J^A,{}[\, J^B,b_i{}\,]{}[ \,J^C,c_i{}\,]\right\rb\cr
&=-\varepsilon^{ABC}\,\sum_i \,a_i\bigl(
	\left\lb\,[J^A,{}[\,J^B,b_i{}\,]\right\rb\,{}[\,J^C,c_i{}\,]
	+{}[\,J^B,b_i{}\,]\,\left[\,J^A,{}[\,J^C,c_i{}\,]\right]\bigr)\cr
	&= -{i\over2}\,\varepsilon^{ABC} \, \sum_i \, a_i  \bigl(
	\varepsilon^{ABD}\,{}[\,J^D,b_i{}\,]\,{}[\,J^C,c_i{}\,]
	+ \varepsilon^{ACD} \, {}[ \,J^B,b_i{}\,] \, {}[\, J^D,c_i{}\,] \bigr)=0
\cr}$$
where we have used eq.~(3.21) and the Jacobi identity.
Thus, $\pi(\delta J^2)$ is
included in $\pi(\uf{1})\,$, and since $\a_0$ is a full matrix algebra, it is
either $0$ or equal to $\pi(\uf{1})\,$. Let $\omega,\eta\,\in\uf{1}$ be
such that $\pi(\omega)=-\one\otimes\psi^A\,$, $\pi(\eta)=0$ and 
$\pi(\delta\eta)=\one\otimes\one\,$. The existence of $\omega$ and $\eta$ is
ensured by eqs.~(3.16) and the fact that $\pi(\delta J^1)=\a_0\,$. We
have $\omega\eta\,\in\uf{2}\,$, $\pi(\omega\eta)=0$ and
$\pi(\delta(\omega\eta))=\one\otimes\psi^A$ as $\pi(\delta\omega)=0\,$. 
This proves that $\pi(\delta J^2)=\pi(\uf{1})$, and we get
$$
	\df{3}\simeq\{\,a\otimes\psi^1\psi^2\psi^3\,|\,a\,\in\a_0\,\}\ .
\eqno(3.22)$$
We proceed with the space of $4$-forms. First, we notice that due to the
Clifford algebra relations, eqs.~(3.4,5,8), we have
$$
	\pi(\uf{4})=\{\,a_{AB}\otimes\psi^A\psi^B\,|\,a_{AB}\,\in\a_0\,\}\ .
\eqno(3.23)$$
Let $\omega\in\uf{1}$ and $\eta\in\uf{2}$ be such that $\pi(\omega)=0\,$, 
$\pi(\delta\omega)=\one\otimes\one\,$, and
$\pi(\eta)=\one\otimes\psi^A\psi^B\,$. The existence of $\omega$ and $\eta$ is
ensured by the fact that $\pi(\delta J^1)=\a_0$ and by eq.~(3.17). We have
$\omega\eta\,\in\uf{3}\,$, $\pi(\omega\eta)=0$ and 
$\pi(\delta(\omega\eta))=\one\otimes\psi^A\psi^B$ as
$\pi(\omega)=0\,$. Since $\a_0$ is a full matrix algebra, this proves that 
$\pi(\uf{4})=\pi(\delta J^3)\,$, and we get $\df{4}=0\,$. Using the fact that
the product of differential forms induces a surjective map
$$
	\df{n}\otimes\df{m}\longrightarrow\df{n+m}
$$
we obtain
$$
	\df{n}=0 \quad \forall\,n>3\,.
\eqno(3.24)$$
Collecting eqs.~(3.16), (3.22) and (3.24), 
we arrive at the following theorem on the structure of differential 
forms over the non-commutative space $(\a_0, \h_0 \otimes W, D, \gamma)$: 
\mn
{\bf Theorem 3.2}\quad The left $\a_0$-modules 
$\df{n}$ are all free and given as follows: 
	\smallskip
\item{0)}$\df{0}=\a_0$ is one-dimensional with basis $\{\one\}\,$;
	\smallskip
\item{1)} $\df{1}$ is three-dimensional with basis 
		$\{\one\otimes\psi^A\}\,$;
	\smallskip
\item{2)} $\df{2}$ is three-dimensional with basis 
$\{\one\otimes\psi^A\psi^{A+1}\}$ (where addition is 
taken modulo $3$);
	\smallskip	
\item{3)} $\df{3}$ is one-dimensional with basis 
		$\{\one\otimes\psi^1\psi^2\psi^3\}\,$;
	\smallskip
\item{4)} $\df{n}=0$  for all $n>3\,$.
\mn 
Notice that the structure of the modules $\df{n}$ is the same as that of 
the spaces of differential forms on SU(2)$\,\simeq S^3\,$. 
\bn
In the following, we compute the action of the exterior differential
$$
	\delta\,:\,\df{n}\longrightarrow\df{n+1}\ .
$$
We introduce the following bases of $\df{1}$ and $\df{2}$
$$\eqalignno{
	e^A&=\one\otimes\psi^A\ \ \in\df{1}\ ,
&(3.25)\cr
	f^A&=\varepsilon^{ABC}\otimes\psi^B\psi^C\ \ \in\df{2}\ ,
&(3.26)\cr}$$
which allows us to identify $\df{1}$ and $\df{2}$ with the standard free module
$\a_0^3\,$, and we decompose their elements with respect to these bases,
$$\eqalignno{
	\omega&=\omega_A \,e^A\quad{\rm for\ }\omega\,\in\df{1}\ ,
&(3.27)\cr
	\omega&=\omega_A \,f^A\quad{\rm for\ }\omega\,\in\df{2}\ .
&(3.28)\cr}$$
It is easily verified that the product of $1$-forms $\omega,\eta\,\in\df{1}$
is given by
$$
	\omega\cdot\eta=\varepsilon^{ABC}\,\omega_B\,\eta_C \,f^A\ .
\eqno(3.29)$$
By the Leibniz rule for the exterior differential $\delta\,$, 
knowledge of the action of $\delta$ on the elements
$a\in\a_0\,$, $e^A$ and $f^A$ fully determines the action of the 
differential on $\df{\bullet}\,$. By definition, we have
$$
	\delta a=\lb \,J^A,a\,\rb\, e^A\ .
\eqno(3.30)$$
Using eq.~(3.30) and the nilpotency of $\delta$ we get
$$
	0=\delta^2 J^A=i\varepsilon^{ABC}\delta(J^C e^B)=
	-\varepsilon^{BAC}\varepsilon^{DCF}J^F e^D e^B +
	i\varepsilon^{BAC}J^C\delta e^B
$$
from which we can successively conclude that 
$$\eqalign{
	&\varepsilon^{BAE}\varepsilon^{DEC}=i\varepsilon^{BAC}\delta e^B\ ,
\cr
	&e^A e^C=i\varepsilon^{BAC}\delta e^B\ .
\cr}$$
With eq.~(3.29), we finally get
$$
	\delta e^A=-if^A\ .
\eqno(3.31)$$
This equation, together with the nilpotency of $\delta$, furthermore 
implies that
$$
	\delta f^A=0\ .
\eqno(3.32)$$
We summarize these results in the following
\mn
{\bf Theorem 3.3}\quad	Let 
$g={1\over 3!}\, \varepsilon^{ABC}\psi^A\psi^B\psi^C$ 
be the basis element of $\df{3}\,$, and $e^A$ and $f^A$ as in eqs.\ (3.25,26). 
Then the algebra structure of $\df{\bullet}$ is given as follows: 
$$\eqalignno{
{\rm a1)}\quad\quad 
&\lb\,a,e^A\,\rb = \lb\,a,f^A\,\rb = \lb\, a,g\,\rb = 0
\quad\ {\rm for\ all}\ \ a\in\a_0 
\phantom{XXXXXXXXXXX}
&(3.33)\cr
{\rm a2)}\quad\quad &e^A e^B = \varepsilon^{ABC}f^C\ ,\quad 
             e^A e^B e^C=\varepsilon^{ABC}\,g\ ,
&(3.34)\cr
&e^A f^B = \delta^{AB}\,g\ .
&(3.35)\cr}$$
The differential structure on $\df{\bullet}$ is given by
$$\eqalignno{
{\rm b1)}\quad\quad   &\delta a= \lb\, J^A,a\,\rb\, e^A\ ,
&(3.36)\cr
{\rm b2)}\quad\quad   &\delta e^A = -if^A\ ,\quad\delta f^A=0\ .
\phantom{XXXXXXXXXXXXXXXXXXXXX}
&(3.37)\cr}$$

\bn\bn
%%%%%%%%%%%%%%%%%%%%%%%%%%%%%%%%%%%%%%%%%%%%%%%%%%%%%%%%%%%%%%%%%%%%%%%%%
{\bf 3.2.2 Cohomology of the de Rham complex}
\bn
Let us now compute the cohomology groups of the de Rham complex
$(\df{\bullet},\delta)\,$ of Theorems 3.2 and 3.3.
\sn
The zeroth cohomology group $H^0$ consists of those elements $a\in\a_0$ 
that are closed, i.e., satisfy $\delta a =0\,$. We have
$$\eqalign{
a\in H^0\ \  &\Longleftrightarrow \ \ \delta a =\lb\, J^A,a\,\rb e^A=0 
\cr
&\Longleftrightarrow \ \ \lb\, J^A,a\,\rb=0\quad{\rm for\ all}\ A
\cr}$$
and it follows that
$$
	H^0=\a_R \equiv\bigoplus_{j=0}^{k/2}
	\one_{V_j^*}\otimes\,{\rm End}\,(V_j)\ ,
\eqno(3.38)$$
and 
$$
	\dim_\C H^0=\sum_{j=0}^{k/2}(2j+1)^2={1\over6}(2k+3)(k+2)(k+1)\ .
\eqno(3.39)$$
In order to compute the first cohomology group, we first determine the closed
$1$-forms. For any $1$-form $\omega=\omega_A e^A \in\df{1}$, relation (3.37)
implies that  
$$
	\delta\omega=\left(\lb\, J^A,\omega_B\,\rb\varepsilon^{ABC}-
	i\omega_C\right)f^C\ , 
$$
and thus $\delta\omega=0$ is equivalent to
$$
	\lb\, J^A,\omega_B\,\rb\varepsilon^{ABC}=i\omega_C\ .
\eqno(3.40)$$
We show that all closed $1$-forms are exact. First, notice that if we view
$\a_0$ as a representation space of su(2), then, for a closed $1$-form,
eq.~(3.40) must hold in all isotypic components. Therefore, there is
no loss of generality in assuming that all coefficients $\omega_A$ transform
under the spin $j$ representation, i.e.,
$$
	\lb\, J^A,{}[\, J^A, \omega_B{}\,]\rb =j(j+1)\,\omega_B\ .
\eqno(3.41)$$
Furthermore, we can assume that $j\neq0$ since otherwise $\omega=0$, as follows
from eq.~(3.40). We define $a(\omega)\in\a_0$ by
$$
	a(\omega)={1\over j(j+1)}\,\lb\, J^A,\omega_A\,\rb
$$
and we compute $\delta a\,$. Using eqs.~(3.40,41) and
the Jacobi identity, we get
$$\eqalign{
	\delta a(\omega) &= {1\over j(j+1)}\,
	\lb\, J^A,{}[\, J^B,\omega_B{}\,]\rb e^A\cr
	&= {1\over j(j+1)}\,
	\left(\,i\varepsilon^{ABC}{}[\, J^C,\omega_B{}\,]+
	\lb\, J^B,{}[\, J^A,\omega_B{}\,]\rb\right)e^A\cr
	&= {1\over j(j+1)}\,
	\lb\, J^B,{}[\, J^B,\omega_A{}\,]\rb e^A=
	\omega_A e^A\ .
\cr}$$
This proves that
$$
	H^1=0\ .
\eqno(3.42)$$
We proceed towards the second cohomology group. The condition for a $2$-form
$\omega = \omega_A f^A$ to be closed reads
$$
	\delta\omega=0\ \ \Longleftrightarrow\ \ \lb\, J^A,\omega_A\,\rb=0\ .
\eqno(3.43)$$
Again, we assume that the components $\omega_A$ belong to a spin $j$
representation of su(2). If $j=0\,$, then setting $\eta_A=i\omega_A$ we get
$$
	\delta(\eta_A e^A)=\omega_A f^A\ ,
$$
proving that $\omega$ is exact. If $j\neq0$,  we set
$$
	\eta_A=-{1\over j(j+1)}\,\varepsilon^{ABC}\lb\, J^B,\omega_C\,\rb\ ,
$$
and one easily verifies that $\delta(\eta_A e^A)=\omega_A f^A\,$. This proves
that
$$ 
	H^2=0\ .
\eqno(3.44)$$
Finally, we compute the third cohomology group. Since all $3$-forms are
closed, we just have to compute the image of the exterior differential in
$\df{3}\,$. For any $2$-form $\omega$ we have
$$
	\delta\omega=\lb\, J^A,\omega_A\,\rb \,g 
$$
with $g$ being the basis element of $\df{3}$ as in Theorem 3.3. 
This means that the image of $\delta$ in $\df{3}$ is given by
$$
	{\rm im}\,\delta\,\Big|_{\df{2}}={\rm span}\Bigr(\bigcup_{A=1}^3
	{\rm im}\,({\rm ad}\,J^A)\Bigl)\, \cdot g\ ,
%	{\rm im}\,\delta\,\Big|_{\df{2}}=\Bigl(\bigcup_{A=1}^3
%	{\rm im}\,({\rm ad}\,J^A)\Bigl)_{\rm lin} \cdot g\ ,
$$
and this space consists of linear combinations of elements of $\a_0$ 
transforming under a spin $j$ representation for $j\neq0\,$, multiplied by
$g\,$. Thus, the quotient $\df{3}/{\rm im}\,\delta$ is given by
$$
	H^3\simeq\a_R \equiv\bigoplus_{j=0}^{k/2}\one_{V^*_j}\otimes
	{\rm End}\,(V_j)\ .
\eqno(3.45)$$
Collecting our results of eqs.\ (3.38,39,42,44) and (3.45), we get 
the following
\bn
{\bf Theorem 3.4}\quad 	The cohomology groups of the de Rham complex 
of Theorem 3.3 are 
$$ H^0\simeq H^3\simeq\a_R\ \ ,\ \quad H^1=H^2=0
$$
with dimensions 
$$
{\rm dim}_{\C} H^0={\rm dim}_{\C} H^3 = {1\over6}(2k+3)(k+2)(k+1)\ .
$$
\bn
This theorem shows that the cohomology groups of the fuzzy 3-sphere -- 
which is the quantum target of the WZW model based on SU(2) \q{FG,Gr} -- 
look very much like those of the classical SU(2) group manifold, except 
for the unexpected dimensions of the spaces $H^0$ and $H^3\,$. 
\hbn 
We observe that in 
the classical setting, the cohomology groups are modules over the ring 
$H^0$ and that, for a connected space, the Betti numbers coincide with the 
dimensions of these modules. We are thus led to the idea that the
dimensions of the cohomology groups over $\C$ may be less relevant than 
their dimensions as modules over $H^0\,$. Of course, it may happen in 
general that some $H^0$-module is not free, and we would, in that case, 
lose the notion of dimension. For the cohomology groups of the de Rham 
complex $(\df{\bullet},\delta)$ we get
$$\eqalign{
	&\dim_{H^0}H^0=\dim_{H^0}H^3=1\ ,\cr
	&\dim_{H^0}H^1=\dim_{H^0}H^2=0\ ,\cr
}$$
which fits perfectly with the classical result. The above proposal is
obviously tailored to make sense of the cohomology groups of Theorem
3.4 and its general relevance remains to be decided by the study of
other examples of non-commutative spaces.
\bn\bn
%%%%%%%%%%%%%%%%%%%%%%%%%%%%%%%%%%%%%%%%%%%%%%%%%%%%%%%%%%%%%%%%%%%%%%%
{\bf 3.3 The geometry of the non-commutative 3-sphere}
\bn
The $N=1$ spectral data $(\a_0, \h_0\otimes W, D, \gamma)$ permit us 
to investigate not only topological but also geometrical aspects of 
the quantized 3-sphere, namely integration of differential forms and 
Hermitian structures, as well as connections and the associated 
Riemann, Ricci and scalar curvatures. For a more detailed account of 
the results of this section, the reader is referred to \q{Gr}.   
\bn
%%%%%%%%%%%%%%%%%%%%%%%%%%%%%%%%%%%%%%%%%%%%%%%%%%%%%%%%%%%%%%%%%%%%%%
{\bf 3.3.1 Integration and Hermitian structures}
\bn
We start with  the canonical scalar product and the Hermitian structures on 
the spaces of differential forms. We use the same notations as in subsections 
2.1.3 -- 2.1.5. 
\sn
Any element $\omega\in\pi(\uf{\bullet})$ can be written uniquely as
$$
	\omega=\omega^0+\omega_A^1 e^A+\omega_A^2 f^A+\omega^3 g
\eqno(3.46)$$
where $\omega^i,\omega_A^i\in\a_0\,$. The integral $\barint\,$, as given in
Definition 2.3, is just the normalized trace on $\h_0\otimes W\,$,
denoted by ${\rm Tr}\,$. Thus, for any element $\omega\,$ as above, we have
$$
	\Barint\omega=\,{\rm Tr\ }\omega^0
\eqno(3.47)$$
It is easy to show that the sesqui-linear form $(\cdot,\cdot)$ associated to
the integral is given by
$$
	(\omega,\eta)=\,{\rm Tr}\, \bigl\lb\,
	\omega^0(\eta^0)^*+\omega_A^1(\eta_A^1)^*+
	\omega_A^2(\eta_A^2)^*+
	\omega^3(\eta^3)^*\bigr\rb\ .
\eqno(3.48)$$
This proves that the kernels $K^i$ of the sesqui-linear form $(\cdot,\cdot)$
equal the kernels $J^i$ of the representation $\pi\,$. Thus, in this example
we also have the equality
$$
	\widetilde{\Omega}_D^p(\a_0)=\df{p}\ .
$$
Furthermore, since $\pi(\delta J^1)=\a_0$ and $\pi(\delta J^2)=\pi(\df{1})\,$,
we see that the decomposition (3.46) gives the canonical
representative $\omega^\bot$ of an arbitrary differential form
$\omega\in\df{\bullet}\,$.
\sn
The Hermitian structure on $\df{p}$ is readily seen to be 
$$
	\langle\omega,\eta\rangle=\omega_A(\eta_A)^*\ ,
	\quad\omega,\eta\in\df{p}\ .
\eqno(3.49)$$
Notice that, in this example, we get a true Hermitian structure on $\df{p}$ 
and not only a {\sl generalized} Hermitian structure on
$\widetilde{\Omega}^p(\a_0)\,$, cf.\ subsection 2.1.5.
\bn\bn
%%%%%%%%%%%%%%%%%%%%%%%%%%%%%%%%%%%%%%%%%%%%%%%%%%%%%%%%%%%%%%%%%%%%%%%%
{\bf 3.3.2 Connections on $\Omeinsnull$}
\bn
This last property makes it possible to regard $\df{1}$ as the cotangent 
bundle of the non-commutative 3-sphere and to study connections on $\df{1}\,$.
\sn
Since the space of $1$-forms $\df{1}$ is a trivial left $\a_0$-module, a
connection $\nabla$ on $\df{1}$ is uniquely determined by the images of the
basis elements, i.e.,
$$
	\nabla e^A = -\omega_{BC}^A \,e^B\otimes e^C
\eqno(3.50)$$
where $\omega_{BC}^A$ are {\sl arbitrary} elements of $\a_0\,$.
\bn
{\bf Proposition 3.5}\quad A connection $\nabla$ is unitary if and only 
if its coefficients	satisfy the Hermiticity condition
	$$
		\omega_{BC}^{A*}=\omega_{BA}^C\ .
\eqno(3.51)$$
\sn
{\kap Proof}: 	It follows from (3.49) that 	
	$\langle e^A,e^B\rangle=\delta^{AB}\,$. Then we have
	for a unitary connection (see Definition 2.12)
	$$
		0=\delta\langle e^A, e^B\rangle=
		- \omega_{CB}^A e^C + e^C\omega_{CA}^{B*}
	\eqno\qed$$
\mn
{\bf Proposition 3.6}\quad The torsion of a connection is given by
$$
	\ttT^A=(-i\delta^{AD}+\omega_{BC}^A\varepsilon^{BCD})\,f^D\ .
$$
\sn
{\kap Proof}: 	Using Definition 2.14 and eqs.\ (2.20), (3.34,37), we get 
	$$
	\ttT(\nabla)\,e^A=-if^A+\omega_{BC}^A\varepsilon^{BCD}f^D\ .
\eqno\qed	$$
\mn
{\bf Proposition 3.7}\quad A connection is torsionless and unitary if and 
only if its	coefficients satisfy the following conditions
$$\eqalignno{
\quad\quad{\rm i)}\quad\quad
  &\omega_{BC}^A-\omega_{BC}^{A*}=i\varepsilon^{ABC}\ ,
\phantom{XXXXXXXXXXXXXXXXXXXXXXXX}
&(3.52)\cr
\quad\quad{\rm ii)}\quad\quad
  &\omega_{BC}^A=\omega_{CB}^{A*}\ ,
&(3.53)\cr
\quad\quad{\rm iii)}\quad\quad 
 &\omega_{BC}^A=\omega_{AB}^C\ .
&(3.54)\cr}$$
In particular, such a connection is uniquely determined by the nine
self-adjoint elements $\omega_{AB}^A\in\a_0$ and the self-adjoint part of
$\omega_{23}^1\,$.
\sn
{\kap Proof}: 
The condition of vanishing torsion,
$$
		\omega_{BC}^A\,\varepsilon^{BCD}=i\delta^{AD}\ ,
$$
can equivalently be written as
$$
		\omega_{BC}^A-\omega_{CB}^A=i\varepsilon^{ABC}\ .
\eqno(3.55)$$
Using alternatively eqs.~(3.55) and the unitarity condition 
eq.~(3.51) we get
$$
		\omega_{BC}^A=i\varepsilon^{ABC}+\omega_{CB}^A=
		i\varepsilon^{ABC}+\omega_{CA}^{B*}=
		\omega_{AC}^{B*}=\omega_{AB}^C\ .
$$
which proves the result. \hfill\qed
\mn
This proposition shows that, as in the classical case,
there are many unitary and torsionless connections. There are two
possibilities to reduce the space of ``natural'' connections further. First,
we can consider {\sl real} connections, i.e., connections whose associated
parallel transport maps real forms to real forms. In the classical setting, a
$1$-form $\omega$ is real if $\omega^*=-\omega$ (the sign comes from the fact 
that the Clifford matrices are anti-Hermitian). Thus, we see that our basis of
$1$-forms consists of {\sl imaginary} $1$-forms, i.e., $e^{A*}=e^A\,$. If the
covariant derivative of an imaginary $1$-form is to be imaginary, then the
connection coefficients $\omega_{BC}^A$ must be anti-Hermitian. We call such a
connection a {\sl real connection}.
\mn
{\bf Corollary 3.8}\quad
	There is a unique real unitary and torsionless connection on the
	cotangent bundle $\df{1}\,$,
	and its coefficients are given by
	$$
		\omega_{BC}^A={i\over2}\,\varepsilon^{ABC}\ .
	$$
\mn
There is another way of reducing the number of
``natural'' connections. If we look at a general unitary and torsionless
connection, we see that it does not have any isotropy property. For example,
the coefficients $\omega_{AA}^A$ are all independent of one another. We hope
that if we require the connection to be invariant under all $1$-parameter
group of isometries (see \q{CFG,Gr}) we shall get relations among these
coefficients. We shall not pursue this route here, but we refer the reader to
\q{Gr} for a detailed analysis.
\mn
We proceed with the computation of the scalar curvature of the {\sl real}
connection $\widetilde\nabla$ of Corollary 3.8. \hfill\break
\noindent
For any connection $\nabla$ with coefficients $\omega_{BC}^A$ as defined in
eq.~(3.50), the curvature tensor is given by (see
Definition {2.11}) 
$$\eqalignno{
	\ttR(\nabla)\,e^A&=-\nabla^2 e^A
&\cr
	=&\left\{{}[ \,J^D,\,\omega_{EC}^A{}]\,\varepsilon^{DEB}
	-i\omega_{BC}^A+\omega_{DE}^A\,\omega_{FC}^E\,\varepsilon^{DFB}\right\}
	f^B\otimes e^C\ .
&(3.56)\cr}$$
In particular, for the {\sl real} connection $\widetilde{\nabla}$ of
Corollary 3.8, the curvature tensor reads
$$
\ttR(\widetilde{\nabla})\,e^A={1\over 4}\,\varepsilon^{ABC}f^B\otimes e^C\ .
\eqno(3.57)$$
In order to compute the Ricci curvature, we use a dual basis to the
generators $e^A\,$, as in subsection 2.1.7, before eq.\ (2.15). 
It is clear that the elements
$\varepsilon_A\in{\df{1}}{}^*$ defined by
$$
	\varepsilon_A(\omega)=\varepsilon_A(\omega_B e^B):=\omega_A
\eqno(3.58)$$
for all $\omega\in\df{1}\,$, form a dual basis to $e^A\,$. Using
eq.~(3.49) it is then easy to verify that the {\sl dual $1$-forms}
$e_A\,$  and their dual maps $e^{\rm{ad}}_A\,$, eq.\ (2.16), are given by
$$
	e_A= e^A\ ,\quad e^{\rm{ad}}_A(f^B)=-\varepsilon^{ABC}e^C\ .
\eqno(3.59)$$
For the real connection $\widetilde\nabla$, we get from eq.~(3.57)
$$
	\ttRic(\widetilde\nabla)=-{1\over2}\,e^A\otimes e^A\ .
\eqno(3.60)$$
We proceed with the computation of the scalar curvature. The 
{\sl right dual maps} $(e_R^A)^{\rm ad}$ to the basis $1$-forms $e^A\,$,
eq.~(2.17), act as 
$$
	(e_R^A)^{\rm ad}(e^B)=\delta^{AB}\ .
\eqno(3.61)$$
The scalar curvature of the real connection $\widetilde\nabla$ follows from
eq.~(3.60) and is given by
$$
	\ttr(\widetilde\nabla)=-{3\over 2}\ .
\eqno(3.62)$$
It is the same as the scalar curvature of the {\sl unique} real unitary
and torsionless connection for the classical SU(2) -- recall that the 
definition of the scalar curvature in the non-commutative setting differs 
from the classical one by a sign, see the remark in Definition 2.16. 
\sn 
This completes our study of the non-commutative 3-sphere in terms of 
$N=1$ spectral data. Our results show that the non-commutative 3-sphere 
has striking similarities with its classical counterpart. As we saw in 
subsections 3.2.1 and 3.2.2, the spaces of differential forms have the 
same structure as left-modules over the algebra of functions, and the 
cohomology groups have the same dimensions as modules over the zeroth 
cohomology group, $H^0\,$. Furthermore, geometric invariants like the 
scalar curvature, too, coincide for the classical and the quantized 
3-sphere.

\bn\bn
%%%%%%%%%%%%%%%%%%%%%%%%%%%%%%%%%%%%%%%%%%%%%%%%%%%%%%%%%%%%%%%%%%%%
{\bf 3.4 Remarks on \Noneone} 
\bn
In the following, we consider $N=(1,1)$ data for the algebra $\a_0$.
The construction of the first subsection starts from the BRST operator 
of the group $G$ and leads to a deformation of the de Rham complex 
for the classical 3-sphere in the form of $N=(1,1)$ data for the 
non-commutative 3-sphere. In the second subsection, we return to the 
two generalized Dirac operators provided by superconformal field theory 
\q{FG}, which lead to a different formulation of $N=(1,1)$ data, 
displaying ``spontaneously broken supersymmetry''. 

\bn\bn
{\bf 3.4.1 \Noneone data from BRST} 
\bn
One way to arrive at $N=(1,1)$ data for the algebra $\a_0$ and at the 
associated (non-commutative) de Rham complex for the quantized 3-sphere 
is to use the action of the group $G$ on the Hilbert 
space ${\cal H}_0$ for introducing a BRST operator (see also section I$\,$2.3). 
\sn
Let $\{\,J_A\,\}$ be the basis of the complexified Lie algebra {\tt g}$^{\C}$ 
of $G$ introduced in eq.\ (3.7). The BRST operator $Q$ for the group $G$ is 
defined as usual: We introduce ghosts $c^A$ and anti-ghosts $b_A$ satisfying 
the ghost algebra 
$$
\{\,c^A,c^B\,\} = \{\,b_A,b_B\,\} = 0\ , \quad \{\,c^A,b_B\,\} = \delta^A_B\ . 
\eqno(3.63)$$
Then the  BRST operator is given by 
$$
Q= c^A J_A - {i\over2}\, f^C_{AB}\,c^A c^B b_C \ ,
\eqno(3.64)$$
and the ghost number operator is 
$$
T= c^A b_A \ .
\eqno(3.65)$$
The Hilbert space of the $N=(1,1)$ data will be of the form ${\cal H}_0 
\otimes W$ where $W$ is a representation space for the ghost algebra. 
\hbn 
In order to obtain $N=(1,1)$ data, we require that the ghost algebra acts 
unitarily on $W$ with respect to the natural ${}^*$-operation, namely
$$c^{A\,*} = g^{AB} b_B\ .
\eqno(3.66)$$
This choice is compatible with positive definiteness of the scalar product 
on $W$, and it renders the ghost number operator $T$ self-adjoint. 
\footnote{$\!\!\!\!{}^1$}{$\!\!${\foon In the context of gauge theories, 
one considers 
representations such that ${\scriptstyle c{}^{A*} = c{}^A\,,\ b_A^* = b_A\,}$. 
These Hermiticity conditions together with the defining relations (3.63) 
imply that the inner product 
of the representation space is not positive definite -- which is why 
${\scriptstyle c^{A}}$ and ${\scriptstyle b_A}$ are called ghosts.}} 
Furthermore, this choice of ${}^*$-operation leads to identifying the ghost 
algebra with the CAR 
$$
\{\,c^A,c^B\,\}  = 0\ , \quad \{\,c^A,c^{B\,*}\,\} = g^{AB}\ ,  
\eqno(3.67)$$
and the BRST operator can be written 
$$
Q= c^A J_A - {i\over2}\, f^C_{AB}\,c^A c^B c^*_C \ ,
\eqno(3.68)$$
where indices are raised and lowered with the metric $g_{AB}$ as usual. 
Under the identifications 
$$
c^A \sim a^{A\,*} := -i\, \theta^A \,\wedge
$$ 
where $\{\,\theta^A\,\}$ is a basis of 1-forms dual to $\{\,\vartheta_A\,\}\,$,  
eq.\ (3.68) for the BRST operator formally coincides 
with the exterior derivative on $G$. This fact was already mentioned 
in section I$\,$2.3. 
\sn
In order to complete our construction of $N=(1,1)$ data, we introduce
the Hodge $*$-operator 
$$
* = {1\over n!}\, \sqrt{g}\; \varepsilon_{A_1\ldots A_n} \,
     (\,c^{A_1} + c^{A_1\,*}\,) \cdots (\,c^{A_n} + c^{A_n\,*}\,)
\eqno(3.69)$$
where $n={\rm dim}\,G$. This operator clearly commutes with the algebra 
$\a_0$ of Theorem 3.1. Moreover, it is easy to verify that $*$ is unitary 
and satisfies 
$$
*^2 = (-1)^{{n(n-1)\over2}}
\eqno(3.70)$$
as well as 
$$
* Q = (-1)^{n-1} Q *\ .
\eqno(3.71)$$
It follows that $(\a,\h, d, \gamma, *)$ with $\a=\a_0$, $\h= \h_0\otimes W$,  
$d= Q$ and where $\gamma$ is 
the modulo 2 reduction of the $\Z$-grading $T$, form a set of $N=(1,1)$ data 
in the sense of Definition 2.20. 
\mn
We refrain from presenting the details of the construction of differential forms 
and of the other geometrical quantities, since the computations are fairly 
straightforward. For example, the space of $k$-forms is given by 
$$
\Omega^k_{d}(\a_0) = \{\, a_{A_1\ldots A_k} c^{A_1}\cdots c^{A_k}\,\vert
          \,a_{A_1\ldots A_k} \in \a_0\,\}\ .
\eqno(3.72)$$
For $G=\,$SU(2), we see that these spaces are isomorphic 
to $\Omega_D^k(\a_0)$ as left $\a_0$-modules. Furthermore, it is easy 
to see that $\Omega^{\bullet}_{d}(\a_0)$ and $\Omega_D^{\bullet}(\a_0)$ are 
isomorphic as complexes, which proves that, in particular, their cohomologies 
coincide. 
\sn 
Of course, the same constructions and results apply to the BRST operator 
associated with the right-action of $G$ on $\h_0$ given by the generators 
$\overline{J}_A$ of eq.\ (3.7).  
\mn
The Hilbert space $\h = \h_0 \otimes W$ can be decomposed into a direct sum 
of eigenspaces of the $\Z$-grading operator $T$, 
$$
\h = \bigoplus_{k=0}^n \h^{(k)}
$$ 
where $\h^{(0)} = \h_0$, $n={\rm dim}\,G$ ($\,=3$ for G$\,=\,$SU(2)). 
The subspaces $\h^{(k)}$ are left-modules for $\a_0$. Furthermore, it 
follows from eqs.\ (3.65) and (3.68) that $d := Q$ maps  $\h^{(k)}$  into 
$\h^{(k+1)}$ for $k=0,\ldots,n$ (with $\h^{(n+1)}:=\{0\}$). Since $d^2 =0$, 
$\h$ is a complex.  
Viewed as linear spaces, the cohomology groups of $\Omega^{\bullet}_d(\a_0)$ 
and $(\h,Q)$ are isomorphic, although the latter do not carry a ring structure. 
\sn
As a side remark, consider an odd operator $H$ on $\h$. Then 
$\tilde d := d + H$ is nilpotent if and only if $\{\,d, H\,\} + H^2 = 0$. 
If $H$ commutes with $\a_0$, then $\Omega^{\bullet}_{\tilde d}(\a_0)$ and 
$\Omega^{\bullet}_d(\a_0)$ are identical complexes. In the next subsection, 
we will meet a conformal field theory motivated example for $\tilde d= d + H$ 
which is nilpotent but for which $H$ does not commute with $\a_0$.

\bn\bn
{\bf 3.4.2 Spontaneously broken supersymmetry} 
\bn
In section 3.1 we introduced two connections $\nabla^{\S}$ and 
$\overline{\nabla}{}^{\S}$ and their associated Dirac operators 
${\cal D}$ and $\overline{\cal D}$, see eqs.\ (3.6-9).  Since these two 
Dirac operators correspond to {\sl different} connections, they 
are not Dirac operators on an $N=(1,1)$ Dirac bundle in the sense 
of Definition I$\,$2.6. It is interesting to notice that ${\cal D}$ and 
$\overline{\cal D}$ nevertheless satisfy the $N=(1,1)$ algebra \q{FG} 
$$
{\cal D}^2 = \overline{\cal D}{}^2\ ,\quad 
\{\,{\cal D}, \overline{\cal D}\,\}=0\ .
\eqno(3.73)$$
The easiest way to prove (3.73) is to verify that the generalized 
exterior derivative 
$$
\tilde d := {1\over2}\,( {\cal D} + i \overline{\cal D} )
\eqno(3.74)$$ 
is nilpotent. Let $\{\,\vartheta_A\,\}$ and $\{\,\theta^A\,\}$ denote 
a basis of the Lie algebra and the dual basis of 1-forms, respectively, 
as before. We define the operators 
$$
a^{A\,*} = \theta^A \,\wedge\ , \quad a_A = \vartheta_A \; \llcorner
$$
as usual, and we can express the fermionic operators $\psi^A$ and 
$\overline{\psi}{}^A$ as 
$$
\psi^A = -i ( a^{A\,*} - a^A ) \ , \quad 
\overline{\psi}{}^A = - ( a^{A\,*} + a^A ) \ ,
\eqno(3.75)$$
where indices are raised and lowered with the metric $g_{AB}$. 
Using eqs.\ (3.9) and (3.74), we can rewrite the operator $\tilde d$ as 
a sum of terms of degree $1, -1$ and $-3$, 
$$
\tilde d= \tilde d_1 + \tilde d_{-1} + \tilde d_{-3}
\eqno(3.76)$$ 
where 
$$\eqalign{
\tilde d_1\  &= a^{A\,*} J_A^+ - {1\over4}\,f_{ABC} a^{A\,*} a^{B\,*} a^C 
\cr
\tilde d_{-1} &= -a^A J_A^- 
\cr
\tilde d_{-3} &=  - {1\over12}\,f_{ABC} a^{A} a^{B} a^C 
\cr}\eqno(3.77)$$
with 
$$J^{\pm}_A = -{i\over2}\,( J_A \pm \overline{J}_A )\ . 
\eqno(3.78)$$
It is then straightforward to show that $\tilde d$ given by eqs.\ (3.76,77) 
satisfies $\tilde d{}^2=0$ and that the associated Laplacian $\triangle = 
\{\,\tilde d,\tilde d{}^*\,\}$ is given by 
$$
\triangle = g^{AB} J_A J_B + {{\rm dim}\,G\over24} = 
g^{AB} \overline{J}_A \overline{J}_B + {{\rm dim}\,G\over24}\ .
\eqno(3.79)$$
Thus, $\triangle$ is a strictly positive operator -- corresponding
to what one calls {\sl spontaneously broken supersymmetry} in 
the context of field theory. This implies that the cohomology of the 
complex $({\cal H},\tilde d)$ is trivial. However, the cohomology
of the complex $\Omega^{\bullet}_{\tilde d}(\a_0)$, as introduced
in Sect.\  2.2, is not trivial. Notice that ${\tilde d}_1$ is 
the BRST operator associated to the generators $J^{+}_A$ and hence
nilpotent. This implies that the BRST cohomology of the fuzzy 3-sphere
can be extracted from $\Omega^{\bullet}_{\tilde d}(\a_0)$.
\bn
\vfill\eject
%\bye
%%%%%%%%%%%%%%%%%%%%%%%%%%%%%%%%%%%%%%%%%%%%%%%%%%%%%%%%%%%%%%%%%%%%%
%%%%%%%%%%%%%%%%%%%%%%%%%%%%%%%%%%%%%%%%%%%%%%%%%%%%%%%%%%%%%%%%%%%%%
\def\hnull{\hskip-3pt \hbox{ $ {\cal H}\kern-5pt \raise7pt
\hbox{$\scriptscriptstyle {\rm o}$}$}\hskip3pt}
%%%%%%%%%%%%%%%%%%%%%%%%%%%%%%%%%%%%%%%%%%%%%%%%%%%%%%%%%%%%%%%%%%%%%%%%%
%%%%%%%%%%%%%%%%%%%%%%%%%%%%%%%%%%%%%%%%%%%%%%%%%%%%%%%%%%%%%%%%%%%%%%%%%
%\pageno=58
\leftline{\bf 4. The non-commutative torus}
%%%%%%%%%%%%%%%%%%%%%%%%%%%%%%%%%%%%%%%%%%%%%%%%%%%%%%%
\bn
As a second, ``classic'' example of non-commutative spaces, we discuss the 
geometry of the non-commutative $2$-torus \q{Ri,Co1,Co5}. 
After a short review of the classical torus in
subsection {4.1}, we analyze the spin geometry ($N=1$) of the
non-commutative torus in subsection 4.2 along the lines 
of \q{FGR2,Gr}. In subsections {4.3} and 4.4, we successively extend the
$N=1$ data to $N=(1,1)$ and $N=(2,2)$ data -- according to the general 
procedure proposed in subsection 2.2.5 above. In these two last subsections, 
we do not give detailed proofs, but merely state the results since the
computations, although straightforward, are tedious and not very illuminating.
\bn\bn
%%%%%%%%%%%%%%%%%%%%%%%%%%%%%%%%%%%%%%%%%%%%%%%%%%%%%%%%%%%%%%%%%%%%%%%
{\bf 4.1 The classical torus}
\bn
To begin with, we describe the $N=1$ data associated to the classical
2-torus ${\T}_0^2$. By Fourier transformation, the algebra of smooth
functions over ${\T}_0^2$ is isomorphic to the Schwarz space 
${\cal A}_0 := {\cal S}({\Z}^2)$ over ${\Z}^2$, endowed
with the (commutative) convolution product:
$$
	(a\, {\scriptstyle{\bullet}}\, b) (p) \ = \ 
	\sum_{q \in {\Z}^2} a (q)\; b(p-q)
\eqno(4.1)$$
where $a, b \in {\cal A}_0$ and $p \in {\Z}^2$. Complex
conjugation 
of functions translates into a ${}^*$-operation: 
$$
	a^* (p) \ = \ \overline{a (-p)} \ ,\quad \ a\;\in\;{\cal A}_0 \ .
\eqno(4.2)$$
If we choose a spin structure over ${\T}_0^2$ in such a way that the
spinors are periodic along the elements of a homology basis, then the
associated spinor bundle is a trivial rank 2 vector bundle. With this
choice, the space of square integrable spinors is given by the direct
sum
$$
	{\cal H}\;=\;l^2 ({\Z}^2)\;\oplus\;l^2 ({\Z}^2) 
\eqno(4.3)$$
where $l^2 ({\Z}^2)$ denotes the space of square summable
functions over ${\Z}^2$. The algebra ${\cal A}_0$ acts
diagonally on ${\cal H}$ by the convolution product. We choose a
flat metric $(g_{\mu\nu})$ on ${\T}_0^2$ and we introduce the
corresponding 2-dimensional gamma matrices
$$
	\left\{ \gamma^\mu, \gamma^\nu \right\} \ = 
	\ -\, 2\,g^{\mu\nu} \ , \quad \gamma^{\mu*} \ = 
	\ -\, \gamma^\mu \ .
\eqno(4.4)$$
Then, the Dirac operator $D$ on ${\cal H}$ is given by 
$$
	( D\,\xi) (p) \ = 
	\ i\,p_\mu\, \gamma^\mu\, \xi(p) \ ,\quad \ \xi\;\in\;{\cal H} \ .
\eqno(4.5)$$
Finally, the ${\Z}_2$-grading on ${\cal H}$, denoted by $\sigma$,
can be written as
$$
	\sigma \ = \ {i\over 2} \ \sqrt{g} \ \varepsilon_{\mu\nu}\;
	\gamma^\mu\,\gamma^\nu 
\eqno(4.6)$$
where $\varepsilon_{\mu\nu}$ is the Levi-Civita tensor. The data
$({\cal A}_0, {\cal H}, D, \sigma)$ are the canonical $N=1$
data associated to the compact spin manifold ${\T}_0^2$, and it is thus
clear that they satisfy all the properties of Definition {2.1}.

\bn\bn
%%%%%%%%%%%%%%%%%%%%%%%%%%%%%%%%%%%%%%%%%%%%%%%%%%%%%%%%%%%%%%%%%%%%%%
{\bf 4.2 Spin geometry ($\,\Neqone\!\!$)}
\bn
The non-commutative torus is obtained by deforming the product of the
algebra ${\cal A}_0$. For each $\alpha \in {\R}\,$, we define
the algebra ${\cal A}_\alpha := {\cal S} ({\Z}^2)$ with
the product 
$$
	(a\, {\scriptstyle{\bullet}}_\alpha \, b)\;(p) \ = \
	\sum_{q\,\in\,{\Z}^2} a(q)\, b(p-q)\; e^{i\pi\alpha\omega (p,q)}
\eqno(4.7)$$
where $\omega$ is the integer-valued anti-symmetric bilinear form on
${\Z}^2 \times {\Z}^2$ 
$$
	\omega (p,q)\;=\; p_1 q_2\, -\, p_2 q_1 \ ,\quad \ p, q\,\in\,
	{\Z}^2\ .
\eqno(4.8)$$
The ${}^*$-operation is defined as before. Alternatively, we could
introduce the algebra ${\cal A}_\alpha$ as the unital ${}^*$-algebra
generated by the elements $U$ and $V$ subject to the relations
$$
	U U^*\;=\;U^*U\;=\;V V^*\;=\;V^*V\;=\;\one\;, \quad \ UV\;=\; e^{- 2\pi
  	i\alpha} \, VU \ .
\eqno(4.9)$$
Having chosen an appropriate closure, the equivalence of the two
descriptions is easily seen if one makes the following
identifications:
$$
	U (p)\;=\;\delta_{p_1,1}\, \delta_{p_2,0} \ , 
	\ V (p)\;=\;\delta_{p_1,0}\, \delta_{p_2,1} \ .
\eqno(4.10)$$
If $\alpha$ is a rational number, $\alpha = {M\over N}$, where $M$ and
$N$ are co-prime integers, then the centre $Z ({\cal A}_\alpha)$ of
${\cal A}_\alpha$ is infinite-dimensional:
$$
	Z({\cal A}_\alpha) \ = \ {\rm span}\left\{ U^{mN} V^{nN} \bigm|
  	m, n \in {\Z} \right\} \ .
\eqno(4.11)$$
Let ${\rm I}_\alpha$ denote the ideal of ${\cal A}_\alpha$ generated by
$Z({\cal A}_\alpha) - \one$. Then it is easy to see that the
quotient ${\cal A}_\alpha / {\rm I}_\alpha$ is isomorphic, as a unital
${}^*$-algebra, to the full matrix algebra $M_N ({\C})$. \hfill\break
\noindent 
If $\alpha$ is irrational, then the centre of ${\cal A}_\alpha$ is 
trivial and ${\cal A}_\alpha$ 
is of type ${\rm II}_1$, the trace being given by the evaluation
at $p=0$. Unless stated differently, we shall only study the case of
irrational $\alpha$.
\sn
We define the non-commutative 2-torus ${\T}_\alpha^2$ by its $N=1$ data
$({\cal A}_\alpha, {\cal H}, D, \sigma)$ where ${\cal H}$,
$D$ and $\sigma$ are as in eqs.\ (4.3), (4.5) and (4.6), 
and ${\cal A}_\alpha$ acts diagonally on ${\cal H}$ by the deformed
product, eq.~(4.7). \hfill\break
\noindent When $\alpha = {M\over N}$ is rational, one may
work with the data $({\cal A}_\alpha / {\rm I}_\alpha, 
M_N({\C}) \otimes \C^2, D_\alpha, \sigma)$, where the Dirac operator
$D_\alpha$ is given by
$$
	D_\alpha \ = \ i\,\gamma^\mu \ { {\rm sin} \left({\pi\over N}\,
    	p_\mu\right) \over {\pi\over N }} \ .
\eqno(4.12)$$
\bn\bn
%%%%%%%%%%%%%%%%%%%%%%%%%%%%%%%%%%%%%%%%%%%%%%%%%%%%%%%%%%%%%%%%%%%%%%%
{\bf 4.2.1 Differential forms}
\bn
Recall that there is a representation $\pi$ of the algebra of
universal forms $\Omega^{\bullet} ({\cal A}_\alpha)$ on ${\cal H}$ 
(see subsection 2.1.2). The images of the homogeneous
subspaces of $\Omega^{\bullet} ({\cal A}_\alpha)$ under $\pi$
are given by
$$\eqalignno{
	\pi\left( \Omega^0 \left({\cal A}_\alpha\right)\right) 
	&= {\cal A}_\alpha \quad {\rm (by \ definition)}
&(4.13)
\cr
	\pi\left( \Omega^{2 k-1}\left({\cal A}_\alpha\right)\right)
	&= \left\{ a_\mu\,\gamma^\mu \bigm| a_\mu\,\in\,{\cal A}_\alpha
	\right\} 
&(4.14)
\cr
	\pi\left( \Omega^{2k} \left({\cal A}_\alpha\right)\right)
	&= \left\{ a\,+\,b\sigma \bigm| a, b\,\in\,{\cal A}_\alpha 
	\right\} 
&(4.15)\cr}$$
for all $k\in\Z_+$. In principle, one should then compute the
kernels $J^n$ of $\pi$ (see eq.~(2.2)), but these are generally huge
and difficult to describe explicitly. To determine the space of
$n$-forms, it is simpler to use the isomorphism 
$$
	\Omega_D^n ({\cal A}_\alpha)\;=\;\Omega^n ({\cal A}_\alpha)
	\big/ \left( J^n + \delta J^{n-1}\right) \;\simeq\; \pi \left(
  	\Omega^n \left( {\cal A}_\alpha\right)\right) \big/ \pi (\delta
	J^{n-1}) \ .
\eqno(4.16)$$
First, we have to compute the spaces of ``auxiliary forms''
$\pi (\delta J^{n-1})$.
\mn
{\bf Lemma 4.1}\quad The spaces $\pi (\delta J^{n-1})$ of auxiliary forms 
      are given by 
$$\eqalignno{
		\pi\left( \delta J^1\right) &= {\cal A}_\alpha 
&(4.17)\cr
		\pi\left( \delta J^{2k}\right)&= \pi\left( \Omega^{2k+1}\left(
    		{\cal A}_\alpha\right)\right) 
&(4.18)\cr
		\pi\left( \delta J^{2k+1}\right) &= \pi\left(
  		\Omega^{2k+2}\left({\cal A}_\alpha\right)\right)
&(4.19)\cr}$$
for all $k\geq 1$.
\sn
{\kap Proof}: 	Let $a_i, b_i \in {\cal A}_\alpha$ be such that the 
	universal $1$-form
	$\eta = \sum_i a_i\delta b_i \in \Omega^1\!\left({\cal A}_\alpha
	\right)$
	satisfies
	$\pi(\eta)=0$. This means that
	$$
		i\sum_{j,q}(p-q)_{\mu}\gamma^{\mu}\,a_j(q)b_j(p-q)\,
		e^{i\pi\alpha\omega(p,q)} = 0
\eqno(4.20)$$
	for all $p\in{\Z}^2$. Using eq.~(4.20), we have
	$$\eqalignno{
		\pi(\delta\eta)&=
		-\sum_{j,q}q_{\mu}(p-q)_{\nu}\gamma^{\mu}\gamma^{\nu}\,
		a_j(q)b_j(p-q)\,e^{i\pi\alpha\omega(p,q)} &\cr
		& = 
		-\sum_{j,q}(q^2-p^2)a_j(q)b_j(p-q)\,e^{i\pi\alpha\omega(p,q)}
		\in {\cal A}_\alpha\ .
&(4.21)
\cr}$$
	This proves that $\pi(\delta J^1)\subset{\cal A}_\alpha$.
	Then, we construct an explicit non-vanishing element of 
	$\pi(\delta J^1)$. We set
	$$\eqalign{
		a_1(p) & =  b_2(p) = \delta_{p_1,-1}\delta_{p_2,0}\ ,
\cr
		a_2(p) & =  b_2(p) = \delta_{p_1,1}\delta_{p_2,0}\ ,
\cr}$$
	and an easy computation shows that the element
	$\eta=\sum_{i=1}^{2}a_i\delta b_i$ satisfies 
	$$
		\pi(\eta)=0\;,\quad \pi(\delta\eta)=-g^{11}\ . 
	$$
	Since $\pi(\delta J^1)$ is an
	${\cal A}_\alpha$-bimodule, eq.~(4.17) follows. \hfill\break
\noindent Let $k\geq3$ and $\eta\in\Omega^k\left({\cal A}_\alpha\right)$.
	Then, using eqs.~(4.14) and (4.15), we see that there
	exists an element $\psi\in\Omega^{k-2}\left({\cal A}_\alpha\right)$
	with $\pi(\eta)=\pi(\psi)$. The first part of the proof ensures
	the existence of an element $\phi\in\Omega^1({\cal A}_\alpha)$
	with $\pi(\phi)=0$ and $\pi(\delta\phi)=\one$. Then we have
	$\phi\psi\in J^{k-1}$, and $\pi(\delta(\phi\psi))=\pi(\psi)=\pi(\eta)$,
	proving that $\eta\in\delta J^{k-1}$, and therefore
	eqs.~(4.18) and (4.19). \hfill\qed
\mn
As a corollary to this lemma, we obtain the following
\sn
{\bf Proposition 4.2}\quad  Up to isomorphism, the spaces of 
differential forms are given by
$$\eqalignno{
	\Omega_D^0 ({\cal A}_\alpha) &= {\cal A}_\alpha\ ,
&(4.22)\cr
		\Omega_D^1 ({\cal A}_\alpha) &\cong \left\{
  		a_\mu\,\gamma^\mu\,\big|\, a_\mu\,\in\,{\cal A}_
		\alpha\right\}\ ,
&(4.23)\cr
		\Omega_D^2 ({\cal A}_\alpha) &\cong \left\{ 
		a\,\sigma \,\big|\,a\,\in\,{\cal A}_\alpha \right\} \ ,
&(4.24)\cr
		\Omega_D^n ({\cal A}_\alpha) &= 0\; \quad \ 
		{\rm for}\; n\,\geq\,3 \ ,
&(4.25)\cr}$$
where we have chosen special representatives on the right hand side.
\mn
Notice that $\Omega_D^1 ({\cal A}_\alpha)$ and 
$\Omega_D^2({\cal A}_\alpha)$ are free left 
${\cal A}_\alpha$-modules of rank 2 and 1, respectively. 
This reflects the fact the bundles of 1- and
2-forms over the 2-torus are trivial and of rank 2 and 1, respectively.

\bn\bn
%%%%%%%%%%%%%%%%%%%%%%%%%%%%%%%%%%%%%%%%%%%%%%%%%%%%%%%%%%%%%%%%%%%%%%%%%
{\bf 4.2.2 Integration and Hermitian structure over 
	$\Omeinsalpha$}
\bn
It follows from eqs.~(4.13-15) that there is an isomorphism
$\pi(\Omega^{\bullet} ({\cal A}_\alpha)) \simeq {\cal
  A}_\alpha \otimes M_2({\C})$. Applying the general
definition of the integral -- see subsection 2.1.3 -- to the
non-commutative torus, one finds for an arbitrary element
$\omega \in \pi (\Omega^{\bullet}({\cal A}_\alpha))\,$,
$$
	\Barint \omega \ = \ {\rm Tr}_{{\C}^2} \left( \omega \left(
    	0\right)\right) 
\eqno(4.26)$$
where ${\rm Tr}_{{\C}^2}$ denotes the normalized trace on 
${\C}^2$.
The cyclicity property, Assumption 2.4 in
subsection 2.1.3, follows directly from the definition of the product
in ${\cal A}_\alpha$ and the cyclicity of the trace on $M_2({\C})$. 
The kernels $K^n$ of the canonical sesqui-linear
form on $\pi (\Omega^{\bullet} ({\cal A}_\alpha))$ -- see
eq.\ (2.5) -- coincide with the kernels $J^n$ of $\pi$, and we get for
all $n \in {\Z}_n$:
$$
	\widetilde{\Omega}^n ({\cal A}_\alpha)\;=\;\Omega^n ({\cal
  	A}_\alpha)\;, \  \quad 
	\widetilde{\Omega}_D^n ({\cal A}_\alpha)\;=\;\Omega_D^n ({\cal
  	A}_\alpha) \ .
\eqno(4.27)$$
Note that the equality $K^n=J^n$ holds in all explicit examples of
non-commutative $N=1$ spaces studied so far. 
It is easy to see that
the canonical representatives $\omega^\bot$ on ${\cal H}$ of
differential forms $[\omega] \in \Omega_D^n ({\cal A}_\alpha)$,
see eq.~(2.10), coincide with the choices already made in
eqs.~(4.22-25). The canonical Hermitian structure on 
$\Omega_D^1({\cal A}_\alpha)$ is simply given by
$$
	\langle \omega, \eta\rangle_D\;=\;\omega_\mu\,g^{\mu\nu}\,\eta_\nu^*
	\in {\cal A}_\alpha
\eqno(4.28)$$
for all $\omega, \eta \,\in\,\Omega_D^1 ({\cal A}_\alpha)$. Note
that this is a true Hermitian metric, i.e., it takes values in
${\cal A}_\alpha$ and not in the weak closure ${\cal
    A}''_\alpha$. Again, this is also the typical situation in other
examples.
\bn\bn
%%%%%%%%%%%%%%%%%%%%%%%%%%%%%%%%%%%%%%%%%%%%%%%%%%%%%%%%%%%%%%%%%%%%%%%%%%%
{\bf 4.2.3 Connections on $\Omeinsalpha$, and
	cohomology}
\bn
Since $\Omega_D^1 ({\cal A}_\alpha)$ is a free left ${\cal
  A}_\alpha$-module, it admits a basis which we can choose to be
$E^\mu := \gamma^\mu$. A connection $\nabla$ on $\Omega_D^1
({\cal A}_\alpha)$ is uniquely specified by its coefficients
$\Gamma_{\mu\nu}^\lambda \in {\cal A}_\alpha$,
$$
\nabla\,E^\mu\;=\;-\,\Gamma_{\nu\lambda}^\mu\,E^\nu \otimes E^\lambda
\ \ \in\   \Omega_D^1 ({\cal A}_\alpha) \otimes_{{\cal A}_\alpha}
	\Omega_D^1 ({\cal A}_\alpha) \ ,
\eqno(4.29)$$
and these coefficients can be chosen arbitrarily. Note that in the
classical case $(\alpha = 0)$ the basis $E^\mu$ consists of real
1-forms. Accordingly, we say that the connection 
$\nabla$ is real if its
coefficients in the basis $E^\mu$ are self-adjoint elements of
${\cal A}_\alpha$. A simple computation shows that there is a
unique real, unitary, torsionless connection $\nabla^{L.C.}$ on
$\Omega_D^1 ({\cal A}_\alpha)$ given by
$$
	\nabla^{L.C.}\, E^\mu \ = \ 0 \ .
\eqno(4.30)$$
In the remainder of this subsection, we determine the de Rham complex and its
cohomology. Let $U$ and $V$ be the elements of ${\cal A}_\alpha$ defined
in eq.~(4.10), then it is easy to verify that the elements 
$E^\mu$ of $\Omega_D^1 ({\cal A}_\alpha)$ given by
$$
	E^1=U^*\delta U\;, \  \quad E^2=V^*\delta V\ ,
\eqno(4.31)$$
form a basis of $\Omega_D^1 ({\cal A}_\alpha)$ and that they are closed,
$$
	\delta E^1=\delta E^2=0 \ .
\eqno(4.32)$$
A word of caution is in order here: Eq.~(4.32) does {\sl not} mean
that $\delta E^\mu$ is zero as an element of $\Omega^2 ({\cal A}_\alpha)$,
but that $\delta E^\mu\in\delta J^1$ which is zero in the quotient
space $\Omega_D^2 ({\cal A}_\alpha)$. As a basis of 
$\Omega_D^2 ({\cal A}_\alpha)$ we choose
$$
	F={1\over 2} \varepsilon_{\mu\nu}\gamma^\mu\gamma^\nu
\eqno(4.33)$$
and we get for the product of basis $1$-forms,
$$
	E^\mu E^\nu = \varepsilon^{\mu\nu} F \ .
\eqno(4.34)$$
This completely specifies the de Rham complex, and we can now compute 
of 
the cohomology groups $H^p$. For  $a\in{\cal A}_\alpha$,  we have the
equivalences 
$$\eqalignno{
	[D,a] = 0\ \  &\Longleftrightarrow\ \  
	i p_\mu\gamma^\mu a(p) = 0 \quad \forall p 
&\cr
	&\Longleftrightarrow\ \  a(p) = \delta_{p,0}\, \tilde a
&(4.35)\cr}$$
for some $\tilde a\in{\C}$. This shows that $H^0\simeq{\C}$. Let 
$a_\mu E^\mu$ be a $1$-form, then we obtain with 
eqs.~(4.32) and (4.34) that 
$$
	\delta(a_\mu E^\mu)=0\ \ \Longleftrightarrow\ \ 
	i{\widehat{p}}_\mu\, a_\nu\,\varepsilon^{\mu\nu}F=0
\eqno(4.36)$$
where ${\widehat{p}}_\mu$ denotes the multiplication operator by
$p_\mu$, i.e.,  $({\widehat{p}}_\mu a)(q)=q_\mu a(q)$. 
Suppose that the $1$-form $a_\mu
E^\mu$ is closed and satisfies $a_\mu(0)=0$, and define the algebra 
element $b$ by $b=\bigl(2i\widehat{p}_\mu\bigr)^{-1}a_\mu$. Using
eq.~(4.36), we see that 
$$
	\delta b=a_\mu E^\mu \ .
\eqno(4.37)$$
This proves that any closed $1$-form is cohomologous to a ``constant''
$1$-form $c_\mu E^\mu$ with $c_\mu\in{\C}$. On the other hand, a 
non-vanishing constant
$1$-form $c_\mu E^\mu$ cannot be exact since the equation
$$
(\delta a)(p)=i p_\mu\, E^\mu\,a(p) = \delta_{p,0}\,c_\mu\,E^\mu
\eqno(4.38)$$
has no solution. Thus, we have $H^1\simeq{\C}^2$. The same argument
shows that a constant $2$-form $cF$, with $c\in{\C}$, is not exact. If a
$2$-form $aF$ satisfies $a(0)=0$, then it is the coboundary of the $1$-form
$i\,\varepsilon_{\mu\nu} \bigl({\widehat p}_{\nu}\bigr)^{-1} a E^\mu$ 
and this proves the following
\mn
{\bf Proposition 4.3}\quad 
	In the basis $\{\,\one,\ E^\mu=\gamma^\mu,\ 
	F={1\over 2} \varepsilon_{\mu\nu}\gamma^\mu\gamma^\nu\,\}$ of
	$\Omega_{D}^{\bullet} ({\cal A}_\alpha)$, the de Rham
	differential algebra is specified by the following relations:
	$$\eqalign{
		&E^\mu E^\nu =\varepsilon^{\mu\nu}F\ , \cr
		&\delta E^\mu = \delta F=0 \ ,\quad\ 
		\delta a= i {\widehat{p}}_\mu E^\mu
		\quad \forall a\in{\cal A}_\alpha\ . 
\cr}$$
Furthermore, the cohomology of the de Rham complex is given by
$$
H^0\simeq H^2\simeq{\C}\;, \ \quad H^1\simeq{\C}^2\ . 
\eqno(4.39)$$
\mn
This completes our study of the $N=1$ data describing the non-commutative
$2$-torus at irrational deformation parameter.
\bn\bn
%%%%%%%%%%%%%%%%%%%%%%%%%%%%%%%%%%%%%%%%%%%%%%%%%%%%%%%%%%%%%%%%%%%%%%
{\bf 4.3 Riemannian geometry (\Noneone\kern -3pt)}
\bn
At the end of our discussion of the non-commutative 3-sphere, in 
subsection 3.4, we have briefly outlined a description in terms of  
``Riemannian'' $N=(1,1)$ data -- with the two generalized Dirac 
operators borrowed from conformal field theory, see \q{FG}. 
In the following, we will treat the non-commutative torus (at 
irrational deformation parameter) as a Riemannian space. Here we can, 
moreover, {\sl construct} a set of $N=(1,1)$ data from the 
Connes spectral triple along the general lines of subsection 2.2.5.  
\sn
Our first task is to find a real structure $J$ on the $N=1$ data
$({\cal A}_\alpha, {\cal H}, D, \sigma)$. To this end, we
introduce the complex conjugation $\kappa : {\cal H} \to {\cal
  H}$, $(\kappa x)(p) := \overline{x} (p) := \overline{x(p)}$, as well as the
charge conjugation matrix $C: {\cal H} \to {\cal H}$ as the
unique (up to a sign) constant matrix such that
$$\eqalignno{
	&C\;\gamma^\mu = -\; \overline{\gamma}^\mu\; C 
&(4.40)\cr
	&C\;=\;C^* = C^{-1} \ .
&(4.41)\cr}$$
Then it is easy to verify that $J=C\kappa$ is a real structure.
\sn
The right actions of ${\cal A}_\alpha$ and
$\Omega_D^1 ({\cal A}_\alpha)$ on ${\cal H}$ (see
subsection {2.2.5}) are given as follows
$$\eqalignno{
	\xi\ {\scriptstyle{\bullet}}\ a &\equiv J a^*\;J^*\xi \;=\;\xi \
	{\scriptstyle{\bullet}}_\alpha \ a^\vee 
&(4.42)\cr
	\xi \ {\scriptstyle{\bullet}} \ \omega &\equiv J \omega^* J^* \xi
	\;=\; \gamma^\mu \xi \ {\scriptstyle{\bullet}}_\alpha 
	\ \omega_\mu^\vee 
&(4.43)\cr}$$
where $\xi \in {\cal H},\ a \in {\cal A}_\alpha$,\ $\omega \in
\Omega_D^1 ({\cal A}_\alpha)$, $\xi \
{\scriptstyle{\bullet}}_\alpha \ a$ \ denotes the
diagonal right action of $a$ on $\xi$ by the deformed product, and
$$
	a^\vee (p) := a(-p)\ .
$$ 
Notice that $(a \ {\scriptstyle{\bullet}}_\alpha \ b)^\vee = a^\vee
{\scriptstyle{\bullet}}_\alpha \  b^\vee$. We
denote by  $\hnull$ the dense
subspace 
${\cal S}({\Z}^2) \oplus {\cal S} ({\Z}^2) \subset
{\cal H}$ of smooth spinors. The space $\hnull$  is a
two-dimensional free left ${\cal A}_\alpha$-module with 
canonical basis $\{ e_1, e_2\}$. Then, any connection $\nabla$ on
$\hnull$
is uniquely determined by its coefficients ${\omega_j}^i \in \Omega_D^1
({\cal A}_\alpha)$: 
$$
	\nabla\,e_i \ = \ {\omega_i}^j \otimes e_j \ = \ {\omega_{\mu i}}^j\,
	\gamma^\mu \otimes e_j \in \Omega_D^1 ({\cal A}_\alpha)
	\otimes_{{\cal A}_\alpha}\hnull
	\ .
\eqno(4.44)$$
The ``associated right connection'' $\overline{\nabla}$ is then given by
$$
	\overline{\nabla}\,e_i = e_j \otimes {\overline{\omega}_i}^j \in\;
	\hnull\otimes_{{\cal A}_\alpha}
	\Omega_D^1 ({\cal A}_\alpha) 
\eqno(4.45)$$
where 
$$
{\overline{\omega}_j}^i \ = \ -\, C_k^i ({\omega_l}^k)^*\,C_j^l \ = \ C_k^i
	({\omega_{\mu\,l}}^k)^*\, C_j^l\,\gamma^\mu \ .
\eqno(4.46)$$
An arbitrary element in $\hnull\otimes_{{\cal A}_\alpha}\hnull$ 
can be written as $e_i \otimes a^{ij} e_j$ where 
$a^{ij} \in {\cal A}_\alpha$. As in subsection 2.2.5, the 
``Dirac operators'' ${\cal D}$ and $\overline{{\cal D}}$ on $\hnull
\otimes_{{\cal A}_\alpha}\hnull$\ \  associated to the connection
$\nabla$ are given by 
$$\eqalignno{
	{\cal D} \left( e_i \otimes a^{ij}\,e_j\right) &=
	e_i \otimes \left(\delta\,a^{ij}\,+\,
	{\overline{\omega}_k}^i\,a^{kj}\,+\,a^{ik}\,{\omega_k}^j\right) \ 
	{\scriptstyle{\bullet}} \ e_j 
&(4.47)\cr
	\overline{{\cal D}} \left( e_i \otimes a^{ij}\,e_j\right) &=
	e_i \ {\scriptstyle{\bullet}} \left(
  	\delta\,a^{ij}\,+\,{\overline{\omega}_k}^i\,a^{kj}\,+\,
 	 a^{ik}\,{\omega_k}^j\right) \otimes \sigma\,e_j \ .
&(4.48)\cr}$$
In order to be able to define a scalar product on 
$\hnull
\otimes_{{\cal A}_\alpha} \hnull $, we need a
Hermitian structure on the {\it right} 
module $\hnull $, 
denoted by $\langle \cdot, \cdot\rangle$, with values in ${\cal
  A}_\alpha$. It is defined by
$$
	\Barint \ \langle \xi, \zeta\rangle\,a \ = \ \left( \xi,\zeta\,a\right)
	\quad \forall\, \xi, \zeta \in 
	\hnull 
	\; , \ \forall \, a \in {\cal A}_\alpha \ .
\eqno(4.49)$$
This Hermitian structure can be written explicitly as 
$$
	\langle \xi,\zeta\rangle \ = \ \sum_{i=1}^2 \overline{\xi^i} \
	{\scriptstyle{\bullet}}_\alpha \ \zeta^{i\vee}\ ,
\eqno(4.50)$$
and it satisfies
$$
	\langle \xi\,a, \zeta\,b\rangle \ = \ a^* \,\langle \xi,
	\zeta\rangle\, b
\eqno(4.51)$$
for all $\xi, \zeta \in \hnull $ and
$a, b \in {\cal A}_\alpha$. Then we define the scalar product on 
$\hnull
\otimes_{{\cal A}_\alpha}\hnull $ as (see \q{Co1})
$$
	\left( \xi_1 \otimes \xi_2,\;\zeta_1\otimes\zeta_2\right) \ = \ 
	\left( \xi_2, \langle \xi_1, \zeta_1\,\rangle\,\zeta_2\right) \ .
\eqno(4.52)$$
This expression can be written in a more suggestive way if one
introduces a Hermitian structure, denoted
$\langle\cdot,\cdot\rangle_L$, on the {\it left} module
$\hnull $: 
$$
	\langle \xi, \zeta \rangle_L \ := \ \langle J\,\xi, J\,\zeta \rangle 
	\ .
$$
This Hermitian structure satisfies
$$
	\langle a\,\xi, b\,\zeta\rangle_L \ = \ a\,\langle 
	\xi, \zeta\rangle_L\,b^*
$$
for all $a, b \in {\cal A}_\alpha$ and $\xi, \zeta \in
\hnull $, and the scalar product on 
$\hnull
\otimes_{{\cal A}_\alpha} \hnull $
can be written as follows
$$
	\left( \xi_1 \otimes \xi_2, \zeta_1 \otimes \zeta_2 \right) \ = \ 
	\Barint \ \langle \xi_1, \zeta_1\rangle\;\langle \zeta_2, 
	\xi_2\rangle_L \ . 
$$
A tedious computation shows that the relations
$$
	{\cal D}^* = {\cal D}\,,\quad \ \overline{{\cal D}}^* 
	= \overline{{\cal D}}\,,\quad \ \{\, {\cal D}, 
	\overline{{\cal D}} \,\} = 0\,,\quad \ {\cal  D}^2 = 
	\overline{{\cal D}}^2
\eqno(4.53)$$
are equivalent to 
$$
	\widetilde{\nabla}\, e_i\otimes e_j \ = \ 0 \quad \forall\,i,\,j \ .
\eqno(4.54)$$
In particular, we see that the original $N=1$ data uniquely determine 
the operators ${\cal D}$ and $\overline{{\cal D}}$ 
satisfying the $N=(1,1)$ algebra -- cf.\ Definition 2.20 -- 
$$
{\cal D}^2 = \overline{{\cal D}}{}^2\ ,\quad\ 
\{\,{\cal D},\overline{{\cal D}}\,\} = 0\ .
$$
\sn
One can prove that there are unique ${\Z}_2$-grading operators
$$
	\gamma=\one\otimes\sigma\;, \ \quad \overline{\gamma}=\sigma\otimes\one
\eqno(4.55)$$
commuting with ${\cal A}_\alpha$ and such that
$$\eqalign{
\{\,{\cal D},\gamma\,\}&=\{\,\overline{{\cal D}},\overline{\gamma}\,\}=0
\cr
	{}[\,{\cal D},\overline{\gamma}{}\,]&=
	{}[\,\overline{{\cal D}},\gamma{}\,]=0\ .
\cr}$$
The combined ${\Z}_2$-grading
$$
	\Gamma=\gamma\overline{\gamma}
$$
together with the Hodge operator
$$
	*=\overline{\gamma}
$$
complete our data to a set of $N=1$ data 
$\left({\cal A}_\alpha,{\cal H}\otimes_{{\cal A}_\alpha}
{\cal H},{\cal D},\overline{{\cal D}},\Gamma, *\right)$.
Furthermore, these data admit a unique ${\Z}$-grading
$$
	{\cal T}={1\over2i}\,g_{\mu\nu}\gamma^\mu\otimes\gamma^\nu\sigma
$$
commuting with ${\cal A}_\alpha$, whose mod $2$ reduction equals $\Gamma$,
and such that
$$
	[\,{\cal T},d\,]=d\ .
$$
\bn\bn
%%%%%%%%%%%%%%%%%%%%%%%%%%%%%%%%%%%%%%%%%%%%%%%%%%%%%%%%%%%%%%%%%%%%%%%%%
{\bf 4.4 K\"ahler geometry (\Ntwotwo\kern -3pt)}
\bn
The classical torus can be regarded as a complex K\"ahler manifold, 
and thus it is  natural to ask whether we can extend the $N=(1,1)$ 
spectral data to $N=(2,2)$ data in the non-commutative case, too. 
The simplest way to determine such an extension is to look for an
anti-selfadjoint operator $I$ commuting with ${\cal A}_\alpha$, $\Gamma$,
$*$, and ${\cal T}$, and then to define a new differential by
$$
	d_I=[\,I,d\,]\ .
\eqno(4.56)$$
The nilpotency of $d_I$ implies further constraints on the operator
$I$. 
The idea behind this construction is to identify $I$ with $i(T-\overline{T})$,
where $T$ and $\overline{T}$ are as in Definition 2.26. In the classical
setting, this operator has clearly the above properties.
\sn
The most general operator $I$ on 
${\cal H} \otimes_{{\cal A}_\alpha} {\cal H}$ 
that commutes with all elements of ${\cal A}_\alpha$ is of
the form
$$
	I \ = \ \sum_{\mu,\nu = 0}^3 \gamma^\mu\,\otimes\,\gamma^\nu \,
	I_{\mu\nu}^R 
\eqno(4.57)$$
where $I_{\mu\nu}^R$ are elements of ${\cal A}_\alpha$ acting on
${\cal H} \otimes_{{\cal A}_\alpha} {\cal H}$ from the
right, and where we have set
$$
	\gamma^0 \ = \ \one \ , \quad \gamma^3 \ = \ \sigma \ .
\eqno(4.58)$$
The vanishing of the commutators of $I$ with $\Gamma$ and
$*$ implies that $I_{\mu\nu}^R=0$ unless $\mu, \nu \in \{
0,3\}$.
The equation $[\,I,{\cal T}\,]=0$ requires $I_{03}^R = I_{30}^R$ and leaves the
coefficients $I_{00}^R$ and $I_{33}^R$ undetermined. Since the
operator $I$ appears only through commutators, its trace part is
irrelevant and we can set $I_{00}^R = 0$. All constraints together
give
$$
	I\;=\;\left( \sigma\,\otimes\,\one\;+\;\one\,\otimes\,\sigma\right)\,
	I_{03}^R \;+\; \left( \sigma\,\otimes\,\sigma\right) \, I_{33}^R 
\eqno(4.59)$$
where $I_{03}^R$ and $I_{33}^R$ are anti-selfadjoint elements of
${\cal A}_\alpha$. We decompose $I$ into two parts
$$\eqalignno{
	I_1 &= \left( \sigma\,\otimes\,\one \ + \ \one\,\otimes\,\sigma\right)
	\; I_{03}^R
&(4.60)\cr
	I_2 &= \left( \sigma\,\otimes\,\sigma\right)\; I_{33}^R
&(4.61)\cr}$$
and we introduce the new differentials according to eq.~(4.56)
$$\eqalignno{
	d_1 &= d \ = \ {1\over2}\,({\cal D}\;-\;i\,\overline{{\cal D}})
&(4.62)\cr
	d_2 &= \left[\, I_1, d\,\right] 
&(4.63)\cr
	d_3 &= \left[\, I_2, d\,\right] \ .
&(4.64)\cr}$$
The nilpotency of $d_2$ and $d_3$ implies that $I_{03}$ and $I_{33}$
are multiples of the identity, and we normalize them as follows
$$\eqalignno{
	I_1 &= {i\over 2} \ \left( \sigma\,\otimes\,\one \ + \
  	\one\,\otimes\,\sigma\right) \ 
&(4.65)\cr
	I_2 &= i\,\left( \sigma\,\otimes\,\sigma\right) \ .
&(4.66)\cr}$$
Comparing eqs.~(4.66) and (4.55), we see that
$$
	I_2 \ = \ i\,\gamma\,\overline{\gamma}
\eqno(4.67)$$
and it follows, using eqs.~(4.62) and (4.64), that
$$
d_3 \ = \ \left[\, I_2, d\,\right] \ = \ 2\,i\,d\,\gamma\,\overline{\gamma} \ .
\eqno(4.68)$$
Thus, the differential $d_3$ is a trivial modification of $d$, and we
discard it. It is then easy to verify that $({\cal A}_\alpha,
{\cal H} \otimes_{{\cal A}_\alpha} {\cal H}, d_1, d_2,
\Gamma, *, {\cal T})$ form a set of $N=(2,2)$ spectral
data together with a ${\Z}$-grading. Furthermore, they are, as we have
shown, canonically determined by 
the original $N=(1,1)$ data. Therefore, a 
Riemannian non-commutative torus (at irrational deformation parameter
$\alpha$) admits a canonical K\"ahler structure. Notice that if we choose
the metric $g^{\mu\nu}=\delta^{\mu\nu}$ in eq.\ (4.4), then 
$\partial = -{1\over2}(d_1+id_2)$ coincides with the holomorphic differential
obtained in \q{Co1} from cyclic cohomology and using the equivalence of 
conformal and complex structures in two dimensions. 
\mn
We have only given the definitions of the spectral data in the
$N=(1,1)$ and the $N=(2,2)$ setting. As a
straightforward application of the general methods described in
section 2, we could compute the associated de~Rham resp.~Dolbeault
complexes, or geometrical quantities like curvature. We do not carry
out these calculations.
\sn
Instead, let us emphasize the following feature: In section 
3, we already say that the topology of ``the'' non-commutative 
3-sphere depends on the spectral data other than the algebra. 
Now, we learn once again that, for rational deformation parameter 
$\alpha = {M \over N}$, the algebra ${\cal  A}_\alpha$ 
does not specify the geometry of the underlying non-commutative space. 
It is only the selection of a specific $K$-cycle $({\cal H}, D)$ that 
allows us to identify this space as a deformed torus.
By choosing different K-cycles $({\cal  H}, D)$ for the same algebra
${\cal A}=M_N({\C})$ (with $N=\sum_{j=1}^{l}j^2$) we are able to describe
{\sl either} a fuzzy three-sphere, as discussed in Sect.\ 3, {\sl or}
a non-commutative torus. In other words, choosing {\sl different} 
spectral data, but keeping the algebra ${\cal A}$ {\sl fixed}, may lead
to {\sl different} non-commutative geometries.
\sn
Yet, it is plausible that the sequence ${\cal A}_N := M_N({\C})$,
$N=1,2,3,\ldots$, of algebras may be associated uniquely with 
non-commutative tori, while the sequence ${\cal A}_N := M_N({\C})$,
$N=\sum_{j=1}^{l}j^2$, $l=1,2,3,\ldots$, may be associated uniquely
with fuzzy three-spheres.

\vfill
\eject

\leftline{\bf 5. Directions for future work}
\bn
In this work and in part I, we have presented an approach to 
(non-commutative) geometry rooted 
in supersymmetric quantum theory. We have classified the various 
types of classical and of non-commutative geometries according 
to the symmetries, or to the ``supersymmetry content'', of their  
associated spectral data. Obviously, many natural 
and important questions remain to be studied. In this 
concluding section, we describe a few of these open problems and sketch, 
once more, some of the physical motivations underlying our work. 
\sn
{\bf(1)}\quad An obvious question is whether one can give a 
complete classification of the possible types of spectral data in terms of 
graded Lie algebras (and, perhaps, $q$-deformed graded Lie algebras). As an 
example, we recall the structure of $N=4^+$ spectral data , describing an 
extension of K\"ahler geometry (see sections 1.2 and 3 of part I). The 
spectral data involve the operators 
$\ttd,\ \ttdst,\ \td,\ \tdst,\ L^3,\ L^+,\ L^-,\ J_0$ and 
$\triangle$, which close under taking (anti-)commutators: They generate 
a graded Lie algebra defined by 
$$\eqalign{
&\lb\,L^3,L^{\pm}\,\rb=\pm2L^{\pm}\ ,\quad
\lb\,L^+,L^{-}\,\rb=L^{3}\ ,\quad 
\lb\,J_0,L^{3}\,\rb=\lb\,J_0,L^{+}\,\rb=0\ ,
\cr
&\lb\,L^3,\ttd\,\rb=\ttd\ ,\quad
\lb\,L^+,\ttd\,\rb=0\ ,\quad
\lb\,L^-,\ttd\,\rb=\tdst\ ,\phantom{NN}
\lb\,J_0,\ttd\,\rb=-i\,\td\ ,    
\cr
&\lb\,L^3,\td\,\rb=\td\ ,\quad
\lb\,L^+,\td\,\rb=0\ ,\quad
\lb\,L^-,\td\,\rb=-\ttdst\ ,\quad
\lb\,J_0,\td\,\rb= i\,\ttd\ ,    
\cr
&\{\,\ttd,\ttd\,\}=\{\,\td,\td\,\}= \{\,\ttd,\td\,\}=
\{\ttd,\tdst\,\}=0\ ,
\cr
&\{\,\ttd,\ttdst\,\}=  \{\,\td,\tdst\,\}=   \triangle \ ,  
\cr}$$
where $\triangle$, the Laplacian, is a central element. The remaining 
(anti-)commutation relations follow by taking adjoints, with the rules 
that $\triangle$, $J_0$ and $L^3$ are self-adjoint, 
and $(L^-)^*=L^+$. \hfill\break
\noindent It would be interesting to determine all graded Lie algebras 
(and their representations) occurring in spectral data of a
(non-commutative) space. In the case of classical geometry, we have given 
a classification up to $N=(4,4)$ spectral data, and there appears to be 
enough information in the literature to settle the problem completely; 
see \q{Bes,HKLR,Joy}.  
In the non-commutative setting, however, further algebraic structures 
might occur, including $q$-deformations of graded Lie algebras.\hfill\break 
\noindent To give a list of all graded Lie algebras that are, in principle, 
admissible, appears possible; see \q{FGR2} for additional 
discussion. However, in view of the classical 
case, where we only found the groups U(1), SU(2), Sp(4) and direct 
products thereof (see part I, section 3) we expect that not all Lie group 
symmetries that may arise in principle are actually realized in 
(non-commutative) geometry. 
\sn
Determining the graded Lie algebras that actually occur in the spectral data 
of geometric spaces is clearly just the first step towards a classification 
of non-commutative spaces. A more difficult problem will be to characterize 
the class of all 
${}^*$-algebras \a\ that admit a given type of spectral data, i.e.\ the 
class of algebras that possess a $K$-cycle $(\h, \ttd_i)$ with a collection of 
differentials $\ttd_i$ generating a certain graded Lie algebra 
such that the ordinary Lie group generators $X_j$ contained 
in the graded Lie algebra commute with the elements of \a. 
\mn
{\bf(2)}\quad Given some non-commutative geometry defined in terms of spectral 
data, it is natural to investigate its symmetries,  i.e.\  to introduce a 
notion of diffeomorphisms. 
For definiteness, we start {}from a set of data $(\a, \h,
{\tt d}, {\tt d}^*, T, *)$  with an $N=2$ structure, cf.\ section 2.2.6. 
To study notions of diffeomorphisms, it is useful to introduce an 
algebra $\Phi_{\rm d}^{\bullet}(\a)$
defined as the smallest ${}^*$-algebra of (unbounded) operators containing 
${\cal B}:= \pi(\Omega^{\bullet}(\a))\lor \pi(\Omega^{\bullet}(\a))^*$ 
and arbitrary graded commutators of \ttd\ and \ttdst\ with elements of 
$\cal B$. Due to the existence of the 
$\Z$-grading $T$, $\Phi_{\rm d}^{\bullet}(\a)$ decomposes into 
a direct sum 
$$
\Phi_{\rm d}^{\bullet}(\a) := \bigoplus_{n\in\Z} \Phi_{\rm d}^{n}(\a)\ ,
\quad\quad \Phi_{\rm d}^{n}(\a) := \bigl\{\,\phi\in\Phi_{\rm d}^{\bullet}
(\a)\,\big\vert\,\lb\,T,\phi\,\rb_g=n\,\phi\,\bigr\}\ .
$$
Note that both positive and negative degrees occur. 
Thus, $\Phi_{\rm d}^{\bullet}(\a)$ is a graded ${}^*$-algebra.
This algebra is quite a natural object to introduce when dealing 
with $N=2$ spectral data, as 
the algebra $\Omega_{\rm d}^{\bullet}(\a)$ of differential forms 
does not have a ${}^*$-representation on \h,  because 
{\tt d} is not self-adjoint.    \hfill\break
\noindent Ignoring operator domain problems arising because the 
(anti-)commutator of \ttd\ with the adjoint of a differential 
form is unbounded, in general, we observe 
that $\Phi_{\rm d}^{\bullet}(\a)$ has the interesting property that  
it forms a complex with respect to the action of \ttd\ by  graded
commutation, and, in view of examples {}from quantum field theory, we 
call it the {\sl field complex} in the following. \hfill\break
\noindent 
For $N=(2,2)$ non-commutative K\"ahler data with 
holomorphic and anti-holomorphic gradings $T$ and $\overline{T}$, see 
Definition 2.26, one may introduce a {\sl bi-graded complex} 
$\Phi_{\partial,\overline{\partial}}^{\bullet,\bullet}(\a)$ in a similar 
way. A slight generalization of such bi-graded field complexes containing 
operators $\phi$ of degree $(n,m)$ with $n$ and $m$ {\sl real}, but 
$n+m \in\Z$, naturally occurs  in $N=(2,2)$ superconformal 
field theory, see e.g.\ \q{FG,FGR2} and references given there. 
\sn
Next, we show how the field complex appears when we attempt to introduce 
a notion of {\sl diffeomorphisms} of a (non-commutative) geometric space 
described in terms of $N=2$ spectral data:  
One possible generalization of the notion of diffeomorphisms 
to non-commutative geometry is to identify them with
${}^*$-automorphisms of the algebra \a\ of ``smooth functions''. 
It may be advantageous, though, to follow concepts from classical 
geometry more closely: An infinitesimal diffeomorphism is then given by a 
derivation $\delta(\cdot) := \lb\,L,\cdot\,\rb$ of \a\ where $L$ is an 
element of $\Phi_{\rm d}^0$ such that $\delta$ commutes with {\tt d}, i.e.\   
$$
\lb\,{\tt d},L\,\rb =0 \ .
$$
The derivation $\delta$ can then be extended to all of $\pi(\Omega_{\rm
d}^{\bullet}(\a))$, and $\delta$ preserves the degree of differential forms 
iff $L$ commutes with $T$, i.e.\ iff $L\in \Phi_{\rm d}^0$.  \hfill\break 
\noindent For a classical manifold $M$, it turns out that each $L$ with 
the above properties can be written as 
$$
L= \{\,{\tt d}, X\,\}
$$ 
for some vector field $X\in\Phi_{\rm d}^{-1}$, i.e.\ $L$ is the Lie 
derivative in the direction of this vector field. 
In the non-commutative situation, however, it might happen  that 
the cohomology of the field complex at the zeroth position is non-trivial. 
In this case, the study of diffeomorphisms of the non-commutative space 
necessitates studying the cohomology of the field complex 
$\Phi_{\rm d}^{\bullet}(\a)$ in degree zero.   
\sn
As in classical differential geometry, it is interesting to investigate 
{\sl special} diffeomorphisms, i.e.\ ones that preserve additional 
structure in the spectral data. As an example, consider derivations  
$\delta(\cdot) = \lb\,L,\cdot\,\rb$ such that $L$ commutes with \ttd\ 
{\sl and} $\ttdst\,$: They generate {\sl isometries} of the non-commutative 
space. For complex spectral data, we may consider derivation not only 
commuting with $\ttd$ but also with $\partial$: They generate one-parameter 
groups of holomorphic diffeomorphisms. In the example of symplectic 
spectral data, we are interested in diffeomorphisms preserving the 
symplectic forms, i.e., in symplectomorphisms. One-parameter 
groups of symplectomorphisms are generated by derivations commuting 
with $\ttd$ and $\tdst$. 
\mn
{\bf(3)}\quad Another important topic in non-commutative geometry is 
{\sl deformation theory}. Given spectral data specified in terms 
of generators $\{\,X_j,\ \ttd,\ \ttd_{\alpha},\ \triangle\,\}$ 
of a graded Lie algebra as in remark (1), we may study 
one-parameter families 
$\{\,X_j^{(t)},\ \ttd,\ \ttd_{\alpha}^{(t)},\ \triangle^{(t)}\,\}_{t\in\R}$ 
of deformations. Here, we choose to keep one generator, $\ttd$, fixed, and we 
require that the graded Lie algebras are isomorphic to one another 
for all $t$. This means that we study deformations of the (non-commutative) 
complex or symplectic structure of a given space \a\ while preserving the 
differential and the de Rham complex. 
Only those deformations of spectral data are of interest which cannot 
be obtained from the original ones by ${}^*$-automorphisms of the algebra 
\a\ commuting with $\ttd$ (i.e.\ by ``diffeomorphisms''). 
In classical geometry, the deformation theory of complex structures is 
well-developed (Kodaira-Spencer theory), and there are non-trivial 
results in the deformation theory of symplectic structures (e.g.\ 
Moser's theorem); but this last topic is still a subject of active 
research. 
\hbn
Next, we consider deformations $\ttd'$ of the differential $\ttd$ of a given 
set of $N=2$ spectral data $(\a, \h, \ttd, \ttdst, T, *)$ which are of the 
form 
$$
{\tt d}' := {\tt d} + \omega \ ,
$$
for some operator $\omega\in \Phi^{\bullet}_{\rm d}(\a)$ of odd degree. 
We require that ${\tt d}'$ again squares to zero, which implies 
that $\omega$ has to satisfy a zero curvature condition 
$$
\omega^2 + \{\,{\tt d}, \omega\,\} = 0\ .
\eqno(5.1)$$
We distinguish between several possibilities: First, we require 
that the deformed data still carry an $N=2$ structure with the same 
$\Z$-grading $T$ as before. Then $\omega$ must be an element of 
$\Phi_{\rm d}^{1}(\a)$ satisfying (5.1), and we can identify it with 
the connection 1-form of a flat connection on some vector bundle; 
for an example, see the discussion of the structure of classical 
$N=(1,1)$ Dirac bundles in section 2.2.3 of part I.    \hfill\break
\noindent  More generally, we only require the deformed data to be of 
$N=(1,1)$ type, with a $\Z_2$-grading $\gamma$ given by the mod$\,2$
reduction of $T$. As a simple example, consider an operator 
$\omega$ in $\Phi^{\bullet}_{\rm d}(\a)$ of degree $2n+1$, with $n\neq 0$. 
Then condition (5.1) implies that 
$$
\omega^2 = 0 \quad {\rm and} \quad \{\,\ttd, \omega\,\} = 0\ .
$$
If $\omega = \lb\,\ttd, \beta\,\rb$ and $\lb\,\beta,\omega\,\rb = 0$ 
then 
$$ \ttd' = e^{-\beta}\, \ttd \,e^{\beta}\ .
$$
We then say that $\ttd$ and $\ttd'$ are equivalent. If $\omega$ represents 
a non-trivial cohomology class of the field complex $\Phi^{\bullet}_{\rm d}(\a)$
then $\ttd$ and $\ttd'$ are inequivalent.

\mn
{\bf (4)}\quad In the introduction to paper I and in \q{FGR2} 
we have remarked that, {}from the point of view of physics, 
it is quite unnatural to attribute 
special importance to the algebra of functions over configuration 
space. The natural algebra in Hamiltonian mechanics is the algebra 
of functions over phase space, and, in quantum mechanics, it is 
a non-commutative deformation thereof, denoted ${\cal F}_{\hbar}$ 
(where $\hbar$ is Planck's constant), which is the natural algebra 
to study. In examples where phase space is given as the cotangent 
bundle $T^*M$ of a smooth manifold $M$, the configuration space, 
one may ask whether there are natural mathematical relations between 
spectral data involving the algebra $\a = C^{\infty}(M)$ and ones 
involving the algebra  ${\cal F}_{\hbar}$. For example, it may be 
possible to represent $\a$ and  ${\cal F}_{\hbar}$ on the same 
Hilbert space $\h$ and consider spectral data $(\a,\h,\ttd,T,*)$ 
and $({\cal F}_{\hbar}, \h, \ttd, T, *)$ with the {\sl same} 
choice of operators $\ttd, T$ and $*$ on \h. It is well known 
that from $(\a,\h,\ttd,T,*)$ configuration space $M$ can be 
reconstructed (Gelfand's theorem and extensions thereof). This 
leads to the natural question whether $M$ can also be reconstructed
{}from $({\cal F}_{\hbar}, \h, \ttd, T, *)$, or whether at least 
some of the topological properties of $M$, e.g.\ its Betti numbers, 
can be determined from these data.
\hbn 
It is known that, in string theory, spectral data generalizing 
 $({\cal F}_{\hbar}, \h, \ttd, T, *)$ do not determine configuration 
space uniquely; this is related to the subject of stringy dualities 
and symmetries, more precisely to $T$ dualities, see e.g.\ \q{GPR} 
and also \q{KS,FG}.
The distinction between ``algebras of functions on configuration space'' 
\a\ and ``algebras of functions on phase space'' ${\cal F}$ remains 
meaningful in many examples of non-commutative spaces. Typically, 
${\cal F}$ arises as a crossed product of \a\ by some group $G$ of 
``diffeomorphisms''. Under what conditions properties of the algebra 
\a\ can be inferred from spectral data $({\cal F}_{\hbar}, \h, \ttd, 
T, *)$ {\sl without} knowing explicitly how the group $G$ acts on 
${\cal F}$ represents a problem of considerable interest in quantum 
theory. 
\hbn
For another perspective concerning the distinction between ``algebras 
of functions on configuration space'' and ``algebras of functions on 
phase space'' see section 2.2.6. 
\sn
It is worth emphasizing that in quantum field theory and string theory, 
where $M$ is an infinite-dimensional space, the analogue of the ``algebra 
of functions on $M$'', i.e.\ of \a, does not exist, while the analogue 
of the ``algebra of functions on phase space $T^*M$'', i.e.\ of ${\cal F}$, 
still makes sense. For additional discussion of these matters see also 
\q{FGR2}. 
\mn
{\bf (5)}\quad A topic in the theory of complex manifold that has 
attracted a lot of interest, recently, is {\sl mirror symmetry}. 
For a definition of mirrors of classical Calabi-Yau manifolds, 
see e.g.\ \q{Y} and references given there, and cf.\ 
the remarks at the end of section I$\,$2.4.3. It is natural to ask whether
one can define mirrors of 
non-commutative spaces, and whether some classical manifolds may have 
non-commutative mirrors. Superconformal field theory with $N=(2,2)$
supersymmetry  suggests how one might define a mirror map in  the context 
of non-commutative geometry (see \q{FG,FGR2}): Assume that two sets of 
$N=(2,2)$ spectral data $(\a_i,\h, \partial_i, \overline{\partial}_i, T_i, 
\overline{T}_i, *_i)$, $i=1,2$, are given, where the algebras $\a_i$ act on 
the Hilbert spaces $\h_i$ which are subspaces of a single Hilbert space 
\h\ on which the operators $\partial_i, \overline{\partial}_i, T_i, 
\overline{T}_i$ and $*_i$ are defined. We say that the space $\a_2$ 
is the {\sl mirror} of $\a_1$ if 
$$\partial_2=\partial_1\ ,\quad \overline{\partial}_2=
\overline{\partial}{}_1^*\ ,\quad 
T_2=T_1\ ,\quad \overline{T}_2=-\overline{T}_1\ ,
$$ 
and if the dimensions 
$b_i^{p,q}$ of the cohomology of the Dolbeault complexes (2.45) 
satisfy  $b_2^{p,q}=b_{1}^{n-p,q}$, where $n$ is the top dimension 
of differential forms (recall that in Definition 2.26 we required $T$ 
and $\overline{T}$ to be bounded operators).  \hfill\break 
\noindent Let \a\ be a non-commutative K\"ahler space with mirror
$\widetilde{\a}$. Within superconformal field theory, there is the 
following additional relation between the two algebras: Viewing \a\ as the 
algebra of functions over a (non-commutative) target $M$, and analogously 
for $\widetilde{\a}$ and $\widetilde{M}$, the phase spaces over the
loop spaces over $M$ and $\widetilde{M}$ coincide. 
\mn
{\bf(6)}\quad The success of the theory presented in this paper will
ultimately be measured in terms of the applications it has to 
concrete problems of geometry and physics. 
In particular, one should try to apply the notions developed here to 
further examples of truly non-commutative spaces such as 
quantum groups, or the non-commutative complex projective spaces (see e.g.\
\q{Ber,Ho,Ma,GKP}), non-commutative Riemann surfaces \q{KL}, and 
non-commutative symmetric
spaces $\lb {\sl BLU,BLR,GP,BBEW}\rb$.  In most of 
these cases, it is natural to ask whether
the ``deformed'' spaces carry a complex or K\"ahler structure in the sense of
section 2.3 above. 
\sn   {}From our point of view, however, the most interesting examples
for the general theory and the strongest motivation to study spectral data
with supersymmetry come {}from string theory: The ``ground states''
of string theory are described by certain $N=(2,2)$ superconformal 
field theories.  They provide the spectral data of the loop space over a
target which is a ``quantization'' of classical space -- or rather of
an internal compact manifold.  
It may happen that the conformal field theory is the quantization of a 
$\sigma$-model of maps from a parameter space into a classical target
manifold. In general, the target space reconstructed from the spectral 
data of the conformal field theory then turns out to be a (non-commutative) 
deformation of the target space of the classical $\sigma$-model. The 
example of the superconformal SU(2) Wess-Zumino-Witten model, which is 
the quantization of a $\sigma$-model with target SU(2), has been studied 
in some detail in \q{FG,Gr,FGR2} and has motivated the results presented 
in section 3. A more interesting class of examples would consist of 
$N=(2,2)$ superconformal field theories which are quantizations of 
$\sigma$-models whose target spaces are given by three-dimensional 
Calabi-Yau manifolds. But one may also apply the methods developed in 
this paper to  superconformal field theories which, at the outset, are 
{\sl not} quantizations of some classical $\sigma$-models. They may 
enable us to reconstruct (typically non-commutative) geometric spaces 
{}from the supersymmetric spectral data of such conformal field 
theories. This leads to the idea that, quite generally, superconformal 
field theories are (quantum) $\sigma$-models, but with target spaces 
that tend to be non-commutative spaces. An interesting family of examples 
of this kind consists of the Gepner models, which are expected to give 
rise to non-commutative deformations of certain Calabi-Yau three-folds. 
For further discussion of these ideas see also \q{FG,FGR2}. 

\sn 
\vfill\eject 
%%%%%%%%%%%%%%%%%%%%%%%%%%%%%%%%%%%%%%%%%%%%%%%%%%%%%%%%%%%%%
\parindent=35pt \vsize=23.5truecm
\halign{#\hfil&\vtop{\parindent=0pt \hsize=35.5em #\strut}\cr
\noalign{\leftline{{\bf References }}} \noalign{\vskip.4cm}
\q{AG} &L.\ Alvarez-Gaum\'e, {\sl Supersymmetry and the Atiyah-Singer 
   index theorem},  Commun.\ Math.\ Phys. {\bf90} (1983) 161-173\cr 
\q{AGF}&L.\ Alvarez-Gaum\'e, D.Z.\ Freedman, {\sl Geometrical structure and 
   ultraviolet finiteness in the supersymmetric $\sigma$-model}, 
   Commun.\ Math.\ Phys. {\bf80} (1981) 443-451\cr 
\q{BC} &P.\ Beazley Cohen, {\sl Structure complexe non commutative et
   superconnexions}, pre\-print MPI f\"ur Mathematik, Bonn, MPI/92-19\cr
\q{Ber} &F.A.\ Berezin, {\sl General concept of quantization},  
   Commun.\ Math.\ Phys.\ {\bf40} (1975) 153-174\cr
\q{Bes} &A.L.\ Besse, {\sl Einstein Manifolds}, Springer Verlag 1987\cr
\q{BLR} &D.\ Borthwick, A.\ Lesniewski, M.\ Rinaldi, {\sl Hermitian 
   symmetric superspaces of type IV}, J.\ Math.\ Phys.\ {\bf343} (1993) 
   4817-4833\cr
\q{BLU} &D.\ Borthwick, A.\ Lesniewski, H.\ Upmeier, {\sl Non-perturbative
   quantization of Cartan domains}, J.\ Funct.\ Anal.\ {\bf113} (1993)
   153-176\cr
\q{BBEW} &M.\ Bordemann, M.\ Brischle, C.\ Emmrich, S.\ Waldmann, 
   {\sl Phase space reduction for star-products: an explicit 
   construction for $\C{\rm P}^n$}, Lett.\ Math.\ Phys.\ {\bf36} 
   (1996) 357-371\cr
\q{CF} &A.H.\ Chamseddine, J.\ Fr\"ohlich, {\sl Some elements of
  Connes' non-commutative geometry, and space-time geometry}, in:\ 
  Chen Ning Yang, a Great Physicist of the Twentieth Century, 
  C.S.\ Liu and S.-T.\ Yau (eds.), International Press 
  1995, pp.\ 10-34\cr
\q{CFF} &A.H.\ Chamseddine, G.\ Felder, J.\ Fr\"ohlich, {\sl Gravity
   in non-commutative geometry}, Commun.\ Math.\ Phys.\ {\bf155}
   (1993) 205-217\cr 
\q{CFG} &A.H.\ Chamseddine, J.\ Fr\"ohlich, O.\ Grandjean, {\sl The
   gravitational sector in the Connes-Lott formulation of the 
   standard model}, J.\ Math.\ Phys.\ {\bf36} (1995) 6255-6275\cr
\q{Co1} &A.\ Connes, {\sl Noncommutative Geometry}, Academic Press
   1994\cr
\q{Co2} &A. Connes, {\sl Noncommutative differential geometry}, 
   Inst.\ Hautes \'Etudes Sci.\ Publ.\ Math.\ {\bf62} (1985) 257-360\cr
\q{Co3} &A.\ Connes, {\sl The action functional in noncommutative 
   geometry}, Commun.\ Math.\ Phys.\ {\bf117} (1988) 673-683\cr 
\q{Co4} &A.\ Connes, {\sl Reality and noncommutative geometry}, 
   J.\ Math.\ Phys.\ {\bf36} (1995) 6194-6231\cr
\q{Co5} &A.\ Connes, {\sl $C^*$-alg\`ebres et g\'eom\'etrie 
   diff\'erentielle}, C.R.\ Acad.\ Sci.\ Paris S\'er.\ A-B {\bf290} 
   (1980) 599-604\cr
\q{CoK} &A.\ Connes, M.\ Karoubi, {\sl Caract\`ere multiplicatif 
   d'un module de Fredholm}, $K$-Theory {\bf2} (1988) 431-463\cr
\q{DFR} &S.\ Doplicher, K.\ Fredenhagen, J.E.\ Roberts, {\sl The quantum 
   structure of space-time at the Planck scale and quantum fields}, 
   Commun.\ Math.\ Phys.\ {\bf172} (1995) 187-220\cr
\q{FG} &J.\ Fr\"ohlich, K.\ \gaw, {\sl Conformal field theory 
   and the geometry of strings}, CRM Proceedings and Lecture Notes 
   Vol.\ {\bf 7} (1994), 57-97\cr
\q{FGR1} &J.\ Fr\"ohlich, O.\ Grandjean, A.\ Recknagel, {\sl Supersymmetric 
   quantum theory and differential geometry}, Commun.\ Math.\ Phys.\ {\bf193} 
   (1998) 527-594\cr 
\q{FGR2} &J.\ Fr\"ohlich, O.\ Grandjean, A.\ Recknagel, {\sl Supersymmetric 
   quantum theory, non-commutative geometry, and gravitation}, Lecture 
   notes for the Les Houches 1995 summer school on ``Quantum Symmetries'', 
   A.\ Connes, K.\ \gaw\ (eds.)\cr 
\q{FGK} &G.\ Felder, K.\ \gaw, A.\ Kupiainen,  {\sl Spectra of 
   Wess-Zumino-Witten models with arbitrary simple groups}, 
   Commun.\ Math.\ Phys.\ {\bf117} (1988) 127-158\cr
\q{FW} &D.\ Friedan, P.\ Windey, {\sl Supersymmetric derivation of the 
   Atiyah-Singer index theorem and the chiral anomaly}, Nucl.\ Phys.\ 
   {\bf B235} (1984) 395-416\cr
\q{GKP} &H.\ Grosse, C.\ Klim\v cik, P.\ Pre\v snajder, {\sl Towards 
   finite quantum field theory in non-commutative geometry}, Int.\ J.\ Theor.\
   Phys.\ {\bf35} (1996) 231-244\cr
\q{GP} &H.\ Grosse, P.\ Pre\v snajder, {\sl The construction of
   non-commutative manifolds using coherent states}, Lett.\ Math.\ Phys.\ 
   {\bf28} (1993) 239-250\cr
\q{GPR} &A.\ Giveon, M.\ Porrati, E.\ Rabinovici, {\sl Target space duality in 
     string theory}, Phys.\ Rep.\ {\bf244} (1994) 77-202%, hep-th/9401139
   \cr
\q{Gr} &O.\ Grandjean, {\sl Non-commutative differential geometry}, Ph.D.\ 
   Thesis, ETH Z\"urich, July 1997\cr
\q{GSW} &M.B.\ Green, J.H.\ Schwarz, E.\ Witten, {\sl Superstring Theory I,II},
   Cambridge University Press 1987\cr
\q{HKLR} &N.J.\ Hitchin, A.\ Karlhede, U.\ Lindstrom, M.\ Rocek, {\sl
   Hyperk\"ahler metrics and supersymmetry}, 
   Commun.\ Math.\ Phys.\ {\bf108} (1987) 535-589\cr
\q{Ho} &J.\ Hoppe, {\sl Quantum theory of a massless relativistic 
   surface and a two-dimensional boundstate problem}, Ph.D.\ Thesis, MIT 1982;
   \quad {\sl Quantum theory of a relativistic surface}, in: Constraint's 
   theory and relativistic dynamics, G.\ Longhi, L.\ Lusanna (eds.), 
   Proceedings Florence 1986, World Scientific\cr
\q{Ja1} &A.\ Jaffe, A.\ Lesniewski, K.\ Osterwalder, {\sl Quantum
   $K$-theory I: The Chern character}, Commun.~Math.~Phys.~{\bf 118}
   (1988) 1-14\cr
\q{Ja2} &A.\ Jaffe, A.\ Lesniewski, K.\ Osterwalder, {\sl 
   On super-KMS functionals and entire cyclic
   cohomology}, $K$-theory~{\bf 2} (1989) 675-682\cr
\q{Ja3} &A.\ Jaffe, K.\ Osterwalder, {\sl Ward identities for non-commutative
   geometry}, Commun.~Math.~Phys.~{\bf 132} (1990) 119-130\cr
\q{Jac} &N.\ Jacobson, {\sl Basic Algebra II}, W.H.\ Freeman and Company  
   1985\cr
\q{Joy} &D.D.\ Joyce, {\sl Compact hypercomplex and quaternionic manifolds}, 
   J.\ Differ.\ Geom.\ {\bf35} (1992) 743-762; 
   {\sl Manifolds with many complex structures}, Q.\ J.\ Math.\ 
   Oxf.\ II.\ Ser.\ {\bf46} (1995) 169-184\cr
\q{Kar} &M.\ Karoubi, {\sl Homologie cyclique et K-th\'eorie}, 
   Soci\'et\'e Math\'ematique de France, Ast\'erisque {\bf 149} (1987)\cr 
\q{KL} &S.\ Klimek, A.\ Lesniewski, {\sl Quantum Riemann surfaces I. The 
   unit disc}, Commun.\ Math.\ Phys.\ {\bf146} (1992) 103-122; \quad
   {\sl Quantum Riemann surfaces II. The discrete series}, Lett.\ Math.\ 
   Phys.\ {\bf24} (1992) 125-139\cr 
\q{KS} &C.\ Klim\v cik, P.\ \v Severa, {\sl Dual non-abelian duality 
   and the Drinfeld double}, Phys.\ Lett.\ B {\bf351} (1995) 455-462\cr
\q{Ma} &J.\ Madore, {\sl The commutative limit of a matrix geometry},
   J.\ Math.\ Phys {\bf32} (1991) 332-335\cr
\q{Pol} &J.\ Polchinski,  {\sl Dirichlet branes and Ramond-Ramond 
   charges}, Phys.\ Rev.\ Lett.\ {\bf 75} (1995) 4724-4727;%, hep-th/9510017; 
   \quad {\sl TASI lectures on D-branes}, hep-th/9611050. \cr
\q{PS} &A.\ Pressley, G.\ Segal, {\sl Loop groups}, Clarendon Press 1986\cr  
\q{Ri} &M.\ Rieffel, {\sl Non-commutative tori -- a case study of 
   non-commutative differentiable manifolds}, Contemp.\ Math.\ {\bf105} 
   (1990) 191-211\cr
\q{Sw} &R.G.\ Swan, {\sl Vector bundles and projective modules}, 
   Trans.\ Amer.\ Math.\ Soc.\ {\bf105} 
   (1962) 264-277\cr
\q{Wi1} &E.\ Witten, {\sl Constraints on supersymmetry breaking}, 
   Nucl.\ Phys.\ {\bf B202} (1982) 253-316\cr
\q{Wi2} &E.\ Witten, {\sl Supersymmetry and Morse theory}, 
   J.\ Diff.\ Geom.\ {\bf17} (1982) 661-692\cr
\q{Wi3} &E.\ Witten, {\sl Non-abelian bosonization in two dimensions}, 
   Commun.\ Math.\ Phys.\ {\bf92} (1984) 455-472\cr
\q{Wi4} &E.\ Witten, {\sl Bound states of strings and D-branes}, 
 Nucl.\ Phys.\ {\bf B460} (1996) 335--350%, hep-th/9510135
  \cr
\q{Y} &{\sl Essays on mirror symmetry}, S.T.\ Yau (ed.), International 
   Press 1992\cr
}
\bye